\documentclass[journal]{IEEEtran}

\usepackage[utf8]{inputenc}

\usepackage{cite}
\usepackage{bm}
\usepackage{amsthm}
\usepackage{extarrows}
\usepackage{amssymb}
\usepackage{graphicx}
\usepackage{color}
\usepackage{enumerate}
\usepackage[hidelinks]{hyperref}  
\usepackage{bookmark} 
\graphicspath{{figure}}
\usepackage{setspace}
\usepackage{tikz}
\usetikzlibrary{matrix}
\usepackage{lipsum} 
\usepackage{framed} 
\usepackage{flushend}

\newcommand{\mr}{\mathrm}

\newcommand{\BE}{\begin{equation}}
\newcommand{\EE}{\end{equation}}
\newcommand{\BS}{\begin{subequations}}
\newcommand{\ES}{\end{subequations}}
\renewcommand{\bf}{\bm}

\DeclareMathOperator*{\argmin}{arg\,min}

\newtheorem{theorem}{Theorem}
\newtheorem{proposition}{Proposition}
\newtheorem{assumption}{Assumption}
\newtheorem{definition}{Definition}

\newtheorem{remark}{Remark}
\newtheorem{lemma}{Lemma}
\newtheorem{corollary}{Corollary}

\title{Sufficient-Statistic Memory AMP}
\author{\IEEEauthorblockN{Lei~Liu, \emph{Member, IEEE}, Shunqi~Huang,  Yuzhi Yang, \\ Zhaoyang Zhang, \emph{Senior Member, IEEE} and  Brian~M.~Kurkoski, \emph{Member, IEEE}}

\thanks{This article has been presented in part at the 2022 IEEE ISIT, Finland, \cite{SS_MAMPISIT}. The source code of this work is publicly available at 
\href{https://sites.google.com/site/leihomepage/research}{sites.google.com/site/leihomepage/research}.}

\thanks{Lei Liu, Yuzhi Yang and Zhaoyang Zhang are with the Zhejiang Provincial Key Laboratory of Information Processing, Communication and Networking, College of Information Science and Electronic Engineering, Zhejiang University, Hangzhou 310007, China. (e-mail: \{lei\_liu, yuzhi\_yang, ning\_ming\}@zju.edu.cn).}

\thanks{Shunqi Huang and Brian~M.~Kurkoski are with the School of Information Science, Japan Institute of Science and Technology (JAIST), Nomi 923-1292, Japan (e-mail: \{shunqi.huang, kurkoski\}@jaist.ac.jp).} 
}

\begin{document}

\maketitle

\begin{abstract}
Approximate message passing (AMP) type algorithms have been widely used in the signal reconstruction of certain large random linear systems. A key feature of the AMP-type algorithms is that their dynamics can be correctly described by state evolution. While state evolution is a useful analytic tool, its convergence is not guaranteed. To solve the convergence problem of the state evolution of AMP-type algorithms in principle, this paper proposes a sufficient-statistic memory AMP (SS-MAMP) algorithm framework under the conditions of right-unitarily invariant sensing matrices, Lipschitz-continuous local processors and the sufficient-statistic constraint (i.e., the current message of each local processor is a sufficient statistic of the signal vector given the current and all preceding messages). We show that the covariance matrices of SS-MAMP are L-banded and convergent, which is an optimal framework (from the local MMSE/LMMSE perspective) for AMP-type algorithms given the Lipschitz-continuous local processors. Given an arbitrary MAMP, we can construct an SS-MAMP by damping, which not only ensures the convergence of the state evolution, but also preserves the orthogonality, i.e., its dynamics can be correctly described by state evolution. As a byproduct, we prove that the Bayes-optimal orthogonal/vector AMP (BO-OAMP/VAMP) is an SS-MAMP. As a result, we recover two interesting properties of BO-OAMP/VAMP for large systems in the literature: 1) the covariance matrices are L-banded and are convergent, and 2) damping and memory are not needed (i.e., do not bring performance improvement). As an example, we construct a sufficient-statistic Bayes-optimal MAMP (SS-BO-MAMP) whose state evolution converges to the minimum (i.e., Bayes-optimal) mean square error (MSE) predicted by replica methods when it has a unique fixed point. In addition, the MSE of SS-BO-MAMP is not worse than the original BO-MAMP.  Finally, simulations are provided to support the theoretical results. 
\end{abstract}


\section{Introduction}
This paper studies the signal recovery of noisy linear systems: $\bf{y}=\bf{Ax}+\bf{n}$, where $\bf{y}\in \mathbb{C}^{M\times1}$ is the observed vector, $\bf{A}\in \mathbb{C}^{M\times N}$ a known matrix, $\bf{x}\in \mathbb{C}^{N\times1}$ the signal vector to be recovered, and $\bf{n}\in \mathbb{C}^{M\times1}$ a Gaussian noise vector. The goal is to recover $\bf{x}$ using  $\bf{y}$ and $\bf{A}$. This paper focuses on large-size systems with $M,N\to\infty$ and a fixed $\delta=M/N$. For non-Gaussian $\bf{x}$, optimal recovery is in general NP-hard \cite{Micciancio2001,verdu1984_1}.   

\subsection{Background}
 Approximate message passing (AMP) is an approach for the signal recovery of high-dimensional noisy linear systems  \cite{Donoho2009}. AMP has three crucial properties. First, the complexity of AMP is as low as $\mathcal{O}(MN)$ per iteration since it suppresses linear interference using a simple matched filter. Second, the dynamics of AMP can be tracked by state evolution  \cite{Bayati2011}. Most importantly, it was proved that AMP is minimum mean square error (MMSE) optimal for uncoded signaling \cite{Barbier2017arxiv, Reeves_TIT2019}  and information-theoretically (i.e., capacity) optimal for coded signaling \cite{LeiTIT}. However, AMP is restricted to independent and identically distributed (IID) $\bf{A}$ \cite{Rangan2015, Vila2014}, which limits the application of AMP. To solve the weakness of AMP, unitarily-transformed AMP (UTAMP) was proposed in \cite{UTAMPa, UTAMPb}, which works well for correlated matrices $\bf{A}$. Orthogonal/vector AMP (OAMP/VAMP) was established for right-unitarily-invariant matrices \cite{Ma2016,Rangan2016}. The correctness of the state evolution of OAMP/VAMP was conjectured in \cite{Ma2016} and proved in \cite{Rangan2016,Takeuchi2017}. The replica MMSE optimality and the replica information-theoretical optimality of OAMP/VAMP were proved respectively in \cite{Ma2016, Tulino2013, Kabashima2006} and \cite{Lei_cap_oampISIT, Lei_cap_oamp}, when the state evolution has a unique fixed point. Recently, the correctness of the heuristic replica method was rigorously proved for some sub-classes of right-unitarily invariant matrices \cite{ Barbier2018b, Li2022random}.  However, the Bayes-optimal OAMP/VAMP requires linear MMSE (LMMSE) estimation whose complexity is as high as $\mathcal{O}(M^2N+M^3)$, which is prohibitive for large-scale systems.
  
To solve the weakness of OAMP/VAMP, low-complexity convolutional AMP (CAMP) was proposed in  \cite{Takeuchi2020CAMP}. However, CAMP may fail to converge, especially for ill-conditioned matrices. Apart from that, the AMP  was also extended to solve the Thouless-Anderson-Palmer (TAP) equations in Ising models for unitarily-invariant matrices \cite{Opper2016}. The results in \cite{Opper2016} were justified via state evolution in \cite{Fan2020arxiv}. More recently, memory AMP (MAMP) was established for unitarily-invariant matrices \cite{Lei2020MAMPTIT}. MAMP has a comparable complexity to AMP since it uses a low-complexity long-memory matched filter to suppress linear interference. In addition, the dynamics of MAMP can be correctly predicted by state evolution. More importantly, the convergence of the state evolution of MAMP is guaranteed by optimized damping. Apart from that, MAMP achieves MMSE optimality (predicted by replica method) when its state evolution has a unique fixed point.

\subsection{Motivation and Related Works}  
A key feature of the AMP-type algorithms \cite{Donoho2009,Ma2016,Rangan2016,Takeuchi2020CAMP,Lei2020MAMPTIT, Rangan2010, Schniter2016, Tian2021GMAMP, Opper2016, Ramji2021RIAMP} is that their dynamics can be rigorously described by state evolution \cite{Rangan2016,Takeuchi2017,Bayati2011, Fan2020arxiv, Pandit2019}. While state evolution is a useful analytic tool, its convergence is not guaranteed \cite{Takeuchi2021OAMP, Lei2016TWC,Lei2019TWC,Gerbelot,RamjiPCA}. Therefore, it is desired to find a new technique to guarantee the convergence of the state evolution of the AMP-type algorithms.

Damping is generally used to improve the convergence of the AMP-type algorithms for finite-size systems \cite{Schniter2019dAMP,  Schniter2021dvamp, Takeuchi2019damp}. However, improper damping may break the orthogonality and Gaussianity, which may result in that the dynamics of damped AMP-type algorithms cannot be described by state evolution anymore \cite{Takeuchi2020CAMP}. Moreover, conventional damping in the literature was heuristic and empirical, and performed on the messages in current and last iterations. In \cite{Lei2020MAMPTIT}, the authors first proposed an analytically optimized vector damping for Bayes-optimal MAMP (BO-MAMP) based on the current and all the preceding messages, which not only solves the convergence problem of MAMP, but also preserves orthogonality and Gaussianity, i.e., the dynamics of damped BO-MAMP can be correctly described by state evolution. Recently, the damping optimization in \cite{Lei2020MAMPTIT} has been used to analyze the convergence of the state evolution of Bayes-optimal OAMP/VAMP (BO-OAMP/VAMP) in \cite{Takeuchi2021OAMP} from a sufficient-statistic perspective. The works in \cite{Lei2020MAMPTIT, Takeuchi2021OAMP}  proposed a novel principle to solve the convergence of the state evolution of AMP-type algorithms.

\subsection{Contributions}  
In this paper, we try to answer a fundamental question about AMP-type algorithms: 
\begin{itemize}
    \item What is the optimal framework (from the local MMSE/LMMSE perspective) for developing AMP-type algorithms given the local processors?
\end{itemize}
Motivated by the damping optimization in \cite{Lei2020MAMPTIT} and its statistical interpretation in \cite{Takeuchi2021OAMP}, this paper proposes a sufficient-statistic memory AMP (SS-MAMP) to solve the convergence problem of state evolution in AMP-type algorithms in principle. 
MAMP uses the current message and a history of preceding messages. The central idea is that the current message of each local processor is a sufficient statistic of $\bf{x}$ given the current and all preceding messages; it is a sufficient statistic in the MMSE/LMMSE sense. Thus, SS-MAMP is a MAMP \cite{Lei2020MAMPTIT} under a sufficient-statistic condition.
We show that SS-MAMP has the following interesting properties.
\begin{itemize}
    \item SS-MAMP can be  defined as a MAMP with L-banded covariance matrices (see Definition \ref{Def:L-baned}) whose elements in each ``L band" are the same. As a result, the original covariance-based state evolution is reduced to a variance-based state evolution. Furthermore, some interesting properties are derived for the determinant and inverse of the L-banded covariance matrices.
    
    \item The covariances in SS-MAMP are monotonically decreasing and converge respectively to a certain value with the increase of the number of iterations. Hence, the state evolution of SS-MAMP is definitely convergent.
    
    \item Damping is not needed, i.e., damping does not bring MSE improvement to SS-MAMP. 
    
    \item Memory is not needed for the Bayes-optimal local processor in SS-MAMP, i.e., joint estimation with preceding messages does not bring MSE improvement to the Bayes-optimal local processor in SS-MAMP.  
\end{itemize}

Given an arbitrary MAMP algorithm, we can construct an SS-MAMP algorithm using the optimal damping in \cite{Lei2020MAMPTIT}. The constructed SS-MAMP has the following interesting properties.
\begin{itemize}
    \item The MMSE/LMMSE sufficient-statistic property (i.e., L-banded and convergent covariance matrices) is guaranteed by the optimal damping at each local processor, which preserves the orthogonality and Gaussianity of the original MAMP. Hence, the dynamics of  the constructed SS-MAMP  can be rigorously described by state evolution.
    
    
    \item The MSE of the constructed SS-MAMP with optimal damping is not worse than the original MAMP.
\end{itemize}
 
As a byproduct, we prove that the BO-OAMP/VAMP is an SS-MAMP using the orthogonality of local MMSE/LMMSE estimators. Therefore, BO-OAMP/VAMP inherits all the properties of SS-MAMP: 1) The covariance matrices of BO-OAMP/VAMP are L-banded and  convergent; 2) Damping is not needed, i.e., damping does not bring MSE improvement; 3) Memory is not needed, i.e., step-by-step non-memory local estimation is optimal and jointly local estimation with preceding messages does not bring MSE improvement. These recover the main statements in \cite{Takeuchi2021OAMP}.

As an example, based on the BO-MAMP in \cite{Lei2020MAMPTIT}, we construct a sufficient-statistic BO-MAMP (SS-BO-MAMP) using optimal damping for outputs of the linear estimator. We show that SS-BO-MAMP is an SS-MAMP, and thus inherits all the properties of SS-MAMP. The state evolution of SS-BO-MAMP is derived. We show that the MSE of SS-BO-MAMP is not worse than the original BO-MAMP in \cite{Lei2020MAMPTIT}, and the state evolution of SS-BO-MAMP converges to the same fixed point as that of BO-OAMP/VAMP. Hence, SS-BO-MAMP is replica Bayes optimal when its state evolution has a unique fixed point. Numerical results are provided to verify that SS-BO-MAMP outperforms the original BO-MAMP in \cite{Lei2020MAMPTIT}.  

Part of the results in this paper has been published in \cite{SS_MAMPISIT}. In this paper, we additionally give detailed proofs, illustrate that BO-OAMP/VAMP is sufficient-statistic,  develop a sufficient-statistic Bayes-optimal memory AMP (SS-BO-MAMP), and provide some numerical results to verify the validity of the theoretical results in this paper.

\subsection{Connection to Existing Work}
\emph{Connection to the BO-MAMP in \cite{Lei2020MAMPTIT}}: The optimized damping technique is used in \cite{Lei2020MAMPTIT} to solve the convergence problem of the state evolution of  BO-MAMP, which is a specific instance of MAMP. However, how to solve the convergence problem of the state evolution for arbitrary MAMP algorithms is still unclear. In this paper, based on the optimized damping technique in \cite{Lei2020MAMPTIT} and the sufficient-statistic technique motivated by \cite{Takeuchi2021OAMP}, we propose an optimal SS-MAMP framework (from the local MMSE/LMMSE perspective) for AMP-type algorithms and solve the convergence problem of the state evolution for arbitrary MAMP algorithms in principle. Additionally, we develop a new SS-BO-MAMP that outperforms the BO-MAMP in \cite{Lei2020MAMPTIT}, demonstrating the suboptimality of the BO-MAMP in \cite{Lei2020MAMPTIT}. 
 
 \emph{Connection to the BO-LM-MP in \cite{Takeuchi2021OAMP}}: The convergence of state evolution recursions for Bayes-optimal long-memory message-passing (BO-LM-MP) is proved in \cite{Takeuchi2021OAMP} via a new statistical interpretation of the optimized damping technique in \cite{Lei2020MAMPTIT}. However, it only is limited to BO-LM-MP and BO-OAMP/VAMP which consist of the LMMSE filter and the Bayes-optimal denoiser. In this paper, we solve the convergence problem of the state evolution of arbitrary MAMP-type algorithms, i.e., the linear estimator is not limited to the LMMSE filter, and the denoiser is not limited to Bayes optimal processors, which includes but is not limited to the BO-LM-MP in \cite{Takeuchi2021OAMP} and BO-OAMP/VAMP in \cite{Ma2016,Rangan2016}. In Section \ref{Sec:BO-OAMP/VAMP}, we prove that BO-OAMP/VAMP is an instance of SS-MAMP based on the orthogonality of local MMSE/LMMSE estimators, and thus recover the main statements in \cite{Takeuchi2021OAMP}, i.e., damping and memory are not needed in BO-OAMP/VAMP, and the covariance matrices in BO-OAMP/VAMP are L-banded and convergent. Most importantly, the proposed SS-BO-MAMP in Section \ref{Sec:SS-BO-MAMP} is a powerful example that goes beyond the findings in \cite{Takeuchi2021OAMP} because SS-BO-MAMP is based on a Bayes-suboptimal long-memory matched filter  rather than the Bayes-optimal LMMSE linear estimator discussed in \cite{Takeuchi2021OAMP}. In the relevant sections, the similarities and differences between the specific results in this paper and those in \cite{Takeuchi2021OAMP} are clarified in detail.
 
 \subsection{Other Related Work}
 To solve a generalized linear model  $\bf{y}=Q(\bf{Ax}, \bf{n})$, where $Q(\cdot)$ can be non-linear such as clipping and quantization, a low-complexity generalized AMP (GAMP) \cite{Rangan2010} was developed for  $\bf{A}$. Generalized VAMP (GVAMP) was proposed for the generalized linear model with  unitarily-invariant $\bf{A}$ \cite{Schniter2016}. The dynamics of GVAMP can be predicted by state evolution, based on which the MMSE optimality (predicted by replica methods) of GVAMP is proved in \cite{Pandit2019}. The information-theoretic limit (i.e., maximum achievable rate) of GVAMP was studied in \cite{Lei2023GOAMP}. Like OAMP/VAMP, GVAMP requires high-complexity LMMSE. Therefore,  a low-complexity  generalized MAMP (GMAMP) was proposed for the generalized linear model with  unitarily-invariant $\bf{A}$ \cite{Tian2021GMAMP}. The dynamics of GMAMP can be described by state evolution. In addition, it was proved that the state evolution fixed point of GMAMP coincides with the MMSE predicted by replica methods. Based on the AMP framework in \cite{Fan2020arxiv, Opper2016}, a rotationally invariant AMP was designed in \cite{Ramji2021RIAMP} for the generalized linear model with unitarily-invariant transformation matrices.

The long-memory AMP algorithms can be classified into two categories. First, the CAMP   \cite{Takeuchi2020CAMP}, long-memory AMP  \cite{Opper2016,Fan2020arxiv} and rotationally invariant AMP \cite{Ramji2021RIAMP} consist of NLEs and a matched filter with Onsager correction terms, whose  structure is similar to that of AMP \cite{Donoho2009} or GAMP \cite{Rangan2010}. Second, the MAMP  \cite{Lei2020MAMPTIT} and GMAMP  \cite{Tian2021GMAMP} were consisted of orthogonal NLEs and an orthogonal long-memory matched filter, whose structure is similar to that of OAMP/VAMP \cite{Ma2016,Rangan2016} or GVAMP \cite{Schniter2016}.

AMP-type algorithms have been successfully extended to handle various statistical signal processing problems such as low-rank matrix recovery \cite{Kabashima2016}, phase retrieval \cite{Schniter2015}, community detection in graphs \cite{Deshpande2016}, general matrices in linear layers \cite{Manoel2017}, and the multiple measurement vector (MMV) problem \cite{Yuwei2018, Yiyao2021,Yiyao2023MMV}. An overview of AMP's applications was provided in \cite{FengAMP2022}. Moreover, these techniques were found in a wide range of engineering applications such as imaging \cite{Fletcher2014}, deep learning \cite{Emami2020}, multiple-input multiple-output detection \cite{Jeon2015} and coded systems \cite{MaTWC, Rush2017, Barbier2017, LeiTIT}.  The advantages of AMP-type algorithms over conventional turbo receivers \cite{Lei20161b, Yuan2014, Wang1999} are demonstrated in \cite{MaTWC, LeiTIT, Lei_cap_oampISIT, Lei_cap_oamp}.

\subsection{Notation}   
Bold upper (lower) letters denote matrices (column vectors). $(\cdot)^{\rm T}$, $(\cdot)^*$  and $(\cdot)^{\rm H}$ denote transpose, conjugate and conjugate transpose,  respectively. $\mathbb{E}\{\cdot\}$ denotes expectation. We let ${\mr{var}}\{\bf{x}|\bf{Y}\} = \tfrac{1}{N} \|\bf{x}- \mathbb{E}\{\bf{x}|\bf{Y}\} \|^2 $ and $\overline{\mr{var}}\{\bf{x}|\bf{Y}\} =  \mathbb{E}\big\{{\mr{var}}\{\bf{x}|\bf{Y}\}   \big\}$. We say that $x=x_{\rm Re}+  x_{\rm Im}\!\cdot\!{\rm i}$, where ${\rm i} =\sqrt{-1}$ represents the imaginary unit, is circularly-symmetric complex Gaussian  if $x_{\rm Re}$ and $x_{\rm Im}$ are two independent Gaussian distributed random variables with $\mathbb{E}\{x_{\rm Re}\} = \mathbb{E}\{x_{\rm Im}\} = 0$ and $ \overline{\mr{var}}\{x_{\rm Re}\} = \overline{\mr{var}}\{x_{\rm Im}\}$. We define $\overline{\mr{var}}\{x\} \equiv \overline{\mr{var}}\{x_{\rm Re}\} + \overline{\mr{var}}\{x_{\rm Im}\}$.  $\mathcal{CN}(\bf{\mu},\bf{\Sigma})$ denotes the complex Gaussian distribution of a vector with mean vector $\bf{\mu}$ and covariance matrix $\bf{\Sigma}$. $\bf{A}_{M\times N}=[a_{i,j}]_{M\times N}$, $\langle\bf{A}_{M\times N} | \bf{B}_{M\times N} \rangle \equiv \frac{1}{N}\bf{A}^{\rm H}_{M\times N}\bf{B}_{M\times N}$, $ \langle \bf{a}\rangle =\frac{1}{N}\sum_1^{N} a_n$ represents the arithmetic mean of the elements, $\|\cdot\|$ denotes the $\ell_2$-norm,  $|\cdot|$ the absolute value or modulus, $X\sim Y$ that $X$ follows the distribution $Y$, and $\overset{\rm a.s.}{=}$ almost sure equivalence (as $N\to \infty$ unless otherwise specified). ${\rm diag}\{\cdot\}$, ${\rm det}\{\cdot\}$ and ${\rm tr}(\cdot)$ respectively denote the diagonal vector, the determinant and the trace of a matrix.  ${\rm min}(\cdot)$ and ${\rm max}(\cdot)$ denotes the minimum and maximum value of a set. $\bf{I}$, $\bf{1}$ and $\bf{0}$ are identity matrix, all-one matrix (or vector), and zero matrix (or vector) with the proper size, respectively. We let$\bf{A}_{I:J}=[\bf{A}_I,\bf{A}_{I+1},\dots, \bf{A}_J]$. A matrix is said to be row-wise IID and column-wise jointly Gaussian if it has IID rows and each row is jointly Gaussian.

\emph{\textbf{Unbiased LMMSE:}} Let $\bf{X}_{t}=[\bf{x}_1,\dots,\bf{x}_t]$ be the estimates of $\bf{x}$ and $\bf{y}=\bf{Ax}+\bf{n}$ be the noisy linear observation of $\bf{x}$. An unbiased LMMSE estimation is defined  as:
\BS\label{Eqn:LMMSE}\BE
    {\rm LMMSE} \{\bf{x}|\bf{y}, \bf{x}_{t}\}= \bf{Q}^{\star} \bf{y} + \textstyle\sum_{i=1}^t \bf{P}^{\star}_{i} \bf{x}_i,
\EE
where \vspace{-2mm}
\BE
    [\bf{Q}^{\star}\!, \bf{P}^{\star}_{1:t}] \!= \!\!\!\! \argmin_{[\bf{Q}, \bf{P}_{1: t}], \bf{QA} + \!\sum_{i=1}^t\!\! \bf{P}_{i}=\bf{I}} \!\!\!\!\mathbb{E}\big\{\! \|\bf{Q} \bf{y} \!+ \!\textstyle\sum_{i=1}^t \!\!\bf{P}^{\star}_{i} \bf{x}_i \!- \!\bf{x}\|^2\!\big\}.
\EE\ES
The MSE of the LMMSE estimation in \eqref{Eqn:LMMSE} is given by
\BE\label{Eqn:V_LMMSE}
    {\rm lmmse} \{\bf{x}|\bf{y}, \bf{x}_{t}\} = \tfrac{1}{N}  \|{\rm LMMSE} \{\bf{x}|\bf{y}, \bf{x}_{t}\} -\bf{x}\|^2,
\EE
and the expectation MSE is given by     
\BE\label{Eqn:EV_LMMSE}
     \overline{\rm lmmse} \{\bf{x}|\bf{y}, \bf{x}_{t}\} = \mathbb{E}\big\{\, {\rm lmmse} \{\bf{x}|\bf{y}, \bf{x}_{t}\} \big\},
\EE

Intuitively, the ${\rm LMMSE}\{\cdot\}$,  ${\rm lmmse}\{\cdot\}$ and $\overline{\rm lmmse}\{\cdot\}$ above can be viewed as a type of \emph{a posteriori} estimation $\mathbb{E}\{\cdot\}$ and \emph{a posteriori} variance ${\rm var}\{\cdot\}$ and the expected  variance $\overline{\rm var}\{\cdot\}$, respectively, under the condition that the \emph{a priori} distribution $p(\bf{x})$ of $\bf{x}$ is unknown. We will show that ${\rm LMMSE}\{\cdot\}$,  ${\rm lmmse}\{\cdot\}$ and $\overline{\rm lmmse}\{\cdot\}$ inherits many properties of $\mathbb{E}\{\cdot\}$,  ${\rm var}\{\cdot\}$ and $\overline{\rm var}\{\cdot\}$, such as the orthogonality and sufficient-statistic properties. See Section \ref{Sec:SS} for more details.

 \subsection{Paper Outline}
This paper is organized as follows. Section \ref{Sec:Pre} gives the preliminaries including problem formulation and memory AMP (MAMP). Section \ref{Sec:SS-L-banded} introduces L-banded matrices. Section \ref{Sec:SS} introduces the sufficient statistic and its properties. Section \ref{Sec:SS-MAMP} proposes a sufficient-statistic memory AMP (SS-MAMP). We prove that BO-OAMP/VAMP is sufficient-statistic in Section \ref{Sec:BO-OAMP/VAMP}. A sufficient-statistic Bayes-optimal memory AMP (SS-BO-MAMP) is developed in Section \ref{Sec:SS-BO-MAMP}.  Numerical results are shown in Section \ref{Sec:Simulation}.

\section{Preliminaries}\label{Sec:Pre}
In this section, we first give the problem formulation and assumptions. Then, we briefly introduce the memory AMP and its properties.

\begin{figure}[t]
  \centering 
  \includegraphics[width=5cm]{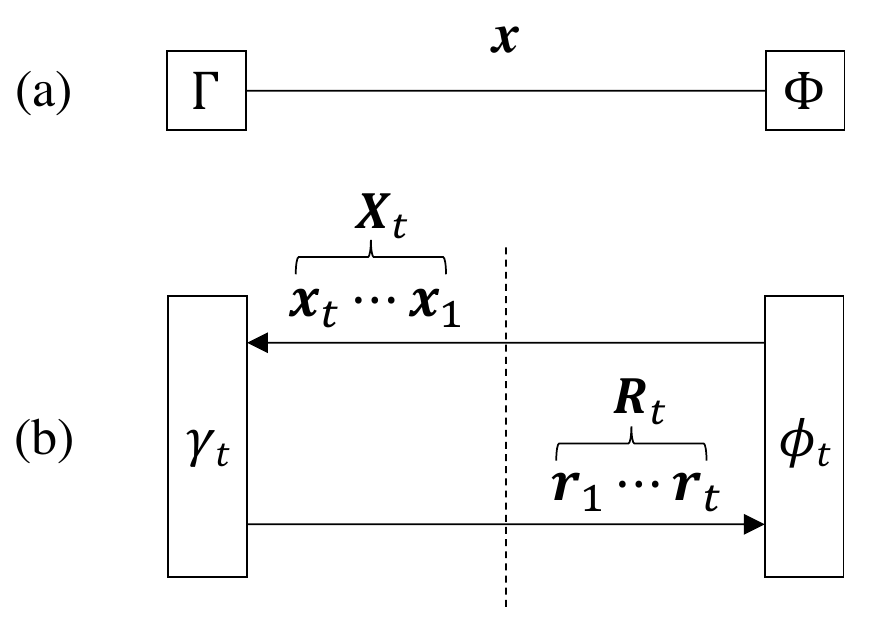}\\ 
  \caption{Graphic illustrations for (a) a system model with two constraints $\Gamma$ and $\Phi$, and (b) a  memory iterative process involving two local processors $\gamma_t$ (for $\Gamma$) and $\phi_t$ (for $\Phi$) \cite{Lei2020MAMPTIT}.}\label{Fig:MAMP}
\end{figure}

\subsection{Problem Formulation} 
Fig. \ref{Fig:MAMP}(a) illustrates the noisy linear system with two constraints: 
\BS\label{Eqn:unitary_sys}\begin{align}
&\Gamma: \quad  \bf{y}=\bf{Ax}+\bf{n},\\
&\Phi: \quad x_i\sim P_x, \;\;\forall i,
\end{align}\ES 
where $\bf{n}\sim \mathcal{CN}(\bf{0},\sigma^2\bf{I})$, $\bf{x}$ is a length-$N$ IID vector following a distribution $P_{x}$, and $\bf{A}$ is an $M\times N$ measurement matrix. In addition, we assume that $M,N\to\infty$ with a fixed $\delta=M/N$.

\begin{assumption}\label{ASS:x}
The entries of $\bf{x}$ are IID with zero mean. The variance of $\bf{x}$ is normalized, i.e., $\lim\limits_{N\to\infty}\frac{1}{N}\mathbb{E}\{\|\bf{x}\|^2\}\overset{\rm a.s.}{=}1$, and the $(2+\epsilon)$th moments of  $\bf{x}$ are finite for some $\epsilon$. 
\end{assumption}
 
 \begin{assumption}\label{ASS:A}
  Let the singular value decomposition (SVD) of $\bf{A}$ be $\bf{A}\!=\!\bf{U}\bf{\Sigma} \bf{V}^{\rm H}$, where $\bf{U}\!\in\! \mathbb{C}^{M\times M}$ and $\bf{V}\!\in\! \mathbb{C}^{N\times N}$ are unitary matrices, and $\bf{\Sigma} $ is a rectangular diagonal matrix.  We assume that $M,\!N\!\to\!\infty$ with a fixed $\delta\!=\!\!M\!/\!N$, and $\bf{A}$ is known and is right-unitarily-invariant, i.e., $\bf{U}\bf{\Sigma}$ and $\bf{V}$ are independent, and $\bf{V}$ is Haar distributed (or equivalently, isotropically random). Furthermore, the empirical eigenvalue distribution of $\bf{AA}^{\rm H}$ converges almost surely to a compactly supported deterministic distribution with unit first moment in
the large system limit, i.e., $ \tfrac{1}{N} {\rm tr}\{\bf{A}\bf{A}^{\rm H}\}\overset{\rm a.s.}{=}1$  \cite{Takeuchi2020CAMP}.  
\end{assumption}\vspace{-3mm}

 \subsection{Memory AMP}\label{Sec:MIP}
\begin{definition}[Memory  Iterative Process]\label{Def:MIP}
 A memory iterative process consists of a memory linear estimator (MLE) and a memory non-linear estimator (MNLE) defined as:  Starting with $t=1$, 
\BS\label{Eqn:MIP}\begin{alignat}{2}
{\rm MLE:}&& \quad \quad  \bf{r}_t &= \gamma_t\!\left(\bf{y},\bf{X}_{t}\right)=  \bf{\mathcal{Q}}_t\bf{y} + \textstyle\sum_{i=1}^t{\bf{\mathcal{P}}}_{t,i} \bf{x}_i,\label{Eqn:MIP_LE}\\
{\rm MNLE:}&& \quad   \bf{x}_{t + 1} &= \phi_t\!\left(\bf{R}_t\right),
\end{alignat}\ES 
 where $\bf{X}_t=[\bf{x}_1\dots\,\bf{x}_t]$, $\bf{R}_t=[\bf{r}_1\dots\,\bf{r}_t]$, $\bf{\mathcal{Q}}_t\bf{A}$ and $\bf{\mathcal{P}}_{t,i}$ are polynomials in $\bf{A}^{\rm H}\bf{A}$. Without loss of generality, we assume that the norms of $\bf{\mathcal{Q}}_t$ and $\{\bf{\mathcal{P}}_{t,i}\}$ are finite, and $\bf{\mathcal{Q}}_t + \textstyle\sum_{i=1}^t\bf{\mathcal{P}}_{t,i}=\bf{I}$ (i.e., $\gamma_t$ is unbiased). Hence, $\gamma_t\!\left(\cdot\right)$ in \eqref{Eqn:MIP_LE} is Lipschitz-continuous \cite{Lei2020MAMPTIT}.
\end{definition} 

Fig. \ref{Fig:MAMP}(b) gives a graphical illustration of a memory  iterative process.  The memory  iterative process is degraded to the conventional non-memory iterative process if $ \gamma_t\!\left(\bf{X}_t\right)= {\gamma}_t \left(\bf{x}_t\right) $ and $\phi_t\!\left(\bf{R}_t\right)={{\phi}}_t \left(\bf{r}_t\right)$. Let
\BS\label{Eqn:errors}\begin{align}
 \bf{R}_t &= \bf{X} + \bf{G}_t, \\
 \bf{X}_t &= \bf{X} + \bf{F}_t, 
\end{align}  \ES
where $\bf{X}=\bf{x}\cdot\bf{1}^{\rm T}$, $\bf{1}$ is an all-ones vector with the proper size,  and  $\bf{G}_t=[\bf{g}_1\dots\,\bf{g}_t]$ and $\bf{F}_t=[\bf{f}_1\dots\,\bf{f}_t]$ indicate the estimation errors with zero means and the covariances as follows:
\BS\label{Eqn:v_gamma_phi} 
\begin{align}
    \bf{V}_t^{\gamma} &\equiv  \langle \bf{G}_t |  \bf{G}_{t}\rangle,\label{Eqn:v_gamma}\\
   \bf{V}_t^{\phi}&\equiv  \langle \bf{F}_t | \bf{F}_{t} \rangle.\label{Eqn:v_bar_phi}
\end{align} 
\ES    
Let $\{v^{\gamma}_{i,j}\}$ and $\{v^{\phi}_{i,j}\}$ be the elements of $ \bf{V}_t^{\gamma}$ and  $\bf{V}_t^{\phi}$, respectively. We define the diagonal vectors of the covariance matrices as
\BS\label{Eqn:vars}\begin{align}
  \bf{v}_t^{\gamma}& \equiv[v^{\gamma}_{1}\dots\,  v^{\gamma}_{t}]^{\rm T},\\
  \bf{v}_t^{\phi}& \equiv[v^{\phi}_{1}\dots\,  v^{\phi}_{t}]^{\rm T},
\end{align}\ES 
where $v^{\gamma}_{i}=v^{\gamma}_{i,i}$ and  $v^{\phi}_{i}=v^{\phi}_{i,i}$.

\begin{definition}[Memory AMP \cite{ Lei2020MAMPTIT}]\label{Def:MAMP}
The memory  iterative process in \eqref{Eqn:MIP} is said to be a memory AMP (MAMP) if the following orthogonal constraint holds for $t\ge 1$:
\BS\label{Eqn:Orth_MAMP}\begin{align} 
  \langle \bf{g}_t | \bf{x} \rangle &\overset{\rm a.s.}{=}  0,  \\
 \langle \bf{g}_t |\bf{F}_t\rangle   & \overset{\rm a.s.}{=}  \bf{0},\\
  \langle \bf{f}_{t+1} | \bf{G}_t \rangle   & \overset{\rm a.s.}{=} \bf{0}. 
 \end{align}\ES      
\end{definition}

The following lemma shows the asymptotically IID Gaussian property of MAMP.

\begin{lemma} [Asymptotically IID Gaussian]\label{Lem:IIDG_MIP}
 Suppose that Assumptions \ref{ASS:x}-\ref{ASS:A} hold, $\tilde{\gamma}_t(\bf{F}_t, \bf{n}, \bf{\lambda})\!\!:\! \!\mathbb{C}^{N\times (t+2)}\!\!\to\! \! \mathbb{C}^N$ is a Lipschitz-continuous function and $\tilde{\phi}_t(\bf{G}_t,\bf{x})\!\!:\!\! \mathbb{C}^{N\times (t+1)}\!\!\to\!\!  \mathbb{C}^N$ is a separable-and-Lipschitz-continuous function. Let $\bar{\bf{G}}_t=[\bar{\bf{g}}_1, \dots, \bar{\bf{g}}_t]$, $\bar{\bf{F}}_t=[\bar{\bf{f}}_1, \dots, \bar{\bf{f}}_t]$, and $\bf{\lambda}$ be the vector of eigenvalues of $\bf{A}^{\rm H}\bf{A}$. Then, we have \cite[Theorem 1]{Takeuchi2020CAMP} 
\BS\label{Eqn:IIDG_MIP}
\begin{align}
\big\langle  \tilde{\gamma}_t(\bf{F}_t, \bf{n}, \bf{\lambda})\big\rangle  & \overset{\rm a.s.}{=} \mathbb{E}\big\{\big\langle \tilde{\gamma}_t(\bar{\bf{F}}_t, \bar{\bf{n}}, \bf{\lambda}) \big\rangle\!\big\},\\ 
\big\langle  \tilde{\phi}_t(\bf{G}_t,\bf{x}) \big\rangle  & \overset{\rm a.s.}{=} \mathbb{E}\big\{\big\langle \tilde{\phi}_t(\bar{\bf{G}}_t,\bar{\bf{x}}) \big\rangle\!\big\}, 
\end{align}\ES 
where $\bar{\bf{n}}$ denotes the noise sampled from the same distribution as that of $\bf{n}$, $\bar{\bf{x}}$ denotes the signal sampled from the same distribution as that of $\bf{x}$, and $\bar{\bf{F}}_t$ and $\bar{\bf{G}}_t$ are row-wise IID, column-wise jointly Gaussian and independent of $\bar{\bf{n}}$ and $\bar{\bf{x}}$, i.e., for $i=1,\dots t$, 
  \BS\begin{align}
        [\bar{f}_{i,1},\dots \bar{f}_{i,t}] \overset{\rm IID}{\sim} \mathcal{CN}{(\bf{0}, \bf{V}_t^\phi)},\\
        [\bar{g}_{i,1},\dots \bar{g}_{i,t}] \overset{\rm IID}{\sim} \mathcal{CN}{(\bf{0}, \bf{V}_t^\gamma)}.
    \end{align}\ES 
\end{lemma} 

If we let $\tilde{\gamma}_t$ and $\tilde{\phi}_t$ be the performance measurement function of $\{{\gamma}_t\}$ and $\{{\phi}_t\}$,  we obtain the following state evolution of MAMP Lemma \ref{Lem:IIDG_MIP}.

\begin{lemma} [Asymptotically IID Gaussian]\label{Lem:SE_MIP}
  Suppose that Assumptions \ref{ASS:x}-\ref{ASS:A} hold. For MAMP with the orthogonality in \eqref{Eqn:Orth_MAMP},  the $\{{\gamma}_t(\cdot)\}$ in \eqref{Eqn:MIP_LE} and separable-and-Lipschitz-continuous $\{\phi_t(\cdot)\}$ \cite{Berthier2020}, the covariance matrices can be calculated by the following state evolution recursion \cite[Theorem 1]{Takeuchi2020CAMP}: $\forall 1\!\le \!t'\!\leq\! t$, 
\BS\label{Eqn:SE_MIP}  \begin{align}
&v_{t,t'}^{\gamma} \! \overset{\rm a.s.}{=} \mathbb{E}\big\{\!\big\langle \gamma_t \big(\bar{\bf{y}},\bar{\bf{X}} \!+\!\bar{\bf{G}}_t\big)\!-\!\bar{\bf{x}}\big|\gamma_{t'} \big(\bar{\bf{y}},\bar{\bf{X}}\!+\!\bar{\bf{G}}_{t'}\big)\!-\!\bar{\bf{x}}\big\rangle\!\big\},\\ 
&v_{t+1,{t'}+1}^{\phi}  \!\overset{\rm a.s.}{=}  \mathbb{E}\big\{ \! \big\langle \phi_t \big(\bar{\bf{X}}\!+\!\bar{\bf{F}}_t \big) \!- \bar{\bf{x}} \big| \phi_{t'} \big(\bar{\bf{X}}\!+\!\bar{\bf{F}}_{t'} \big) \!-\bar{\bf{x}}\big\rangle\!\big\},  
\end{align}\ES 
where $\bar{\bf{X}}=\bar{\bf{x}}\cdot\bf{1}^{\rm T}$,  $\bar{\bf{y}} = \bf{A}\bar{\bf{x}} + \bar{\bf{n}}$, $\bar{\bf{n}}$ denotes the noise sampled from the same distribution as that of $\bf{n}$, $\bar{\bf{x}}$ denotes the signal sampled from the same distribution as that of $\bf{x}$, and $\bar{\bf{F}}_t$ and $\bar{\bf{G}}_t$ are row-wise IID, column-wise jointly Gaussian and independent of $\bar{\bf{n}}$ and $\bar{\bf{x}}$, i.e., for $i=1,\dots t$, 
  \BS\begin{align}
        [\bar{f}_{i,1},\dots \bar{f}_{i,t}] \overset{\rm IID}{\sim} \mathcal{CN}{(\bf{0}, \bf{V}_t^\phi)},\\
        [\bar{g}_{i,1},\dots \bar{g}_{i,t}] \overset{\rm IID}{\sim} \mathcal{CN}{(\bf{0}, \bf{V}_t^\gamma)}.
    \end{align}\ES 
\end{lemma}
 
Note that we can not strictly claim that the error matrices $\bf{G}_t$ and $\bf{F}_t$ are row-wise IID and column-wise jointly Gaussian, which is too strong. However, as it is shown in Lemma \ref{Lem:SE_MIP}, when evaluating the covariance matrices of the MAMP iterates, we can replace $\bf{G}_t$ and $\bf{F}_t$ by row-wise IID and column-wise jointly Gaussian matrices $\bar{\bf{G}}_t$ and $\bar{\bf{F}}_t$ with the same covariance matrices, respectively. To avoid confusion, we use different notations to refer to the MAMP iterates (i.e., $\{\bf{x}, \bf{y}, \bf{X}_t,\bf{R}_t, \bf{F}_t,\bf{G}_t, \bf{n}\}$) and the state evolution random variables (i.e., $\{\bar{\bf{x}}, \bar{\bf{y}},  \bar{\bf{X}}_t,\bar{\bf{R}}_t, \bar{\bf{F}}_t,\bar{\bf{G}}_t, \bar{\bf{n}}\}$).

 \begin{figure}[b!]
  \centering 
  \includegraphics[width=9cm]{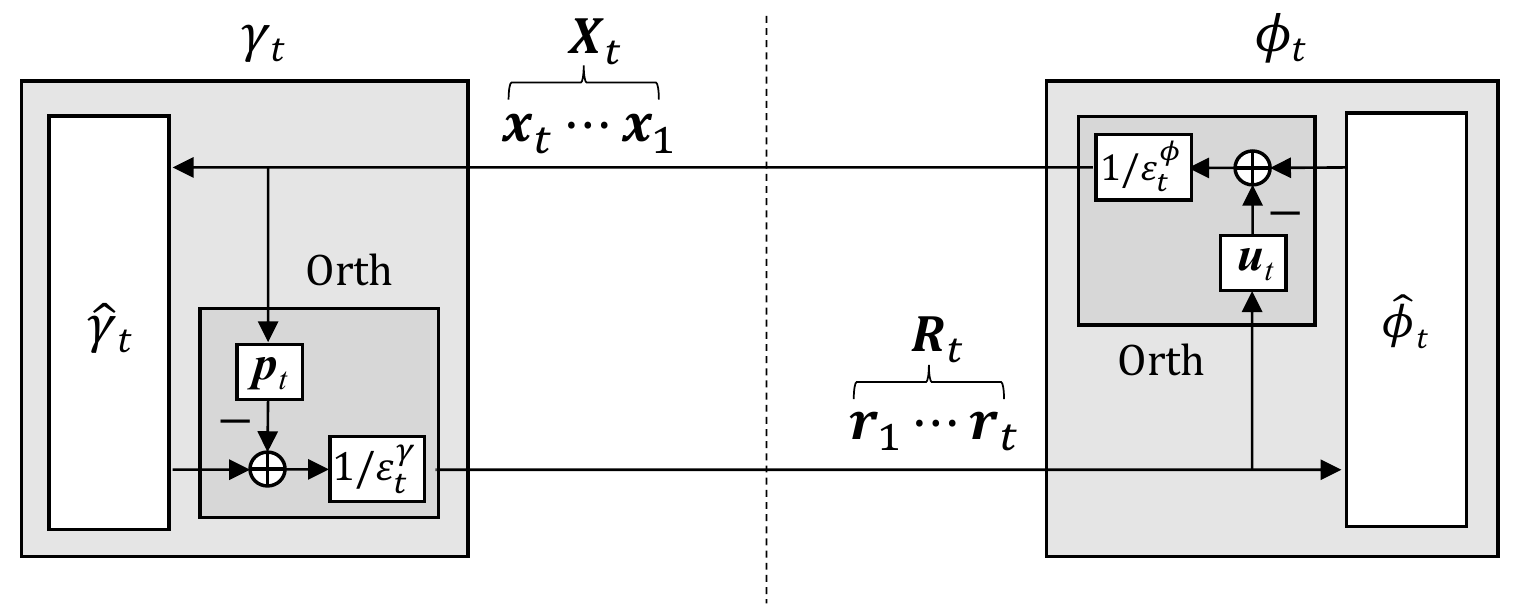}\\ 
  \caption{Graphic illustration of MAMP involving two local orthogonal processors $\gamma_t$ (for $\Gamma$) and $\phi_t$ (for $\Phi$), which are realized by orthogonalization for $\hat{\gamma}_t$ and $\hat{\phi}_t$  (see \eqref{Eqn:MAMP}) \cite{Lei2020MAMPTIT}.}\label{Fig:MAMP_Framework}
\end{figure}

 \begin{lemma}[MAMP Construction \cite{Lei2020MAMPTIT}]
 Given an arbitrary differentiable, separable and Lipschitz-continuous $\hat\phi_t(\bf{R}_t)$, and a general $\{ \hat{\gamma}_t \}$ below
\BE\label{Eqn:orth_MLE_build}
    \hat{\gamma}_t(\bf{y}, \bf{X}_t) = {\bf{Q}}_t\bf{y} + \textstyle\sum_{i=1}^t{{\bf{P}}}_{t,i} \bf{x}_i,
\EE
where $\bf{Q}_t\bf{A}$ and $\bf{P}_{t,i}$ are polynomials in $\bf{A}^{\rm H}\bf{A}$, we can construct $\{ {\gamma}_t \}$ and $\{{\phi}_t \}$ for MAMP via orthogonalization:\vspace{-4mm}
 \BS\label{Eqn:MAMP}\begin{align} 
 &{\rm Orth:}&& {\gamma}_t(\bf{y}, \bf{X}_t) =  \tfrac{1}{\varepsilon_t^\gamma}\big( \hat{\gamma}_t(\bf{y},\bf{X}_t) -  \bf{X}_t\bf{p}_t \big), \label{Eqn:gamma_build}\\
 &{\rm Orth:}&& \phi_t(\bf{R}_t) =  \tfrac{1}{\varepsilon_t^\phi} \big( \hat\phi_t( \bf{R}_t) - \bf{R}_t \bf{u}_t \big),\label{Eqn:phi_build}  
   \end{align} 
where 
\begin{align}
   &\varepsilon_t^\gamma = \tfrac{1}{N} {\rm tr} \big\{{\bf{Q}}_t\bf{A} \big\},\\
   & \bf{p}_t=\big[\tfrac{1}{N}{\rm tr}  \{ {\bf{P}}_{t,1} \}\,\cdots\,  \tfrac{1}{N}{\rm tr}  \{ {\bf{P}}_{t,t} \}\big]^{\rm T},\\
  &  \bf{u}_t =   \big[\mathbb{E}\big\{\tfrac{\partial \hat{\phi}_t } {\partial  r_1}\big\} \,\cdots \, \mathbb{E}\big\{\tfrac{\partial \hat{\phi}_t } {\partial  r_t}\big\} \big]^{\rm T}, 
\end{align}\ES 
and $\varepsilon_t^\phi$ is an arbitrary constant and generally is determined by minimizing the MSE of $\phi_t(\cdot)$. 
\end{lemma}

Note that $\{\gamma_t(\cdot)\}$ and  $\{\phi_t(\cdot)\}$ in \eqref{Eqn:gamma_build} and \eqref{Eqn:phi_build} satisfies the orthogonality in \eqref{Eqn:Orth_MAMP}. Hence, \eqref{Eqn:MAMP} is a MAMP. Fig. \ref{Fig:MAMP_Framework} gives a graphic illustration of MAMP constructed by \eqref{Eqn:MAMP}.

The error covariance matrices of MAMP can be tracked by the following state evolution: Starting with $v_{1,1}^{\phi}=1$,
\BS\label{SE_MAMP}\begin{align} 
{\rm MLE:} \;\qquad \quad v_{t,1:t}^\gamma &= \gamma_t^{\mr{SE}}(\bf{V}_t^{{\phi}}),\label{SE_MAMP_MLE}\\
{\rm NLE:} \; \quad  v_{t+1,1:t+1}^{\phi}  &= {\phi}_t^{\mr{SE}}(\bf{V}_t^{\gamma}),\label{SE_MAMP_NLE}
\end{align}\ES
\hspace{-1mm}where $v_{t,1:t}^\gamma\!\!=\![{v}_{t,1}^\gamma,\dots,{v}_{t,t}^\gamma], v_{t+1,1:t+1}^\gamma\!\!=\![{v}_{t+1,1}^\gamma,\dots,{v}_{t+1,t+1}^\phi]$,   and $\gamma_t^{\mr{SE}}(\cdot)$ and ${\phi}_t^{\mr{SE}}(\cdot)$ are the MSE transfer functions given in \eqref{Eqn:SE_MIP} corresponds to $\gamma_t(\cdot)$ and $\phi_t(\cdot)$, respectively. Without loss of generality, we assume that the transfer functions $\gamma_t^{\rm SE}$ and $\phi_t^{\rm SE}$ satisfy the following assumption. 

 \begin{assumption}\label{Ass:SE_funs}
 For any finite integer $I$, we have
\BS\label{Eqn:SE_assump}\begin{align} 
   \lim_{t\to\infty} v_{t,t}^\gamma &\overset{\rm a.s.}{=} [\gamma_t^{\mr{SE}}(\bf{V}_{I\to t}^{{\phi}})]_{t,t}, \\
   \lim_{t\to\infty} v_{t+1,t+1}^\phi &\overset{\rm a.s.}{=} [{\phi}_t^{\mr{SE}}(\bf{V}_{I\to t}^{\gamma})]_{t+1,t+1},   
\end{align}\ES  
where $\bf{V}_{I\to t}^{\phi}$ and $\bf{V}_{I\to t}^{\gamma}$ are the covariance matrices of $[\bf{x}_I \dots \bf{x}_t]$ and $[\bf{r}_I \dots \bf{r}_t]$, respectively.  
\end{assumption}

The intuition behind Assumption \ref{Ass:SE_funs} is as follows: When the number of iterations is very large, the initial finite estimates have a negligible effect on the current estimation. For example, in the Bayes-optimal OAMP/VAMP \cite{Rangan2016, Ma2016}: $v_{t+1,t+1}^\phi=\phi_t^{\rm SE}(v_{t,t}^{\gamma})$ and $v_{t+1,t+1}^\phi=\gamma_t^{\rm SE}(v_{t,t}^{\phi})$ since both the NLE and LE are non-memory (See Section \ref{Sec:OAMP/VAMP_SS} for more details). Therefore, Assumption \ref{Ass:SE_funs}  holds for the Bayes-optimal OAMP/VAMP. Apart from that, in Bayes-optimal MAMP \cite{Lei2020MAMPTIT}: $v_{t+1,t+1}^\phi=\phi_t^{\rm SE}(v_{t,t}^{\gamma})$ since the NLE is non-memory, and $\lim_{t\to\infty} v_{t,t}^\gamma \overset{\rm a.s.}{=} [{\gamma}_t^{\mr{SE}}(\bf{V}_{I\to t}^{\phi})]_{t,t}$ since the inputs $[\bf{x}_1 \dots \bf{x}_I]$ of MLE correspond to the high-order (i.e., $t-1$ to $t-I$) terms of the Taylor series, which tend to zero as $t\to \infty$ (see Appendix G-B in \cite{Lei2020MAMPTIT}). That is, the transfer functions of Bayes-optimal MAMP in \cite{ Lei2020MAMPTIT} satisfy Assumption \ref{Ass:SE_funs}.
 
\section{L-Banded Matrices}\label{Sec:SS-L-banded}
In this section, we introduce the L-banded covariance, L-banded matrix, and their properties that will be used. 
 
\subsection{Properties of L-Banded Covariance} 

The following lemma is motivated by the optimal long-memory damping in \cite{Lei2020MAMPTIT}. A similar result  was provided in \cite[Lemma 4]{Takeuchi2021OAMP} independently.

  \begin{lemma}[Damping is not Needed for L-Banded  Covariance]\label{Lem:useless_damp}
Let $\bf{s}_t = x\bf{1}_{t\times 1}  + \bf{n}_t$, where  $\bf{n}_t$ is zero mean with covariance matrix $\bf{V}_t$ (can be singular). If the elements in the last row and the last column of $\bf{V}_t$ are the same, i.e.,  
\BE
     v_{t,i}=v_{i,t}=v_{t,t}, \quad \forall 1\le i\le t,
\EE 
damping is then not needed for $\bf{s}_t$ (i.e., does not improve the MSE), i.e.,  
\BE
    [0\dots 0\;1]^{\rm T} = \argmin_{\bf{\zeta}_{t}:\bf{\zeta}_{t}^{\rm T}\bf{1}=1} \mathbb{E} \big\{\|\bf{\zeta}_t^{\rm T}\bf{s}_t -{x} \|^2\big\}.
\EE
\end{lemma} 
 
 \begin{IEEEproof}
    See Appendix \ref{APP:damp_useless}.  
 \end{IEEEproof}

 \begin{lemma}[Inequality of L-Banded  Covariance]\label{Lem:decreas_V}
 Let $\bf{V}_t$ be a $t\times t$ covariance matrix. If the elements in last row and last column of $\bf{V}_t$ are the same, i.e.,  
\BE
    v_{t,i}=v_{i,t}=v_{t,t}, \quad \forall 1\le i\le t,
\EE
we then have 
\begin{align}
   v_{t,t}  &\le v_{i,i},\quad \forall 1\le i\le t.
 \end{align} 
\end{lemma}

\begin{IEEEproof}
   Following the covariance inequality,  we have
    \BE 
       |v_{t,i}| \le \sqrt{v_{t,t} v_{i,i}}. 
    \EE
    Since $v_{t,i} =v_{t,t}, \forall i\le t$, we then have
     \BE 
       v_{t,t} \le  v_{i,i}, \quad \forall i \le t. 
    \EE
    Therefore, we complete the proof of Lemma \ref{Lem:decreas_V}. 
   \end{IEEEproof}

Lemma \ref{Lem:useless_damp} and Lemma \ref{Lem:decreas_V} apply to any covariance matrices for which only the elements in the last row and last column are the same, even if they are not invertible. Hence they are more general than \cite[Lemma 4]{Takeuchi2021OAMP}, which is limited to invertible L-banded covariance matrices (i.e., the elements in each ``L band" of the matrix are the same (see Definition \ref{Def:L-baned})). In addition,  \cite[Lemma 4]{Takeuchi2021OAMP} was proved via its positive determinate, while the proof of Lemma \ref{Lem:decreas_V} is based on the covariance inequality.

 \subsection{L-Banded  Matrix} 
 
 {
    The analytically optimized vector damping (See Section \ref{Sec:SS_construct}) was first proposed in \cite{Lei2020MAMPTIT}, and plays a significant role in BO-MAMP. It was found that the covariance matrix of damped estimates exhibits a special structure where the entries in each ``L band" are identical. Meanwhile, the covariance matrix with this special structure was also implicitly found by Takeuchi independently for the convergence proof of OAMP/VAMP in \cite{Takeuchi2021OAMP}. We refer to any matrix (not necessarily covariance matrix) with the special structure as an L-banded matrix.
 }
 
 \begin{definition}[L-Banded Matrix \cite{Lei2020MAMPTIT,Takeuchi2021OAMP}]\label{Def:L-baned}
 An $t\times t$  matrix $\bf{V}_t=\{v_{i,j}\}$ is said to be L-banded if  the elements in each ``L band" of the matrix are the same, i.e.,
\BE
   v_{i,j}  =  v_{\max(i,j)}, \quad \forall 1\le i\le t, \forall 1\le j\le t. \label{Eqn:SS_cov}  
\EE  
That is,
 \BE
   \bf{V}_t= \begin{tikzpicture}[baseline=-\dimexpr-.0mm\relax]
  \matrix [matrix of math nodes, left delimiter=(,
 right delimiter=), row sep=1.5mm,column sep=1.5mm, ampersand replacement=\&] (M) {
   v_1 \& v_2 \& \cdots \& v_t          \\ 
   v_2 \& v_2 \& \cdots \& v_t          \\ 
   \vdots \& \vdots \& \ddots \& \vdots   \\
   v_t \& v_t \& \cdots  \& v_t             \\ 
   };
   \draw[](M-1-2.north east)--(M-2-2.south east)--(M-2-1.south west)--(M-2-1.north west)--(M-2-2.north west)--(M-1-2.north west)--(M-1-2.north east);
   \draw[](M-1-4.north east)--(M-4-4.south east)--(M-4-1.south west)--(M-4-1.north west)--(M-4-4.north west)--(M-1-4.north west)--(M-1-4.north east);
\end{tikzpicture}.
 \EE 
\end{definition}

{
The following lemma provides the analytic expression of the inverse of an L-banded matrix, which can avoid the high-complexity matrix inverse.
}
 
 \begin{lemma}[Inverse of L-Banded  Matrix
 ]\label{Lem:SS_matrix_inv}
 Let $\bf{V}_t$ be an $t\times t$ invertible L-banded  matrix with diagonal elements $\{v_1,\dots, v_t\}$, and $\bf{V}_t^{-1}=\{v_{i,j}^\dagger\}$ be the  inverse of $\bf{V}_t$. Let $v_0\equiv \infty$, $v_{t+1}\equiv0$ and $\delta_i={v_{i}-v_{i+1}}$. Then,  $\bf{V}_t^{-1}$ is a tridiagonal matrix given by  
 \BE\label{Eqn:SS_V_inv}
    v_{i,j}^\dagger = \left\{ \begin{array}{ll} 
        \delta_{i-1}^{-1} + \delta_i^{-1}, & {\rm if}\;  1\le i=j\le t \vspace{3mm}\\ 
        -\delta^{-1}_{\min(i,j)}, & {\rm if}\;  |i-j|=1 \vspace{3mm}\\
        0, & {\rm otherwise}
    \end{array}\right.,
 \EE
 i.e.,
 \BE\label{Eqn:inverse_V}
  \!\! \!\! \bf{V}_t^{-1}\! \!=\!  \!\!\begin{pmatrix}\!\!   
        \delta^{-1}_1 \!\!\!&\!\!\! -\delta^{-1}_1 &\!\!\! & & \!\!\! \\[2pt]
      \!\!   -\delta^{-1}_1 \!\!\!& \delta^{-1}_1\!\!+\!\delta^{-1}_2 &\!\!\! -\delta^{-1}_2  &\qquad\textbf{ \Large 0}  &\!\!\!\\[2pt]
      \!\! \!\!  \!\!\! & \!\!\!\ddots &\!\!\! \ddots & \ddots &\!\!\!   \\[2pt]
      \!\! \!\!   \!\!\!&\!\!\!\textbf{ \Large 0}\qquad  &\!\!\! -\delta^{-1}_{t-2} & \delta^{-1}_{t-2}\!\!+\!\delta^{-1}_{t-1}  &\!\!\! -\delta^{-1}_{t-1}  \\[2pt]
      \!\! \!\!   \!\!\!&\!\!\! &\!\!\! & -\delta^{-1}_{t-1}&\!\!\! \delta^{-1}_{t-1} \!\!+\! v_t^{-1}
   \!\! \end{pmatrix}\!\!.
 \EE
  Furthermore, 
 \BS\label{Eqn:SSV_inv_sum}\begin{align}
     \bf{1}^{\rm T} \bf{V}_t^{-1}& = [ \bf{V}^{-1}\bf{1}]^{\rm T}  = \big[0, \dots, 0, v_t^{-1}\big],\\
     \bf{1}^{\rm T}\bf{V}_t^{-1}\bf{1} & = v_t^{-1},\\
    \det(\bf{V}_t) &= \prod_{i=1}^t\delta_i.\label{Eqn:SSV_det}
 \end{align} \ES 
 From \eqref{Eqn:SSV_det}, $\bf{V}_t$ is invertible if and only if 
\BE
  v_i\neq v_{i+1}, \quad i=1,\dots,t.
\EE 
 \end{lemma}
 
 \begin{IEEEproof}
 See Appendix \ref{APP:SS_matrix_inv}.  
 \end{IEEEproof}

{ Appendix \ref{APP:SS_matrix_inv} proves the inverse part 
in \eqref{Eqn:inverse_V} by verifying that $\bf{V}_t^{-1}\bf{V}_t=\bf{I}$. We also provided an inductive proof in \cite[Appendix G]{Huang2023}. The analytic expressions in \eqref{Eqn:SSV_inv_sum} were also independently proposed in \cite[Lemma 3]{Takeuchi2021OAMP}. In this paper, we obtain \eqref{Eqn:SSV_inv_sum} straightforwardly from \eqref{Eqn:inverse_V}, while [31, Lemma 3] derived them by statistics and linear algebra.} The following lemma gives the monotonicity and convergence of L-banded covariance matrices.
  
\begin{lemma}[Monotony and Convergence of L-Banded Covariance Matrices]\label{Lem:converg_SS_V}
 Suppose that $\bf{V}_t$ is an $t\times t$  L-banded covariance matrix with diagonal elements $\{v_1,\dots, v_t\}$. Then, the sequence $(v_1,\dots, v_t)$ is monotonically decreasing and converges to a certain value $v_{\star}$, i.e.,
 \BS\begin{align}
   v_{i}  &\le v_{j},\quad {\rm if} \; j\le i\le t,\\
   v_t  &\to v_{\star},\quad {\rm if} \;t\to\infty. 
 \end{align} \ES 
\end{lemma}

\begin{IEEEproof} 
    Since $\bf{V}_t$ is L-banded, following Lemma \ref{Lem:decreas_V},
     \BE 
       v_{i} \le  v_{j}, \quad \forall j \le i. 
    \EE
    That is, $\{v_t\}$ is monotonically decreasing with $t$. Furthermore, $v_{i}>0, \forall i \le t$, i.e., sequence $(v_1,\dots, v_t)$ has a lower bound 0. Hence, if $t\to\infty$, sequence $\{v_{t}\}$ converges to a certain value, i.e., $\lim_{t\to\infty} v_{t} \to v_{\star}$. Therefore, we complete the proof of Lemma \ref{Lem:converg_SS_V}. 
   \end{IEEEproof}

Similar to Lemma \ref{Lem:decreas_V}, since it applies to any L-banded covariance matrices, Lemma \ref{Lem:converg_SS_V} is more general than \cite[Lemma 3]{Takeuchi2021OAMP}, which is limited to invertible L-banded covariance matrices. Moreover, the convergence of L-banded covariance matrices (i.e., $v_t\to v^{\star}, {\rm if}\; t\to\infty$) in Lemma \ref{Lem:converg_SS_V} was not provided in \cite[Lemma 3]{Takeuchi2021OAMP}. More fundamental and interesting properties of L-banded matrices, such as definiteness, LDL decomposition, Cholesky decomposition, minors and cofactors, were provided in \cite{Huang2023}.

 \subsection{Construction of L-Banded Error Covariance Matrix}\label{Sec:SS_construct}

 \subsubsection{Invertible Error Covariance Matrix} 
Given a Gaussian observation sequence with an invertible covariance matrix, the lemma below constructs a new sequence with an L-banded error covariance matrix, which was also provided in \cite[Eqs. (5)-(9)]{Takeuchi2021OAMP} motivated by the optimal long-memory damping in \cite{Lei2020MAMPTIT}.

 \begin{lemma}[Damping for L-Banded Covariance Matrices]\label{Lem:L-banded_seq}
Let $\bf{z}_t = x\bf{1}_{t\times 1}  + \tilde{\bf{n}}_t$, where  $\tilde{\bf{n}}_t$ is zero mean with invertible covariance matrix $\tilde{\bf{V}}_t$, which is not necessarily L-banded.  We can construct a new sequence: For $1\le i\le t$,
\BS\label{Eqn:damp_seq}\BE
    s_i = \bf{\zeta}_i^{\rm T} \bf{z}_i,  
\EE
where 
\BE\label{Eqn:zeta}
    \bf{\zeta}_i =     \dfrac{  \tilde{\bf{V}}_{i}^{{-1}} \bf{1}}{\bf{1}^{\rm T} \tilde{\bf{V}}_{i}^{{-1}}\bf{1}}.  
\EE\ES 
Then, we have ${\bf{s}}_t = x\bf{1}_{t\times 1}  + {\bf{n}}_t$, where  ${\bf{n}}_t$ is zero mean with L-banded covariance matrix ${\bf{V}}_t$.
\end{lemma} 
 
 
 \begin{IEEEproof}
      {We prove Lemma \ref{Lem:L-banded_seq} by induction. 
     \begin{itemize}
         \item When $t=1$, $\tilde{\bf{V}}_1$ is invertible since $v_{1,1}>0$. From \eqref{Eqn:damp_seq}, we have $s_1=z_1$. Therefore,  ${\bf{V}}_1= \tilde{v}_{1,1}$ is L-banded.
         \item For $1< \tau\le t$,  assume that ${\bf{V}}_{\tau-1}$ is L-banded, we then prove that ${\bf{V}}_{\tau}$ is  L-banded.  From \eqref{Eqn:damp_seq}, for $i\le \tau$,
         \BS\label{Eqn:v_tau_i}\begin{align}
             v_{\tau, i}  
             &= \mathbb{E}\{n_\tau^{\rm *}  {n}_i \} \\
             &= \mathbb{E}\{ (\bf{\zeta}_\tau^{\rm T}\tilde{\bf{n}}_\tau)^{\rm *} \tilde{\bf{n}}_i^{\rm T}\bf{\zeta}_i  \} \\
             &= \bf{\zeta}_\tau^{\rm H} \mathbb{E}\{ \tilde{\bf{n}}_\tau^{\rm *} \tilde{\bf{n}}_i^{\rm T} \}\bf{\zeta}_i \\
             &= \bf{\zeta}_\tau^{\rm H} [\tilde{\bf{V}}_\tau]_{1:\tau, 1:i} \bf{\zeta}_i \\
             &= v_{\tau, \tau} \bf{1}^{\rm T} \bf{\zeta}_i\label{Eqn:v_tau_i_c}\\ 
             &= v_{\tau, \tau}, \label{Eqn:v_tau_i_d}
         \end{align} \ES
         where \eqref{Eqn:v_tau_i_c} follows \eqref{Eqn:damp_seq} and Lemma \ref{Lem:SS_damp} (which implies $\bf{\zeta}_\tau^{\rm H} [\tilde{\bf{V}}_\tau]_{1:\tau, 1:i}= v_{\tau, \tau} \bf{1}^{\rm T}$), and \eqref{Eqn:v_tau_i_d} follows $\bf{\zeta}_i^{\rm T}\bf{1}=1$. Since ${\bf{V}}_{\tau-1}$ is L-banded and $v_{\tau, i} =v_{i,\tau}=v_{\tau, \tau}$ (see \eqref{Eqn:v_tau_i}),  ${\bf{V}}_t$ is L-banded.
     \end{itemize}  
     Therefore, we complete the proof of Lemma \ref{Lem:L-banded_seq}.}
 \end{IEEEproof}

Intuitively, the $i$-th damping $s_i =\bf{\zeta}_i^{\rm T} \bf{z}_i $ in \eqref{Eqn:damp_seq} ensures that the elements in the $i$-th L-band of $\bf{V}_t$ are the same. Hence,  the step-by-step damping in \eqref{Eqn:damp_seq} results in an L-banded $\bf{V}_t$.  \vspace{3mm}

 \subsubsection{Singular Error Covariance Matrix}   In Lemma \ref{Lem:L-banded_seq}, it assumes that $\tilde{\bf{V}}_{t}$ is invertible. When $\tilde{\bf{V}}_{t}$ is singular and $\tilde{v}_{i,i}>0, \forall 1\le i\le t$, we can construct $\bf{s}_t$ with an L-banded error covariance matrix as follows.

For simplicity, we let $\mathcal{I}_t$ be an index set for effective elements whose covariance matrix is invertible, 
and its complementary set $\bar{\mathcal{I}}_t$ be the index set of the trivial elements. Hence, we have $\mathcal{I}_t \cap \bar{\mathcal{I}}_t=\emptyset$ and $\mathcal{I}_t \cup \bar{\mathcal{I}}_t=\{1,\dots, t\}$. We can obtain ${\mathcal{I}}_t$ using the following recursion: Starting from  ${\mathcal{I}}_{1}=\{1\}$ and $i=2$, 
\BE\label{Eqn:eff_index}
  {\mathcal{I}}_{i} = \left\{\!\!\!\begin{array}{ll}
  \{{\mathcal{I}}_{i-1}, i\}, &\;\; {\rm if}\;   \tilde{\bf{V}}_{\{{\mathcal{I}}_{i-1}, i\}} {\rm \;is\; invertible} \vspace{2mm}\\ 
  {\mathcal{I}}_{i-1}, & \;\; {\rm otherwise} 
  \end{array}\right., 
\EE
 where $\tilde{\bf{V}}_{\mathcal{S}}$ denotes the covariance matrix of $\{z_i,i\in\mathcal{S}\}$. For example, consider $[z_1, \dots, z_4]$ and assume that $\tilde{\bf{V}}_{1}$ and $\tilde{\bf{V}}_{\{1,3\}}$ are invertible, and  $\tilde{\bf{V}}_{\{1, 2\}}$ and $\tilde{\bf{V}}_{\{1,3,4\}}$ are singular. We have $\mathcal{I}_4 =\{1, 3\}$ and $\bar{\mathcal{I}}_4=\{2,4\}$.  

We do not allow the trivial elements to join the damping process as they do not improve the MSE performance. Hence, $\{{\zeta}_{t, i} = 0,  i\in \bar{\mathcal{I}}_{t}\}$, i.e., $\bf{\zeta}_{\bar{\mathcal{I}}_{t}}=\bf{0}$. Therefore, it only needs to optimize  $\bf{\zeta}_{\mathcal{I}_t} =[ \zeta_{t, i}, i\in {\mathcal{I}}_{t}]$ for the effective elements. From \eqref{Eqn:eff_index}, $\tilde{\bf{V}}_{{\mathcal{I}}_{t}}$ is invertible. Similarly to \eqref{Eqn:zeta}, we then have $\bf{\zeta}_{\mathcal{I}_t} = {  \tilde{\bf{V}}_{{\mathcal{I}}_{t}}^{-1} \bf{1}}/\big[{\bf{1}^{\rm T} \tilde{\bf{V}}_{{\mathcal{I}}_{t}}^{-1}\bf{1}}\big]$. Therefore, we obtain the following lemma.

 \begin{lemma}[Damping for L-Banded Covariance Matrices]\label{Lem:L-banded_sing}
Let $\bf{z}_t = x\bf{1}_{t\times 1}  + \tilde{\bf{n}}_t$, where  $\tilde{\bf{n}}_t$ is zero mean with covariance matrix $\tilde{\bf{V}}_t$ and $\tilde{v}_{i,i}>0, \forall 1\le i\le t$.  We can construct a new sequence: For $1\le i\le t$,
\BS\label{Eqn:damp_seq_sing}\BE
    s_i = \bf{\zeta}_i^{\rm T} \bf{z}_i,  
\EE
where 
\BE\label{Eqn:zeta_sing}
     \bf{\zeta}_{\mathcal{I}_t} =     \dfrac{  \tilde{\bf{V}}_{{\mathcal{I}}_{t}}^{-1} \bf{1}}{\bf{1}^{\rm T} \tilde{\bf{V}}_{{\mathcal{I}}_{t}}^{-1}\bf{1}}, \qquad \bf{\zeta}_{\bar{\mathcal{I}}_{t}}=\bf{0}.
\EE\ES 
Then, we have ${\bf{s}}_t = x\bf{1}_{t\times 1}  + {\bf{n}}_t$, where  ${\bf{n}}_t$ is zero mean with L-banded covariance matrix ${\bf{V}}_t$. Furthermore, following \eqref{Eqn:damp_seq_sing}, for all $1\le i\le t-1$, 
\BE\label{Eq:back-off}
   s_{i+1}  = s_{i}, \quad \!{\rm if}\;   {\bf{V}}^{\phi}_{\{{\mathcal{I}}^{\phi}_{i}, i+1\}} {\rm \;is\; singular}.
\EE

\end{lemma}

\section{Sufficient Statistic}\label{Sec:SS}
In this section, we introduce the sufficient statistic and its properties.

\subsection{Sufficient Statistic}\label{Sec:SS_sub}  
 \begin{definition}[Sufficient Statistic]
  For any $\bf{s}_t = [s_{1} \dots s_{t}]^{\rm T}$, $s_{t}$ is a sufficient statistic of $x$ given $\bf{s}_t$, if $\bf{s}_{t-1}$ --- $s_t$ --- $x$ is a Markov chain, i.e.,
  \begin{align} 
    p(x|\bf{s}_t) &=p(x|s_t).\label{Eqn:SS_p}
\end{align}  
 \end{definition}  
 
  A sufficient statistic has the following straightforward proposition.  
  
 \begin{proposition}\label{Cor:SS_Est}
 We assume that $s_{t}$ is a sufficient statistic of $x$ given $\bf{s}_t$, i.e., \eqref{Eqn:SS_p} holds for any $\bf{s}_t$. Then, we have
\BE
    \mathbb{E}\{x|\bf{s}_t\}  =\mathbb{E}\{x|s_t\}.\label{Eqn:SS_E} 
\EE
 \end{proposition}

 
   Proposition \ref{Cor:SS_Est} shows that if $s_t$ is a sufficient statistic of $x$ given the Gaussian observations $\bf{s}_t$, the \emph{a-posterior} (i.e., MMSE) estimation of $x$ based on $\bf{s}_t$ is equivalent to the \emph{a-posterior} estimation based on $s_t$.
  
   The following lemma provides a sufficient statistic of Gaussian observations.
   
 \begin{lemma}[L-Banded  Covariance of Sufficient Statistics]\label{Lem:SS_Est} 
Let $\bf{s}_t = x\bf{1}_{t\times 1}  + \bf{n}_t$, where $\bf{s}_t = [s_{1} \dots s_{t}]^{\rm T}$ and $\bf{n}_t\sim \mathcal{CN}(\bf{0},\bf{V}_t)$ with invertible $\bf{V}_t$. For any $\bf{s}_t$, $s_{t}$ is a sufficient statistic of $x$ given $\bf{s}_t$ if and only if  
\BE
    v_{t,i}=v_{i,t}=v_{t,t}, \quad \forall 1\le i\le t.
\EE
That is, the elements in the last row and the last column of $\bf{V}_t$ are the same.
\end{lemma} 
  
 \begin{IEEEproof}
 See Appendix \ref{APP:SS_Est}.
 \end{IEEEproof} 

\cite[Lemma 4]{Takeuchi2021OAMP} also provides the sufficiency for the whole L-banded covariance matrix $\bf{V}_t$ (see Definition \ref{Def:L-baned}).  In contrast, Lemma \ref{Lem:SS_Est} provides both sufficiency and necessity for a looser condition that only the elements in the last row and last column of $\bf{V}_t$ are identical.

The following lemma gives the construction of  a sufficient statistic given the Gaussian observations, which was also provided in \cite[Eqs. (5)-(9)]{Takeuchi2021OAMP} motivated by the optimal long-memory damping in \cite{Lei2020MAMPTIT}.
 
  \begin{lemma}[Damping for a Sufficient Statistic] \label{Lem:SS_damp}
Let $\bf{z}_t = x\bf{1}_{t\times 1}  + \bf{n}_t$, where $\bf{z}_t = [z_{1} \dots z_{t}]^{\rm T}$ and $\bf{n}_t\sim \mathcal{CN}(\bf{0},\bf{V}_t)$ with invertible $\bf{V}_t$. Then,  a sufficient statistic of $x$ given $\bf{z}_t$ can be constructed by optimal damping \cite{Lei2020MAMPTIT}:
\BS\label{Eqn:damp}\BE
   s_{t} = \bf{\zeta}_t^{\rm T}\bf{z}_t,\label{Eqn:damp_x}
\EE
where 
\BE\label{Eqn:damp_zeta}
    \bf{\zeta}_t = \tfrac{\bf{V}_t^{-1}\bf{1}}{\bf{1}^{\rm T}\bf{V}_t^{-1}\bf{1}},
\EE
because for $i=1,\dots, t$,
   \BE \label{Eqn:damp_v}
       \!\!\!\!\! \mathbb{E}\{ (s_t-x)^{\rm *}{n}_i\} \! =\! \mathbb{E}\{ (s_t-x)^{\rm *}(s_t-x)\}  
         \! =\! \tfrac{1}{\bf{1}^{\rm T}\bf{V}_t^{-1}\bf{1}}.
    \EE  \ES 
\end{lemma} 
 
 \begin{IEEEproof}
Eqn. \eqref{Eqn:damp_v} is derived by
    \BS\label{Eqn:damp_pro}\begin{align}
        \mathbb{E}\{(s_t-x)^{\rm *}\bf{n}_t^{\rm T}\}& = \bf{\zeta}_t^{\rm H} \mathbb{E}\{\bf{n}^*_t\bf{n}_t^{\rm T}\} \label{Eqn:damp_pro1}\\
        &= \bf{\zeta}_t^{\rm H} \bf{V}_t   \label{Eqn:damp_pro2}\\
        &=\tfrac{\bf{1}^{\rm T}}{\bf{1}^{\rm T}\bf{V}_t^{-1}\bf{1}},\label{Eqn:damp_pro3}\\
        \mathbb{E}\{ (s_t-x)^{\rm *}(s_t-x)\} & = \bf{\zeta}_t^{\rm H} \mathbb{E}\{\bf{n}^*_t\bf{n}_t^{\rm T}\}\bf{\zeta}_t \label{Eqn:damp_pro4}\\
        &= \bf{\zeta}_t^{\rm H} \bf{V}_t\bf{\zeta}_t   \label{Eqn:damp_pro5}\\
        &=\tfrac{1}{\bf{1}^{\rm T}\bf{V}_t^{-1}\bf{1}},\label{Eqn:damp_pro6}
    \end{align}
    \ES
    where \eqref{Eqn:damp_pro1} and \eqref{Eqn:damp_pro4}  follow \eqref{Eqn:damp_x}, \eqref{Eqn:damp_pro2}  and \eqref{Eqn:damp_pro5}  follow $\mathbb{E}\{\bf{n}^*_t\bf{n}_t^{\rm T}\}=\bf{V}_t$, \eqref{Eqn:damp_pro3} and \eqref{Eqn:damp_pro6} follow \eqref{Eqn:damp_zeta}. From Lemma \ref{Lem:SS_Est} and \eqref{Eqn:damp_v},  $s_{t}$ is a sufficient statistic of $\bf{x}$ given $\bf{z}_t$.
 \end{IEEEproof}

{\subsection{MMSE/LMMSE Sufficient Statistic}\label{Sec:MMSE-ss} 
 The rigorous sufficient statistic in \eqref{Eqn:SS_p} is defined on the exact \emph{a posteriori} probabilities, which are generally unavailable in the iterative process including the AMP-type algorithms. To solve this issue, we generalize the sufficient statistic as follows from the MMSE and LMMSE perspectives.

\begin{definition}[MMSE Sufficient Statistic]\label{Def:MMSE-SS}
  Suppose that Assumptions \ref{ASS:x}-\ref{ASS:A} hold. We say $\bf{r}_t$ is an \textbf{MMSE sufficient statistic} of $\bf{x}$ given $\bf{R}_t$ if  
    \BE
          {\mr{var}}\{\bf{x}|\bf{R}_t\}\overset{\rm a.s.}{=} {\mr{var}}\{\bf{x}|\bf{r}_t\}.
    \EE 
\end{definition}

\begin{definition}[LMMSE Sufficient Statistic]\label{Def:LMMSE-SS}
  Suppose that Assumptions \ref{ASS:x}-\ref{ASS:A} hold.  Let $\bf{y}=\bf{Ax}+\bf{n}$. We say $\bf{x}_t$ is an \textbf{LMMSE sufficient statistic} of $\bf{x}$ given $\bf{X}_t$ if 
    \BE
          {\mr{lmmse}}\{\bf{x}|\bf{y}, \bf{X}_t\}\overset{\rm a.s.}{=} {\mr{lmmse}}\{\bf{x}|\bf{y},\bf{x}_t\},
    \EE
     where $\mr{lmmse}(\cdot)$ is defined in \eqref{Eqn:V_LMMSE}.
\end{definition}
 
Typically, the MMSE sufficient statistic is employed when the asymptotic joint distribution $p(\bf{x}, \bf{G}_t)$ is approximately known, which is the case when $\bf{R}_t$ is the input of the non-linear estimator $\phi_t$. The LMMSE sufficient statistic is commonly used when we only know the mean and covariance matrix of $\bf{F}_t$ while the  distributions (or joint distribution) of $\bf{x}$ and $\bf{F}_t$ are unknown, which is the case when $\bf{X}_t$ is the input of the linear estimator $\gamma_t$.

Lemma \ref{Lem:MMSE-SS} and Lemma \ref{Lem:LMMSE-SS} below show that Lemma \ref{Lem:SS_Est} also applies to MMSE/LMMSE sufficient statistic, i.e., L-banded covariance is a sufficient condition for MMSE/LMMSE sufficient statistic.

\begin{lemma}\label{Lem:MMSE-SS}
    Suppose that Assumptions \ref{ASS:x}-\ref{ASS:A} hold. In MAMP, ${\bf{r}}_t$ is an MMSE sufficient statistic of ${\bf{x}}$ given ${\bf{R}}_t$ if
    \BE
        v^\gamma_{t,i}=v^\gamma_{i,t}=v^\gamma_{t,t}, \quad \forall 1\le i\le t.
    \EE  
\end{lemma}

\begin{IEEEproof}
Following Lemma \ref{Lem:IIDG_MIP}, we have
\BS\label{Eqn:ite2SE_gamma}\begin{align}
    {\mr{var}}\{{\bf{x}}|{\bf{R}}_t\} \overset{\rm a.s.}{=} \overline{\mr{var}}\{\bar{\bf{x}}|\bar{\bf{R}}_t\} \\
    {\mr{var}}\{{\bf{x}}|{\bf{r}}_t\} \overset{\rm a.s.}{=} \overline{\mr{var}}\{\bar{\bf{x}}|\bar{\bf{r}}_t\}
\end{align}\ES
where $\bar{\bf{R}}_t=\bar{\bf{X}}+\bar{\bf{G}}_t$,  and $\bar{\bf{G}}_t$ is row-wise IID, column-wise jointly Gaussian and independent of $\bar{\bf{x}}$, i.e.,  
    \BE
        [\bar{g}_{i,1},\dots \bar{g}_{i,t}] \overset{\rm IID}{\sim} \mathcal{CN}{(\bf{0}, \bf{V}_t^\gamma)}, \;\; i=1,\dots t.
    \EE   
Since both $\bar{\bf{x}}$ and $\bar{\bf{G}}_t$ are row-wise IID, we can apply Lemma \ref{Lem:SS_Est} to $\bar{\bf{R}}_t$ row-by-row. Therefore, based on $\{v^\gamma_{t,i}=v^\gamma_{i,t}=v^\gamma_{t,t},  \forall 1\le i\le t\}$, we have $\overline{\mr{var}}\{\bar{\bf{x}}|\bar{\bf{R}}_t\}\overset{\rm a.s.}{=} \overline{\mr{var}}\{\bar{\bf{x}}|\bar{\bf{r}}_t\}$. Furthermore, following \eqref{Eqn:ite2SE_gamma}, we have ${\mr{var}}\{{\bf{x}}|{\bf{R}}_t\}\overset{\rm a.s.}{=} {\mr{var}}\{{\bf{x}}|{\bf{r}}_t\}$. Then, following Definition \ref{Def:MMSE-SS}, ${\bf{r}}_t$ is an MMSE sufficient statistic of ${\bf{x}}$ given ${\bf{R}}_t$. 
\end{IEEEproof}

\begin{lemma}\label{Lem:LMMSE-SS}   
Suppose that Assumptions \ref{ASS:x}-\ref{ASS:A} hold. In MAMP, ${\bf{x}}_t$ is an LMMSE sufficient statistic of ${\bf{x}}$ given ${\bf{X}}_t$ if
    \BE
        v^\phi_{t,i}=v^\phi_{i,t}=v^\phi_{t,t}, \quad \forall 1\le i\le t.
    \EE 
\end{lemma}

\begin{IEEEproof} 
    See Appendix \ref{APP:LMMSE-SS}.
\end{IEEEproof}

Intuitively,  $\overline{\rm lmmse}\{\cdot\}$ can be viewed as a type of  \emph{a posteriori} variance $\overline{\rm var}\{\cdot\}$ under the condition that the \emph{a priori} distribution $p({\bf{x}})$ of $\bf{x}$ is unknown. Therefore, $\overline{\rm lmmse}\{\cdot\}$ inherits the sufficient-statistic property of $\overline{\rm var}\{\cdot\}$.

The following Lemma shows that Lemma \ref{Lem:SS_damp} also applies to the MMSE/LMMSE sufficient statistic, i.e., an MMSE/LMMSE sufficient statistic can be constructed by optimal damping.

\begin{lemma}\label{Lem:MMSE-damp}
Suppose that Assumptions \ref{ASS:x}-\ref{ASS:A} hold, and $\bf{V}_t^\gamma$ and $\bf{V}_t^\phi$ are invertible. In MAMP, an MMSE sufficient statistic of $\bf{x}$ given $\bf{R}_t$ can be constructed by optimal damping 
    \cite{Lei2020MAMPTIT}:
\BE
   \bf{r}^{\rm ss}_{t} = \bf{R}_t\bf{\zeta}^\gamma_t, \quad \bf{\zeta}^\gamma_t = \tfrac{[\bf{V}^\gamma_t]^{-1}\bf{1}}{\bf{1}^{\rm T}[\bf{V}^\gamma_t]^{-1}\bf{1}},
\EE
and an LMMSE sufficient statistic of $\bf{x}$ given $\bf{X}_t$ can be constructed by optimal damping \cite{Lei2020MAMPTIT}:
\BE
   \bf{x}^{\rm ss}_{t} = \bf{X}_t\bf{\zeta}^\phi_t, \quad \bf{\zeta}^\phi_t = \tfrac{[\bf{V}^\phi_t]^{-1}\bf{1}}{\bf{1}^{\rm T}[\bf{V}^\phi_t]^{-1}\bf{1}}.
\EE
\end{lemma}
\begin{IEEEproof}
    The proof is omitted since it is the same as the proof of Lemma \ref{Lem:SS_damp}.
\end{IEEEproof}

\subsection{Orthogonality of MMSE/LMMSE Estimators}

Lemma \ref{Lem:MMSE_orth} and Lemma \ref{Lem:LMMSE_orth} below show the orthogonality of MMSE/LMMSE estimation for the sufficient-statistic messages in MAMP.

 \begin{lemma}[Orthogonality of MMSE Estimation]\label{Lem:MMSE_orth}
          Suppose that Assumptions \ref{ASS:x}-\ref{ASS:A} hold.  In MAMP, if $\bf{r}_t$ is an MMSE sufficient statistic of $\bf{x}$ given $\bf{R}_t$, for any Lipschitz-continuou $\phi(\cdot)$, we have
            \BE\label{Eqn:orth_mmse_f} 
                \big\langle   \mr{E}\{\bf{x}|\bf{r}_{t}\} -\bf{x}| \phi(\bf{R}_t)  \big\rangle 
                \overset{\rm a.s.}{=} 0,
           \EE     
           which implies 
            \BE
              \big\langle   \mr{E}\{\bf{x}|\bf{r}_{t}\} -\bf{x}| \mr{E}\{\bf{x}|\bf{r}_{t}\} - \phi(\bf{R}_t)  \big\rangle 
                \overset{\rm a.s.}{=} 0.
           \EE   
\end{lemma}
        
    \begin{IEEEproof}   
    We prove \eqref{Eqn:orth_mmse_f} as follows.
         \BS \label{Eqn:MMSE_orth}\begin{align}
           &\big\langle  \mathbb{E}\{\bf{x}|\bf{r}_t\} - \bf{x}\big|  \phi(\bf{R}_t)  \big \rangle \nonumber \\ 
             &\overset{\rm a.s.}{=}  \mathbb{E}\big\{ \big\langle   \mathbb{E}\{\bar{\bf{x}}|\bar{\bf{x}}+\bar{\bf{g}}_t\} - \bar{\bf{x}}\big|   \phi(\bar{\bf{X}}+\bar{\bf{G}}_t)  \big\rangle\big\} \label{Eqn:MMSE_orth4}\\
             &\overset{\rm a.s.}{=} \mathbb{E}\big\{ \big\langle   \mathbb{E}\{\bar{\bf{x}}|\bar{\bf{X}}+\bar{\bf{G}}_t\} - \bar{\bf{x}}\big| \phi(\bar{\bf{X}}+\bar{\bf{G}}_t) \big \rangle\big\} \label{Eqn:MMSE_orth5}\\
            &\overset{\rm a.s.}{=}0,\label{Eqn:MMSE_orth6}
        \end{align}\ES  
     where \eqref{Eqn:MMSE_orth4} follows Lemma~\ref{Lem:IIDG_MIP},  \eqref{Eqn:MMSE_orth5} follows Proposition \ref{Cor:SS_Est} and Lemma \ref{Lem:SS_Est} and the fact that $\bf{r}_{t}$ is an MMSE sufficient statistic of $\bf{x}$ given $\bf{R}_t$, \eqref{Eqn:MMSE_orth6} follows the orthogonal property of MMSE estimation.  
    \end{IEEEproof}

 \begin{lemma}[Orthogonality of LMMSE Estimation]\label{Lem:LMMSE_orth}
          Suppose that Assumptions \ref{ASS:x}-\ref{ASS:A} hold.  In MAMP, if $\bf{x}_t$ is an LMMSE sufficient statistic of $\bf{x}$ given $\bf{X}_t$, for any linear estimation $\gamma(\cdot)$ in \eqref{Eqn:MIP_LE}, we have
           \BE\label{Eqn:orth_mmse_g}
               \big\langle   \mr{LMMSE}\{\bf{x}|\bf{y},\bf{x}_{t}\} \! - \!\bf{x}\big|  \mr{LMMSE}\{\bf{x}|\bf{y},\bf{x}_{t}\} \!- \! \gamma(\bf{y}, \bf{X}_t) \!\big\rangle \!\overset{\rm a.s.}{=} \!0.
           \EE  
\end{lemma}

    \begin{IEEEproof}   
    See Appendix \ref{APP:orth_LMMSE}.  
    \end{IEEEproof}}

\begin{remark}
Lemma \ref{Lem:MMSE_orth} and Lemma \ref{Lem:LMMSE_orth} may not be limited to the MMSE/LMMSE estimation, which is only one of the sufficient conditions of the orthogonality. For MAMP with other estimators, Lemma \ref{Lem:MMSE_orth} and Lemma \ref{Lem:LMMSE_orth} may also hold. 
\end{remark}


\section{Sufficient-Statistic MAMP (SS-MAMP)}\label{Sec:SS-MAMP}
The state evolution of MAMP may not converge or even diverge if it is not properly designed. In this section, using the sufficient-statistic technique in Section \ref{Sec:SS}, we construct an SS-MAMP to solve the convergence problem of the state evolution of MAMP in principle.
 
\subsection{Definition  and Properties of SS-MAMP}
We define a sufficient-statistic memory AMP as follows.

\begin{definition}[SS-MAMP]
Suppose that Assumptions \ref{ASS:x}-\ref{ASS:A} hold. A MAMP is said to be sufficient-statistic if $\forall t$,
 \BS\label{Eqn:SS_mmse}\begin{align} 
   {\mr{var}}\{\bf{x}|\bf{R}_t\}&\overset{\rm a.s.}{=} {\mr{var}}\{\bf{x}|\bf{r}_t\},\\
    { \mr{lmmse}}\{\bf{x}|\bf{y}, \bf{X}_t\}&\overset{\rm a.s.}{=} {\mr{lmmse}}\{\bf{x}|\bf{y},\bf{x}_t\}.
\end{align} \ES
\hspace{-3mm}That is,  in SS-MAMP, $\bf{r}_t$ and $\bf{x}_t$ are respectively sufficient statistics of $\bf{x}$ given $\bf{R}_t$ and $\bf{X}_t$ from the MMSE/LMMSE perspective.  
\end{definition}

\begin{remark}
 Notice that \eqref{Eqn:SS_mmse} is a weak sufficient statistic from the MMSE/LMMSE perspective (see Section \ref{Sec:MMSE-ss}), which is different from the strict sufficient statistic defined on the probability in \eqref{Eqn:SS_p}. It should be emphasized that even though the SS-MAMP is defined from the local MMSE/LMMSE perspective, the local MNLE and MLE can be arbitrary Lipschitz-continuous processors and are not limited to local MMSE/LMMSE estimators. Therefore, the SS-MAMP in the paper is more general than the long-memory OAMP in \cite{Takeuchi2021OAMP}, which focused on the local Bayes-optimal (i.e., LMMSE/MMSE) estimators. Consequently, the results in this section (e.g., Lemmas \ref{Lem:no_damp_MMSE}-\ref{Lem:SS_MAMP_MMSE} and Theorem \ref{The:SS-MAMP_con}) are more general than those in \cite{Takeuchi2021OAMP}.
\end{remark}

In SS-MAMP, the MSE of the local MMSE/LMMSE estimation using the current message is the same as that using the current and preceding messages. In other words, jointly estimation with preceding messages  (called memory) does not bring improvement in the MSE of local MMSE/LMMSE estimation, i.e., memory is not needed in local MMSE/LMMSE  estimators in SS-MAMP. Therefore, we have the following lemma.

\begin{lemma}[Memory is not Needed in SS-MAMP]\label{Lem:no_damp_MMSE}  
Suppose that Assumptions \ref{ASS:x}-\ref{ASS:A} hold. In SS-MAMP, memory is not needed for the Bayes-optimal local estimator (e.g., LMMSE estimator $\hat{\gamma}_t$ or MMSE estimator $\hat{\phi}_t$), i.e., joint estimation with preceding messages does not bring improvement in MSE. 
\end{lemma}

If the MLE and MNLE in SS-MAMP are both locally Bayes-optimal, i.e., LMMSE $\hat{\gamma}_t$ and MMSE $\hat{\phi}_t$ (e.g., the OAMP/VAMP in Section \ref{Sec:BO-OAMP/VAMP}), then Lemma \ref{Lem:no_damp_MMSE} degrades into the long-memory OAMP in \cite{Takeuchi2021OAMP}. Nevertheless, Lemma \ref{Lem:no_damp_MMSE} also applies to the SS-MAMP in which one local estimator is Bayes-optimal (MMSE/LMMSE) but the other is not (e.g., the SS-BO-MAMP in Section \ref{Sec:SS-BO-MAMP}), which goes beyond the results in \cite{Takeuchi2021OAMP}.


Following Lemma \ref{Lem:MMSE-SS} and Lemma \ref{Lem:LMMSE-SS}, we give a sufficient condition of SS-MAMP in the following.

\begin{lemma}[Sufficient Conditions of SS-MAMP]\label{Lem:SS} 
Suppose that Assumptions \ref{ASS:x}-\ref{ASS:A} hold. A MAMP is sufficient-statistic if the covariance matrices $\bf{V}^{\gamma}_t$ (of $\bf{R}_t$) and $\bf{V}^{\phi}_t$  (of $\bf{X}_t$) are L-banded, i.e., for all $t$, $1\le i\le t$ and $1\le j\le t$,
\BS\label{Eqn:SS}\begin{align} 
  &    v^{\gamma}_{i,j}  =  v^{\gamma}_{\max(i,j)},\label{Eqn:SS_gamma} \\
&   v^{{\phi}}_{i,j} = v^{{\phi}}_{\max(i,j)},\label{Eqn:SS_phi}
\end{align} \ES   
and for all  $1\le \tau\le t-1$
\BS\label{Eqn:rep}\begin{align} 
  &    \bf{r}_{\tau+1}  = \bf{r}_{\tau}, \quad {\rm if}\;    {\bf{V}}^{\gamma}_{\{{\mathcal{I}}^{\gamma}_{\tau}, \tau+1\}} {\rm \;is\; singular},\\
&   \bf{x}_{\tau+1}  = \bf{x}_{\tau}, \quad \!{\rm if}\;   {\bf{V}}^{\phi}_{\{{\mathcal{I}}^{\phi}_{\tau}, \tau+1\}} {\rm \;is\; singular},
\end{align} \ES
where ${\mathcal{I}}^{\gamma}_{\tau}$ and ${\mathcal{I}}^{\phi}_{\tau}$ are defined the same as that in \eqref{Eqn:eff_index}.
\end{lemma}

\begin{IEEEproof} 
Intuitively, under the condition \eqref{Eqn:rep}, following Lemma \ref{Lem:MMSE-SS} and Lemma \ref{Lem:LMMSE-SS}, in $i$-th iteration, $\bf{r}_i$ (or $\bf{x}_i$) is an MMSE (or LMMSE) sufficient statistic of $\bf{x}$ given $\bf{R}_i$ (or $\bf{X}_i$) if the elements in the $i$-th L-band of $\bf{V}_t^\gamma$ (or $\bf{V}_t^\phi$) are the same. Since the sufficient-statistic property holds for all iterations in SS-MAMP, it is equivalent to that the elements in each L-band of $\bf{V}_t^\gamma$ (or $\bf{V}_t^\phi$) are the same.
\end{IEEEproof}

In general, \eqref{Eqn:SS} and \eqref{Eqn:rep} are sufficient (not necessary) conditions of SS-MAMP. However, in this paper, we only focus on the specific SS-MAMP under \eqref{Eqn:SS} and \eqref{Eqn:rep}. For convenience, in the rest of this paper, the specific SS-MAMP under \eqref{Eqn:SS} and \eqref{Eqn:rep} is simply referred to as SS-MAMP.

\begin{lemma}[State Evolution of SS-MAMP]\label{Lem:SE}
Suppose that Assumptions \ref{ASS:x}-\ref{ASS:A} hold. In SS-MAMP, the covariance matrices $\bf{V}_t^{\gamma}$ and $\bf{V}_t^{\phi}$ are determined by their  diagonal sequences $  \bf{v}_t^{\gamma} =[v^{\gamma}_{1}\dots\,  v^{\gamma}_{t}]^{\rm T}$ and $\bf{v}_t^{\phi} =[v^{\phi}_{1}\dots\,  v^{\phi}_{t}]^{\rm T}$, respectively. The state evolution of SS-MAMP can be simplified to: Starting with $t=1$ and $v^\phi_{1}=1$, 
{\BS\begin{align}
     v^\gamma_{t} &=\gamma_t^{\mr{SE}}(\bf{v}^\phi_t)\overset{\rm a.s.}{=} \tfrac{1}{N}\mathbb{E}\big\{\|\gamma_t \big(\bar{\bf{y}},\bar{\bf{X}}+\bar{\bf{F}}_t\big)-\bar{\bf{x}}\|^2\big\},  \label{Eqn:SE_MAMP_LE}  \\
      {v}^\phi_{t+1} &=  \phi_t^{\mr{SE}}(\bf{v}^\gamma_t)\overset{\rm a.s.}{=} \tfrac{1}{N}\mathbb{E}\big\{\|\phi_t \big(\bar{\bf{X}}+\bar{\bf{G}}_t \big)-\bar{\bf{x}}\|^2\big\}, \label{Eqn:SE_MAMP_NLE}
    \end{align}\ES 
    where  $\bar{\bf{X}}=\bar{\bf{x}}\cdot\bf{1}^{\rm T}$,   $\bar{\bf{y}} = \bf{A}\bar{\bf{x}} + \bar{\bf{n}}$, $\bar{\bf{n}}$ denotes the noise sampled from the same distribution as that of $\bf{n}$, $\bar{\bf{x}}$ denotes the signal sampled from the same distribution as that of $\bf{x}$, and $\bar{\bf{F}}_t$ and $\bar{\bf{G}}_t$ are row-wise IID, column-wise jointly Gaussian and  independent of $\bar{\bf{n}}$ and $\bar{\bf{x}}$, i.e., for $i=1,\dots t$,  
  \BS\begin{align}
        [\bar{f}_{i,1},\dots \bar{f}_{i,t}] \overset{\rm IID}{\sim} \mathcal{CN}{(\bf{0}, \bf{V}_t^\phi)},\\
        [\bar{g}_{i,1},\dots \bar{g}_{i,t}] \overset{\rm IID}{\sim} \mathcal{CN}{(\bf{0}, \bf{V}_t^\gamma)}.
    \end{align}\ES}  
\end{lemma}

 The lemma below follows Lemma \ref{Lem:converg_SS_V} and Lemma \ref{Lem:SS}.

\begin{lemma}[Convergence of SS-MAMP] \label{Lem:converge}
 The diagonal sequences $[v^{{\gamma}}_1,v^{{\gamma}}_2,\dots]$ and $[v^{{\phi}}_1,v^{{\phi}}_2,\dots]$ in SS-MAMP are monotonically decreasing and converge respectively to a certain value, i.e., 
   \BS\begin{align}
   v^{{\gamma}}_{t} &\le v^{{\gamma}}_{t'}, \;  \forall t' \le t, &
        \lim\limits_{t\to\infty} v^{{\gamma}}_{t} &\to v^{{\gamma}}_{\star}, \label{Eqn:mon_v_g}\\
       v^{{\phi}}_{t} &\le v^{{\phi}}_{t'}, \;  \forall t' \le t, &
        \lim\limits_{t\to\infty} v^{{\phi}}_{t} &\to v^{{\phi}}_{\star}.\label{Eqn:mon_v_f}
   \end{align}\ES
   Therefore, the state evolution of SS-MAMP is convergent.  
\end{lemma}
 
\begin{remark}
 Note that one-side sufficient-statistic outputs (of $\{\gamma_t\}$ or $\{\phi_t\}$, e.g., $\bf{V}_t^\gamma$ or $\bf{V}_t^\phi$ is L-banded) are sufficient to guarantee the convergence of the state evolution of MAMP.  BO-MAMP \cite{Lei2020MAMPTIT} is such an example whose convergence is guaranteed by the sufficient-statistic outputs of NLE $\phi_t$ using damping, while the outputs of MLE $\gamma_t$ in BO-MAMP are not sufficient statistics. However, as shown in Section \ref{Sec:SS-BO-MAMP} (see also Section \ref{Sec:sim_SS-BO-MAMP}), the performance or convergence speed of MAMP can be further improved if the outputs of $\gamma_t$ and $\phi_t$ are both sufficient statistics.
\end{remark}

 The lemma below follows Lemma \ref{Lem:useless_damp} and Lemma \ref{Lem:SS}.

\begin{lemma}[Damping is not Needed in SS-MAMP]\label{Lem:damp_useless}
 Damping is not needed (i.e., has no MSE improvement) in SS-MAMP. Specifically,  the optimal damping of $\bf{R}_t$ (or $\bf{X}_t$) has the same MSE as that of $\bf{r}_t$ (or $\bf{x}_t$), i.e.,
    \BS\label{Eqn:unnec_damp}\begin{align}
         v^\gamma_t = \min_{\bf{\zeta}_{t}:\bf{\zeta}_{t}^{\rm T}\bf{1}=1} \tfrac{1}{N}\|\bf{R}_t  \bf{\zeta}_{t}-\bf{x} \|^2,\label{Eqn:unnec_damp1}\\
        v_t^\phi =\min_{\bf{\zeta}_{t}:\bf{\zeta}_{t}^{\rm T}\bf{1}=1} \tfrac{1}{N}\|\bf{X}_t  \bf{\zeta}_{t}-\bf{x} \|^2.\label{Eqn:unnec_damp2} 
    \end{align}   \ES 
\end{lemma}

\subsection{Construction of SS-MAMP}
Given an arbitrary MAMP, the following theorem constructs an SS-MAMP using damping.

 \begin{theorem}[SS-MAMP Construction]\label{The:SS-MAMP_con}
Suppose that Assumptions \ref{ASS:x}-\ref{ASS:A} hold. Given an arbitrary MAMP: 
\BS\label{Eqn:org_MAMP}\begin{align} 
  \tilde{\bf{r}}_t &= \gamma_t\!\left(\bf{y},\tilde{\bf{X}}_{t}\right),\label{Eqn:MAMP_g}\\
  \tilde{\bf{x}}_{t + 1} &= \phi_t\!\left(\tilde{\bf{R}}_t\right),\label{Eqn:MAMP_f}
\end{align}\ES  
where $\gamma_t(\cdot)$ is given in \eqref{Eqn:MIP_LE}, $\phi_t(\cdot)$ is a separable-and-Lipschitz-continuous, and the orthogonality in \eqref{Eqn:Orth_MAMP} holds. Let $\tilde{\bf{R}}_{t}=[\tilde{\bf{r}}_{1}\dots \tilde{\bf{r}}_{t}]$ and $\tilde{\bf{X}}_{t+1}=[\tilde{\bf{x}}_{1}\dots \tilde{\bf{x}}_{t+1}]$. Then, following Lemmas  \ref{Lem:L-banded_sing} and \ref{Lem:SS} and based on  \eqref{Eqn:org_MAMP}, we can always construct an SS-MAMP via  damping: Starting with $t\!=\!1$ and $\bf{X}_{1}\!=\!\bf{0}$,
\BS\label{Eqn:SS_MAMP}\begin{alignat}{2}
  \bf{r}_t &= \gamma^{\rm ss}_t \left( \bf{y},\bf{X}_{t}\right) = \tilde{\bf{R}}_{t} \bf{\zeta}^\gamma_t,\label{Eqn:MAMP_g_con}\\
   \bf{x}_{t + 1} &= \phi^{\rm ss}_t \left(\bf{R}_t\right) = \tilde{\bf{X}}_{t}   \bf{\zeta}^\phi_{t+1},\label{Eqn:MAMP_f_con}
\end{alignat} 
where  
\begin{align} 
  &\bf{\zeta}^\gamma_{\mathcal{I}^\gamma_t} = \frac{  \big[\tilde{\bf{V}}^{\gamma}_{\mathcal{I}^\gamma_t}\big]^{-1} \bf{1}}{\bf{1}^{\rm T} \big[\tilde{\bf{V}}^{\gamma}_{\mathcal{I}^\gamma_t}\big]^{-1}\bf{1}}, \quad\;\qquad \bf{\zeta}^\gamma_{\bar{\mathcal{I}}^\gamma_t}=\bf{0}, \label{Eqn:dam_g} \\
 & \bf{\zeta}^\phi_{\mathcal{I}^\phi_{t+1}} = \frac{  \big[\tilde{\bf{V}}^{\phi}_{\mathcal{I}^\phi_{t+1}}\big]^{-1} \bf{1}}{\bf{1}^{\rm T} \big[\tilde{\bf{V}}^{\phi}_{\mathcal{I}^\phi_{t+1}}\big]^{-1}\bf{1}}, \qquad \bf{\zeta}^\phi_{\bar{\mathcal{I}}^\phi_{t+1}}=\bf{0}, \label{Eqn:dam_f}\\
 & \tilde{\bf{V}}^{\gamma}_{\mathcal{I}^\gamma_t} =  \big\langle   \tilde{\bf{R}}_{\mathcal{I}^\gamma_t}-\bf{X} \big| \tilde{\bf{R}}_{\mathcal{I}^\gamma_t} -\bf{X}  \big\rangle,\\
 & \tilde{\bf{V}}^{\phi}_{\mathcal{I}^\phi_{t+1}} = \big\langle  \tilde{\bf{X}}_{\mathcal{I}^\phi_{t+1}} -\bf{X} \big| \tilde{\bf{X}}_{\mathcal{I}^\phi_{t+1}} -\bf{X}   \big\rangle,
\end{align}\ES   
$\tilde{\bf{R}}_{\mathcal{I}^\gamma_t}=[\tilde{\bf{r}}_i, i\in {\mathcal{I}^\gamma_t}]$ and $\tilde{\bf{X}}_{\mathcal{I}^\phi_{t+1}}=[\tilde{\bf{x}}_i, i\in {\mathcal{I}^\phi_{t+1}}]$. Fig. \ref{Fig:SS_MAMP_Framework} gives a graphic illustration of SS-MAMP constructed by \eqref{Eqn:SS_MAMP}.
In addition, the following hold for the SS-MAMP in \eqref{Eqn:SS_MAMP}.
\begin{enumerate}[(a)] 
    \item The SS-MAMP in \eqref{Eqn:SS_MAMP} is a memory iterative process given in Definition \ref{Def:MIP}. That is, the $\gamma_t^{\rm ss}$ and $\phi_t^{\rm ss}$ in \eqref{Eqn:SS_MAMP} can be written in the form of \eqref{Eqn:MIP}.
    Furthermore, damping preserves the orthogonality, i.e., the following orthogonality holds: $\forall t\ge 1$,
    \BE\label{Eqn:orth_SS_MAMP} 
    \langle \bf{g}_t | \bf{x} \rangle  \overset{\rm a.s.}{=}  0, \;\;\; 
        \langle \bf{g}_t| \bf{F}_t\rangle      \overset{\rm a.s.}{=}  \bf{0},   \;\;\;
       \langle \bf{f}_{t+1}| \bf{G}_t \rangle       \overset{\rm a.s.}{=} \bf{0}. 
     \EE
    That is, the SS-MAMP in \eqref{Eqn:SS_MAMP} is still a MAMP. \vspace{1mm} 
    

\begin{figure}[t]
  \centering 
  \includegraphics[width=8.5cm]{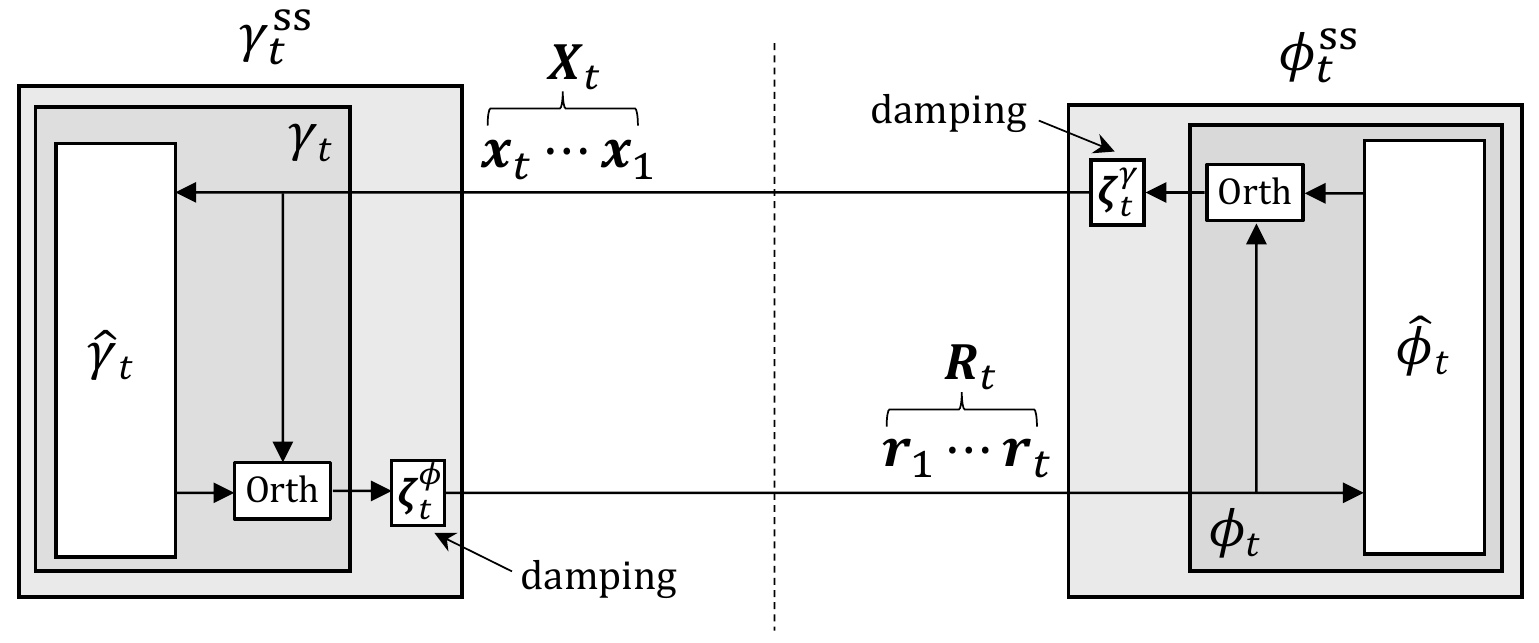}\\ 
  \caption{Graphic illustration of SS-MAMP involving two local sufficient-statistic orthogonal processors $\gamma^{\rm ss}_t$ (for $\Gamma$) and $\phi^{\rm ss}_t$ (for $\Phi$), which are realized by damping for ${\gamma}_t$ and  ${\phi}_t$ (see \eqref{Eqn:SS_MAMP}).}\label{Fig:SS_MAMP_Framework}
\end{figure}

    \item The covariance matrices $\bf{V}^{\gamma}_t$ and $\bf{V}^{\phi}_t$ of the SS-MAMP in \eqref{Eqn:SS_MAMP} are L-banded and are determined by sequences $  \bf{v}_t^{\gamma} =[v^{\gamma}_{1}\dots\,  v^{\gamma}_{t}]^{\rm T}$ and $\bf{v}_t^{\phi} =[v^{\phi}_{1}\dots\,  v^{\phi}_{t}]^{\rm T}$, respectively. In detail, 
        \BS\label{Eqn:damped_v}\begin{align}
          \!\!\!v^{\gamma}_{t} &=  
                  \dfrac{1}{\bf{1}^{\rm T} \big[\tilde{\bf{V}}^{\gamma}_{\mathcal{I}^\gamma_t}\big]^{-1}\bf{1}},  \label{Eqn:damped_v_g}\\
         \!\!\!  v^{\phi}_{t+1} &=  
                  \dfrac{1}{\bf{1}^{\rm T} \big[\tilde{\bf{V}}^{\phi}_{\mathcal{I}^\phi_{t+1}}\big]^{-1}\bf{1}}.\label{Eqn:damped_v_f}
        \end{align}\ES    
    Furthermore, the state evolution of the SS-MAMP in \eqref{Eqn:SS_MAMP} converges to a fixed point, i.e., 
    \BS\label{Eqn:V_convg}\begin{align}
        \lim_{t\to\infty} \;v^{\gamma}_{t} \overset{\rm a.s.}{=} v^{\gamma}_{\star},\\
        \lim_{t\to\infty} \; v^{\phi}_{t} \overset{\rm a.s.}{=} v^{\gamma}_{\star}.
    \end{align}\ES

    \item Damping is not needed (i.e., has no MSE improvement) for the SS-MAMP in \eqref{Eqn:SS_MAMP}. In other words,  the SS-MAMP in \eqref{Eqn:SS_MAMP} obtains the lowest MSE among all the damped MAMP given $\{{\gamma}_t\}$ and $\{{\phi}_t\}$. 
    \vspace{2mm} 
    
    \item Suppose that Assumption \ref{Ass:SE_funs} holds for $\gamma^{\rm ss}_t(\cdot)$ and $\phi^{\rm ss}_t(\cdot)$. Then, the state-evolution fixed points  of the SS-MAMP in \eqref{Eqn:SS_MAMP} are given by
    \BS\label{Eqn:fixed_point} \begin{align} 
      v^{\gamma}_{\star}& \overset{\rm a.s.}{=} \lim_{t\to \infty} [\gamma_t^{\mr{SE-ss}}(v^{\phi}_{\star}\bf{1})]_{t,t}, \\
      v^{\phi}_{\star}  & \overset{\rm a.s.}{=} \lim_{t\to \infty} [{\phi}_t^{\mr{SE-ss}}(v^{\gamma}_{\star}\bf{1})]_{t+1,t+1},   
    \end{align}\ES
     where $\bf{1}$ is an all-one matrix with the proper size, and $\gamma_t^{\mr{SE-ss}}(\cdot)$ and $\phi_t^{\mr{SE-ss}}(\cdot)$ are the MSE transfer functions corresponds to $\gamma^{\rm ss}_t(\cdot)$ and $\phi^{\rm ss}_t(\cdot)$, respectively.\vspace{2mm}

    \item The MSE of the SS-MAMP in \eqref{Eqn:SS_MAMP}  is not worse than that of the original MAMP in \eqref{Eqn:org_MAMP}.

\end{enumerate}
\end{theorem}
 
\begin{IEEEproof}
We prove the points (a)-(e) one by one as follows.
\begin{enumerate}[(a)]
    \item {From \eqref{Eqn:SS_MAMP}, we have
    \BS\label{Eqn:ss-unfold} \begin{align}
        \gamma^{\rm ss}_t \left(\bf{y}, \bf{X}_{t}\right) &\!=\!  [ \gamma_1 \left(\bf{y},\bf{X}_{1}\right),  \dots, \gamma_t \left(\bf{y}, \bf{X}_{t}\right) ]\bf{\zeta}_t^\gamma, \\
        \phi^{\rm ss}_t \left(\bf{R}_t\right) &\!=\! [\bf{x}_1, \phi_1 \left( \bf{R}_{1}\right),  \dots, \phi_t \left( \bf{R}_{t}\right) ]\bf{\zeta}_{t+1}^\phi,
    \end{align}\ES
    where $\bf{x}_1$ is an given initialization. Since $\{\gamma_i\}$ and $\{\phi_i\}$ are the local processors in MAMP (i.e., they can be written in the form of memory iterative process), following \eqref{Eqn:org_MAMP} and \eqref{Eqn:ss-unfold}, it is obvious that the $\gamma_t^{\rm ss}$ and $\phi_t^{\rm ss}$ in \eqref{Eqn:SS_MAMP} can be written in the form of \eqref{Eqn:MIP}. Therefore, the SS-MAMP in \eqref{Eqn:SS_MAMP} is a memory iterative process given in Definition \ref{Def:MIP}.} Since $\{\gamma_t(\cdot)\}$ and $\{\phi_t(\cdot)\}$ come from MAMP, from Definition \ref{Def:MAMP},  we have
        \BE\label{Eqn:orth_tau} 
         \langle \tilde{\bf{g}}_t | \bf{x} \rangle  \overset{\rm a.s.}{=}  0,  \;\;\;
        \big \langle \tilde{\bf{g}}_t |\tilde{\bf{F}}_t\rangle     \overset{\rm a.s.}{=}  \bf{0},  \;\;\;
         \big\langle \tilde{\bf{f}}_{t+1}| \tilde{\bf{G}}_t\rangle      \overset{\rm a.s.}{=}  \bf{0}.
      \EE 
      Since $\bf{\zeta}^\gamma_{t}\bf{1}=1$ and $\bf{\zeta}^\phi_{t+1}\bf{1}=1$, we have
      \BS\label{Eqn:error_dam}\begin{align} 
            \bf{g}_{t} &=\tilde{\bf{G}}_t\bf{\zeta}^\gamma_{t},\\
            \bf{f}_{t+1} &=\tilde{\bf{F}}_{t+1} \bf{\zeta}^\phi_{t+1}.
        \end{align}\ES
     That is, $\{\bf{g}_{t}\}$ and $\{\bf{f}_{t}\}$ are the linear combination of $\{\tilde{\bf{g}}_t\}$ and $\{\tilde{\bf{f}}_t\}$, respectively. Since linear combination preserves the orthogonality in \eqref{Eqn:orth_tau}, we have the orthogonality in \eqref{Eqn:orth_SS_MAMP},  i.e., damping preserves the orthogonality. That is, the SS-MAMP in \eqref{Eqn:SS_MAMP} is a MAMP.\vspace{2mm}

    \item The L-banded $\bf{V}^{\gamma}_t$ and $\bf{V}^{\phi}_t$ follows Lemma \ref{Lem:L-banded_seq}, \eqref{Eqn:damped_v} follows Lemma \ref{Lem:MMSE-damp}, and \eqref{Eqn:V_convg} follows Lemma \ref{Lem:converge}.\vspace{2mm}

    \item  See Lemma \ref{Lem:damp_useless}. \vspace{2mm} 

    \item Suppose that the SS-MAMP has converged at $I$-th iteration. Following \eqref{Eqn:damped_v} and \eqref{Eqn:V_convg}, the covariance matrices in SS-MAMP satisfy $\lim_{t\to \infty} \bf{V}_{I\to t}^\gamma \overset{\rm a.s.}{=} v^\gamma_{\star}\bf{1}$ and $\lim_{t\to \infty} \bf{V}_{I\to t}^\phi \overset{\rm a.s.}{=} v^\phi_{\star}\bf{1}$. Therefore, following Assumption \ref{Ass:SE_funs}, the state evolution fixed point of SS-MAMP can be solved by \eqref{Eqn:fixed_point}.\vspace{2mm}

    \item Following (a) and (b),  the algorithm in \eqref{Eqn:SS_MAMP} is SS-MAMP. Then, following Lemma \ref{Lem:damp_useless}, damping is not needed for the SS-MAMP in \eqref{Eqn:SS_MAMP}. That is, the damping used in \eqref{Eqn:SS_MAMP} is optimal, which is not worse than the undamped MAMP in \eqref{Eqn:org_MAMP} (e.g., $\bf{\zeta}_t^{\gamma}=[0,\dots, 0, 1]$ and $\bf{\zeta}_{t+1}^{\phi}=[0,\dots, 0, 1]$). Therefore, the MSE of the SS-MAMP in \eqref{Eqn:SS_MAMP} is not worse than the MAMP in \eqref{Eqn:org_MAMP}. 
     
\end{enumerate}
 Hence, we complete the proof of Theorem \ref{The:SS-MAMP_con}.
 \end{IEEEproof}\vspace{-3mm}

\begin{remark}
Intuitively, in each iteration, the damping operation in \eqref{Eqn:SS_MAMP} scales the previous messages with a damping factor of less than one. When the number of iterations is very large, the initial finite estimates are scaled by the product of an infinite number of numbers less than 1 and thus have a negligible effect on the current estimation. Therefore, Assumption \ref{Ass:SE_funs} applies generically to SS-MAMP.  
\end{remark}
   
\begin{remark}
The damping operation in \eqref{Eqn:SS_MAMP} follows the discussions  in  Section \ref{Sec:SS_construct}, motivated by the damping optimization in \cite{Lei2020MAMPTIT}. That is, damping is performed on the effective inputs $\tilde{\bf{R}}_{\mathcal{I}^\gamma_t}=[\tilde{\bf{r}}_i, i\in {\mathcal{I}^\gamma_t}]$ and $\tilde{\bf{X}}_{\mathcal{I}^\phi_{t+1}}=[\tilde{\bf{x}}_i, i\in {\mathcal{I}^\phi_{t+1}}]$, whose covariance matrices $\tilde{\bf{V}}^{\gamma}_{\mathcal{I}^\gamma_t}$ and $\tilde{\bf{V}}^{\phi}_{\mathcal{I}^\phi_{t+1}}$ are always positive. 
\end{remark}

The following lemma shows that the damping is unnecessary if the local processor in the original MAMP is MMSE.    

\begin{lemma}[Orthogonal MMSE/LMMSE Functions are MMSE/LMMSE Sufficient-Statistic]\label{Lem:SS_MAMP_MMSE}
Suppose that Assumptions \ref{ASS:x}-\ref{ASS:A} hold. 
\begin{itemize} 
\item Consider the following $\phi_t(\cdot)$ in MAMP:
\BE\label{Eqn:MAMP_MMSE_f}
    \phi_t(\bf{R}_t) = \phi_t(\bf{r}_t) = c_t^\phi  \hat\phi_t(\bf{r}_{t}) + (1 - c_t^\phi) \bf{r}_{t},
\EE
where $\hat\phi_t(\bf{r}_{t}) = \mr{E}\{\bf{x}|\bf{r}_{t}\}$  and $c_t^\phi=[1-\overline{\mr{var}}\{\bf{x}|\bf{r}_{t}\}/v_t^\gamma]^{-1}$. Then, damping is unnecessary for the $\phi^{\rm ss}_t(\cdot)$ (see \eqref{Eqn:MAMP_f_con}) in SS-MAMP, i.e.,
 \BS\label{Eqn:SS_MMSE_f}\BE
   \phi^{\rm ss}_t(\bf{r}_{t}) = \phi_t(\bf{r}_t),
\EE
and the variance in \eqref{Eqn:damped_v_f} is reduced to
\BE
   v^{\phi}_{t+1} \overset{\rm a.s.}{=} \left[\frac{1}{\overline{\mr{var}}\{\bf{x}|\bf{x} + \bar{\bf{g}}_{t} \}} - \frac{1}{v_{t}^{\gamma}}\right]^{-1},
\EE\ES
where $\bar{\bf{g}}_t\!\!\sim\!\mathcal{CN}(\bf{0},v_{t}^{\gamma}\bf{I})$ is independent of $\bf{x}$. 

\item  Consider the following $\gamma_t(\cdot)$ in MAMP:
\BE\label{Eqn:MAMP_MMSE_g}
    \gamma_t(\bf{y},\bf{X}_t) = \gamma_t(\bf{y},\bf{x}_t) = c_t^\gamma  \hat\gamma_t(\bf{y},\bf{x}_t)  + (1 - c_t^\gamma) \bf{x}_{t},
\EE
where $\hat\gamma_t(\bf{y},\bf{x}_{t}) = {\rm LMMSE} \{\bf{x}|\bf{y}, \bf{x}_{t}\}$ and $c_t^\gamma =[1-\overline{\rm lmmse} \{\bf{x}|\bf{y}, \bf{x}_{t}\}/v_t^\phi]^{-1}$. Then, damping is unnecessary for the $\gamma^{\rm ss}_t(\cdot)$ (see \eqref{Eqn:MAMP_g_con}) in SS-MAMP, i.e.,
\BS\label{Eqn:SS_MMSE_g} \BE
   \gamma^{\rm ss}_t(\bf{y},\bf{x}_{t}) = \gamma_t(\bf{y},\bf{x}_t),
\EE
and the variance in \eqref{Eqn:damped_v_g} is reduced to
\BE
   v^{\gamma}_{t} \overset{\rm a.s.}{=} \left[\frac{1}{\overline{\rm lmmse}\{\bf{x}|\bf{y}, \bf{x} + \bar{\bf{f}}_{t}\}} - \frac{1}{v_{t}^{\phi}}\right]^{-1},
\EE\ES
where $\bar{\bf{f}}_t\!\!\sim\!\mathcal{CN}(\bf{0},v_{t}^{\phi}\bf{I})$ is independent of $\bf{x}$. 
\end{itemize}
\end{lemma}

\begin{IEEEproof}
See Appendix \ref{APP:SS_MAMP_MMSE}. 
\end{IEEEproof}

If the $\hat{\gamma}_t$ and $\hat{\phi}_t$ are both Bayes-optimal (e.g., the BO-OAMP/VAMP in Section \ref{Sec:BO-OAMP/VAMP}), then Lemma \ref{Lem:SS_MAMP_MMSE} degrades into the results in \cite{Takeuchi2021OAMP}. Nevertheless, Lemma \ref{Lem:SS_MAMP_MMSE} also applies to the SS-MAMP in which one local estimator is Bayes-optimal but the other is not (e.g., the SS-BO-MAMP in Section \ref{Sec:SS-BO-MAMP}), which goes beyond the long-memory OAMP in \cite{Takeuchi2021OAMP}.

\subsection{Low-Complexity Calculation of \texorpdfstring{$\big[\tilde{\bf{V}}^{\gamma}_{\mathcal{I}^\gamma_t}\big]^{-1}$ and $\big[\tilde{\bf{V}}^{\phi}_{\mathcal{I}^\phi_t}\big]^{-1}$ }{TEXT}}
The optimal damping in \eqref{Eqn:dam_g} and \eqref{Eqn:dam_f} need to calculate $\big[\tilde{\bf{V}}^{\gamma}_{\mathcal{I}^\gamma_t}\big]^{-1}$ and $\big[\tilde{\bf{V}}^{\phi}_{\mathcal{I}^\phi_t}\big]^{-1}$, respectively. It will cost high complexity if we calculate them straightforwardly.   Using real block matrix inversion  
\BE\label{Eqn:bm_inv}
    \left[ \begin{array}{cc}
      \bf{A} & \bf{b} \vspace{2mm}\\
        \bf{b}^{\rm H} & c
      \end{array}\right]^{-1} = \left[ \begin{array}{cc}
      \bf{A}^{-1} + \alpha \bf{\beta}\bf{\beta}^{\rm H}  & -\alpha \bf{\beta}  \vspace{2mm}\\
        -\alpha \bf{\beta}^{\rm H}& \alpha 
      \end{array}\right], 
\EE
where $\alpha=(c-\bf{b}^{\rm H} \bf{A}^{-1}\bf{b})^{-1}$ and $\bf{\beta} = \bf{A}^{-1}\bf{b}$, we can calculate $\big[\tilde{\bf{V}}^{\gamma}_{\mathcal{I}^\gamma_t}\big]^{-1}$ and $\big[\tilde{\bf{V}}^{\phi}_{\mathcal{I}^\phi_t}\big]^{-1}$ based on  $\big[\tilde{\bf{V}}^{\gamma}_{\mathcal{I}^\gamma_{t-1}}\big]^{-1}$ and $\big[\tilde{\bf{V}}^{\phi}_{\mathcal{I}^\phi_{t-1}}\big]^{-1}$ calculated in the last iteration. Therefore, the matrix inverses  only cost time complexity $\mathcal{O}(t^2)$ per iteration.

\section{BO-OAMP/VAMP is Sufficient-Statistic}\label{Sec:BO-OAMP/VAMP}
Bayes-optimal orthogonal/vector AMP (BO-OAMP/VAMP)  \cite{Ma2016, Rangan2016} is a non-memory iterative process that solves the problem in \eqref{Eqn:unitary_sys} with a right-unitarily-invariant matrix. In this section, suppose that Assumptions \ref{ASS:x}-\ref{ASS:A} hold, we answer the following fundamental questions about BO-OAMP/VAMP:
\begin{itemize}
    \item If damping is performed on each local processor's outputs, can we further improve the MSE of BO-OAMP/VAMP?
    
    \item  If the preceding messages (i.e., memory) are  used in  local estimation, can we further improve the MSE of BO-OAMP/VAMP?
    
    \item Are the covariance matrices in BO-OAMP/VAMP convergent?
\end{itemize}
In this section, by proving that BO-OAMP/VAMP is sufficient-statistic, we recover the main statements in \cite{Takeuchi2021OAMP}, i.e., damping and memory are not needed in BO-OAMP/VAMP, and the covariance matrices in BO-OAMP/VAMP are L-banded and convergent. Different from the linear algebra proof in \cite{Takeuchi2021OAMP}, we prove the sufficient-statistic property of BO-OAMP/VAMP by the  orthogonality of local MMSE/LMMSE estimators in Lemma \ref{Lem:SS_MAMP_MMSE} (see also Lemmas \ref{Lem:MMSE_orth} and \ref{Lem:LMMSE_orth}).

\subsection{Review of BO-OAMP/VAMP}
The following is the BO-OAMP/VAMP algorithm.
 
 \begin{framed} 
\emph{BO-OAMP/VAMP\cite{Ma2016, Rangan2016}:} Let $s_t=v_{t,t}^\phi/\sigma^2$, and $\hat{\gamma}_t(\cdot)$ and $\hat\phi_t(\cdot)$ be estimates of $\bf{x}$ at $t$-th iteration:   
\BS\label{Eqn:BO-OAMP/VAMP} \begin{align}
   \hat{\gamma}_t \left(\bf{y},\bf{x}_t\right) & \!\equiv\!{\rm LMMSE} \{\bf{x}|\bf{y}, \bf{x}_{t}\} \nonumber\\
   &=    \big( s_t\bf{I} \!+\!   \bf{A}^{\mr H}\!\bf{A}\big)^{-1}(s_t\bf{A}^{\mr H}\bf{y} \!+\!  \bf{x}_t ),\label{Eqn:LMMSE_ori}\\
    \hat\phi_t(\bf{r}_{t})& \equiv \mr{E}\{\bf{x}|\bf{r}_{t} \}. \label{Eqn:MMSE_ori}
\end{align}
A BO-OAMP/VAMP is then defined as: Starting with $t=1$, $v_{1}^\phi=1$ and $\bf{x}_1=\bf{0}$, 
\label{Eqn:OAMP/VAMP}\begin{align}
&{\mr{LE:}} \;\; \bf{r}_t  =  \gamma_t \left(\bf{y},\bf{x}_t\right)\equiv  \tfrac{1}{1- \hat{\gamma}'_t } \big[ \hat{\gamma}_t \left(\bf{y},\bf{x}_t\right) -\hat{\gamma}'_t \bf{x}_t \big], \label{Eqn:OAMP/VAMP_LE}\\
&{\mr{NLE:}} \;\; \bf{x}_{t+1}  = \phi_t \left( \bf{r}_t \right)\equiv  \tfrac{1}{1-\hat\phi'_t } \big[  \hat\phi_t(\bf{r}_t) -\hat\phi'_t  \bf{r}_t \big],\label{Eqn:OAMP/VAMP_NLE}
\end{align}
where \vspace{-3mm}
\begin{align}
  \hat{\gamma}'_t  &\!=\! \tfrac{1}{N}{\mr {tr}} \big\{\big(s_t\bf{I} +   \bf{A}\bf{A}^{\mr H}\big)^{\!-1}\big\}, \\ v^\gamma_t &\!=\!\gamma_t^{\mr{SE}}(v_{t}^\phi) \!\equiv \! {v}^\phi_{t}  [(\hat{\gamma}'_t) ^{-1}\!\! - \!1],  \label{Eqn:OAMP_vara}  \\
 \hat\phi'_t  &\!= \!\tfrac{1}{N v^\gamma_{t}} \mathbb{E}\big\{ \|\hat\phi_t(\bf{x}+\!\sqrt{v^\gamma_{t}} \bf{\eta} ) -\bf{x}\|^2\big\}, \\
 {v}^\phi_{t+1} &\!= \! \phi_t^{\mr{SE}}(v^\gamma_{t}) \!\equiv \!v^\gamma_{t} [(\hat\phi'_t ) ^{-1} \!\!-\!1],\label{Eqn:OAMP_varb}
\end{align}\ES
and $\bf{\eta}\sim {\cal{CN}}(\bf{0}, \bf{I})$ is independent of $\bf{x}$. The final estimation is $\hat\phi_t(\bf{r}_t)$.
 \end{framed}

\begin{assumption}\label{Ass:bouned}
 Each posterior variance $\mr{E}\{|x_n|^2|r_{t,n}\} - |\hat\phi_t(r_{t,n})|^2, \forall n$ is almost surely bounded, where $x_n, r_{t,n}$ and $\Phi_n$ denote the $n$-th entry/constraint of $\bf{x}, \bf{r}_t$ and $\Phi$, respectively \cite{Takeuchi2017}. 
\end{assumption}

The asymptotic IID Gaussian property of a non-memory iterative process  was conjectured in \cite{Ma2016} and proved in \cite{Takeuchi2017,Rangan2016} based on the  error orthogonality below.

\begin{lemma} [Orthogonality and Asymptotically IID Gaussian]\label{Lem:IIDG}
Suppose that Assumptions \ref{ASS:x}-\ref{ASS:A} and \ref{Ass:bouned} hold. Then, the orthogonality below holds for BO-OAMP/VAMP: For all $1\le t'\le t$,  
\BE\label{Eqn:error_orth} 
 \langle \bf{g}_t| \bf{x} \rangle   \overset{\rm a.s.}{=}  0,\;\;\; 
   \langle \bf{g}_t |\bf{F}_t\rangle       \overset{\rm a.s.}{=}  \bf{0},\;\;\;
  \langle \bf{f}_{t+1}| \bf{G}_t \rangle     \overset{\rm a.s.}{=} \bf{0}.
\EE
Furthermore, we have \cite[Theorem 1]{Takeuchi2020CAMP}: $\forall 1\!\le \!t'\!\leq\! t$, 
\BS\label{Eqn:IIDG_OAMP}  \begin{align}
 &v_{t,t'}^{\gamma} \! \overset{\rm a.s.}{=}\!  \mathbb{E}\big\{ \big\langle  \gamma_t\!\big(\bar{\bf{y}},\bar{\bf{x}}\!+\!\bar{\bf{f}}_{t}\big)\!-\!\bar{\bf{x}},\big| \gamma_{t'}\!\big(\bar{\bf{y}},\bar{\bf{x}}+\bar{\bf{f}}_{t'}\big)\!-\!\bar{\bf{x}} \big\rangle\big\},\\ 
& v_{t+1,{t'}+1}^{\phi} \!\overset{\rm a.s.}{=} \!    \mathbb{E}\big\{ \big \langle  \phi_t\!\big(\bar{\bf{x}}\!+\!\bar{\bf{g}}_{t} \big)\!-\!\bar{\bf{x}}\big|\phi_{t'}\!\big(\bar{\bf{x}}\!+\!\bar{\bf{g}}_{t'} \big)\!-\!\bar{\bf{x}} \big\rangle\big\},  
\end{align}\ES 
where  $\bar{\bf{y}} = \bf{A}\bar{\bf{x}} + \bar{\bf{n}}$, $\bar{\bf{n}}$ denotes the noise sampled from the same distribution as that of $\bf{n}$,  $\bar{\bf{x}}$ denotes the signal sampled from the same distribution as that of $\bf{x}$, and $\bar{\bf{G}}_{t}=[\bar{\bf{g}}_{1}\dots\, \bar{\bf{g}}_{t}]$ and  $\bar{\bf{F}}_{t}=[\bar{\bf{f}}_{1} \dots\, \bar{\bf{f}}_{t}]$ are row-wise IID, column-wise jointly Gaussian and independent of $\bar{\bf{n}}$ and $\bar{\bf{x}}$, i.e., for $i=1,\dots t$,  
  \BS\begin{align}
        [\bar{f}_{i,1},\dots \bar{f}_{i,t}] \overset{\rm IID}{\sim} \mathcal{CN}{(\bf{0}, \bf{V}_t^\phi)},\\
        [\bar{g}_{i,1},\dots \bar{g}_{i,t}] \overset{\rm IID}{\sim} \mathcal{CN}{(\bf{0}, \bf{V}_t^\gamma)}.
    \end{align}\ES
\end{lemma}

\emph{State Evolution of BO-OAMP/VAMP:} The iterative performance of BO-OAMP/VAMP can be predicted by state evolution as follows: Starting with $t=1$ and $v^\phi_{1}=1$,%
\BS\begin{align}
{\mr{LE:}} \qquad  v^\gamma_{t} &=\gamma_t^{\mr{SE}}(v^\phi_{t}),  \label{Eqn:SE_OAMP_LE}  \\
{\mr{NLE:}} \quad   {v}^\phi_{t+1}  &=  \phi_t^{\mr{SE}}(v^\gamma_{t}), \label{Eqn:SE_OAMP_NLE}
\end{align}\ES 
where $\gamma_t^{\mr{SE}}(\cdot)$ and $\phi_t^{\mr{SE}}(\cdot)$ are defined in \eqref{Eqn:OAMP_vara} and \eqref{Eqn:OAMP_varb}, respectively.

\subsection{BO-OAMP/VAMP is Sufficient-Statistic}\label{Sec:OAMP/VAMP_SS}

Since the local processors in BO-OAMP/VAMP are MMSE, we have the theorem below following Lemma \ref{Lem:SS_MAMP_MMSE}.

\begin{theorem}[BO-OAMP/VAMP is Sufficient-Statistic]\label{The:OAMP_properties}
   Suppose that Assumptions \ref{ASS:x}-\ref{ASS:A} and \ref{Ass:bouned} hold. The BO-OAMP/VAMP in \eqref{Eqn:BO-OAMP/VAMP} is an SS-MAMP.  
\end{theorem}

\begin{IEEEproof}
 Using Taylor series expansion, $\big(s_t \bf{I} + \bf{A}^{\mr H}\bf{A}\big)^{-1}$ in \eqref{Eqn:LMMSE_ori} can be rewritten to a polynomial in $\bf{A}^{\rm H}\bf{A}$. Therefore, the BO-OAMP/VAMP-LE  in \eqref{Eqn:OAMP/VAMP_LE} is a special case of the MLE in \eqref{Eqn:MIP_LE}.  In addition, the orthogonal MMSE NLE in \eqref{Eqn:OAMP/VAMP_NLE} is Lipschitz-continuous for $\bf{x}$ under Assumption \ref{ASS:x} and an equivalent Gaussian observation\footnote{In BO-OAMP/VAMP, $\bf{r}_t$ can be treated as Gaussian observation of $\bf{x}$, which can be guaranteed by the orthogonality in \eqref{Eqn:error_orth}  (see Lemma \ref{Lem:IIDG}).}  $\bf{r}_t$ (see \cite[Lemma 2]{Takeuchi2017}).  Following Definitions \ref{Def:MIP}-\ref{Def:MAMP} and \eqref{Eqn:error_orth}, BO-OAMP/VAMP is a special instance of MAMP. Furthermore, in BO-OAMP/VAMP, the LE $\gamma_t(\cdot)$ in \eqref{Eqn:LMMSE_ori} coincides with that in \eqref{Eqn:MAMP_MMSE_g}, and the NLE $\phi_t(\cdot)$ in \eqref{Eqn:OAMP/VAMP_NLE} coincides with that in \eqref{Eqn:MAMP_MMSE_f}. Then, from Lemma \ref{Lem:SS_MAMP_MMSE}, 
 \BS\label{Eqn:se-OAMP}\begin{align}
    \gamma^{\rm ss}_t(\bf{y},\bf{x}_{t}) &= \gamma_t(\bf{y},\bf{x}_t),\\ 
    \phi^{\rm ss}_t(\bf{r}_{t}) &= \phi_t(\bf{r}_t).
 \end{align}\ES
    That is, BO-OAMP/VAMP is an SS-MAMP. 
\end{IEEEproof}
 
Therefore, BO-OAMP/VAMP inherits all the properties of SS-MAMP. The following is a corollary of Theorem \ref{The:OAMP_properties}.

\begin{corollary}[Memory is not Needed in BO-OAMP/VAMP]\label{Cor1:OAMP}
 Suppose that Assumptions \ref{ASS:x}-\ref{ASS:A} and \ref{Ass:bouned} hold. Then, memory is not needed in BO-OAMP/VAMP, i.e., for any $t$, $\bf{r}_t$ and  $\bf{x}_t$  are sufficient statistics of $\bf{x}$ given $\bf{R}_t$ and $\bf{X}_t$ (from the MMSE/LMMSE perspective), respectively:
            \BS\label{Eqn:ss_OAMP}\begin{align} 
               v^\gamma_{t}&=  {\mr{lmmse}}\{\bf{x}|\bf{y},\bf{x}_t\} \overset{\rm a.s.}{=} {\mr{lmmse}}\{\bf{x}|\bf{y},\bf{X}_t\},\\
               v^\phi_{t+1}&= {\mr{var}}\{\bf{x}|\bf{r}_t\}\overset{\rm a.s.}{=} {\mr{var}}\{\bf{x}|\bf{R}_t \}.
                \end{align} \ES   
\end{corollary}

The following is a corollary of Lemma \ref{Lem:SS},  Lemma \ref{Lem:converge} and Theorem \ref{The:OAMP_properties}.

\begin{corollary}[Monotony and Convergence of BO-OAMP/VAMP]
 Suppose that Assumptions \ref{ASS:x}-\ref{ASS:A} and \ref{Ass:bouned} hold. Then, the covariance matrices $\{\bf{V}^{{\gamma}}_{t}\}$ and $\{\bf{V}^{{\phi}}_{t}\}$ in BO-OAMP/VAMP are L-banded (see Definition \ref{Def:L-baned}). In addition, the state evolution of BO-OAMP/VAMP is convergent. In detail,  $\{v^{{\gamma}}_{t}\}$ and $\{v^{{\phi}}_{t}\}$ in BO-OAMP/VAMP are respectively monotonically decreasing and converge respectively to a certain value, i.e., 
           \BS\label{Eqn:conv_OAMP}\begin{align}
           v^{{\gamma}}_{t} &\le v^{{\gamma}}_{t'}, \;  \forall t' \le t, &
                \lim\limits_{t\to\infty} v^{{\gamma}}_{t} &\to v^{{\gamma}}_{\star},  \\
               v^{{\phi}}_{t} &\le v^{{\phi}}_{t'}, \;  \forall t' \le t, &
                \lim\limits_{t\to\infty} v^{{\phi}}_{t} &\to v^{{\phi}}_{\star}. 
           \end{align}\ES
\end{corollary}

The following is a corollary of  Lemma \ref{Lem:damp_useless} and Theorem \ref{The:OAMP_properties}.

\begin{corollary}[Damping is is not Needed in BO-OAMP/VAMP]\label{Cor:OAMP_damp}
 Suppose that Assumptions \ref{ASS:x}-\ref{ASS:A} and \ref{Ass:bouned} hold. Then, damping is not needed in BO-OAMP/VAMP. Specifically,  the optimal damping of $\bf{R}_t$ and $\bf{X}_t$ have the same MSE as that of $\bf{r}_t$ and $\bf{x}_t$, respectively, i.e.,
        \BS \label{Eqn:dam_OAMP}\begin{align}
            v^{{\gamma}}_{t} = \min_{\bf{\zeta}_{t}:\bf{\zeta}_{t}^{\rm T}\bf{1}=1} \tfrac{1}{N}\|\bf{R}_t  \bf{\zeta}_{t}-\bf{x} \|^2, \\
            v^{{\phi}}_{t} = \min_{\bf{\zeta}_{t}:\bf{\zeta}_{t}^{\rm T}\bf{1}=1} \tfrac{1}{N}\|\bf{X}_t  \bf{\zeta}_{t}-\bf{x} \|^2. 
        \end{align}   \ES
\end{corollary}

\begin{remark}
Theorem \ref{The:OAMP_properties} is based on the assumption that the system size is infinite. For finite-size matrix $\bf{A}$,   Assumption \ref{ASS:A} does not hold. As a result, the IID Gaussian property in \eqref{Eqn:IIDG_OAMP}, the state evolution in \eqref{Eqn:se-OAMP} and Corollary \ref{Cor1:OAMP} do not rigorously hold. Therefore, Corollary \ref{Cor:OAMP_damp} also does not rigorously hold, i.e., damping may have significant improvement in the MSE of AMP and OAMP/VAMP for finite-size matrix $\bf{A}$ \cite{Takeuchi2019damp, Schniter2019dAMP, Schniter2021dvamp}.
\end{remark}

\section{Sufficient-Statistic Bayes-Optimal MAMP}\label{Sec:SS-BO-MAMP}
In this section, based on the Bayes-Optimal MAMP (BO-MAMP) in \cite{Lei2020MAMPTIT}, we design a sufficient statistic BO-MAMP (SS-BO-MAMP) using the sufficient-statistic technique discussed in Section \ref{Sec:SS-MAMP}. Different from the BO-MAMP in \cite{Lei2020MAMPTIT} that  considers optimal damping only for the NLE outputs, the SS-BO-MAMP in this paper considers optimal damping only for the MLE outputs. SS-BO-MAMP obtains a faster convergence speed than the original BO-MAMP in \cite{Lei2020MAMPTIT}. In addition, the state evolution of SS-BO-MAMP is further simplified using the sufficient-statistic property. Most significantly, the proposed SS-BO-MAMP in this section is a powerful illustration of the fact that the key findings in this paper go beyond those of the literature \cite{Takeuchi2021OAMP} because the SS-BO-MAMP in this section is based on a Bayes-suboptimal long-memory matched filter, while the results in \cite{Takeuchi2021OAMP} are limited to the Bayes-optimal LMMSE linear estimator.  

\subsection{SS-BO-MAMP Algorithm}  
\begin{assumption}\label{Ass:lambda}
 Let $\lambda_t=\tfrac{1}{N} {\rm tr}\{(\bf{A}\bf{A}^{\rm H})^t\}$ and ${\lambda}^\dag\equiv[ \lambda_{\max}+ \lambda_{\min}]/2$, where $\lambda_{\min}$ and $\lambda_{\max}$ denote the minimal and maximal eigenvalues of $\bf{A}\bf{A}^{\rm H}$, respectively. Without loss of generality, we assume that $\{\lambda_{\min}, \lambda_{\max}\}$ and $\{\lambda_t, \forall t\le 2\mathcal{T}\}$ are known, where $\mathcal{T}$ is the maximum number of iterations. 
\end{assumption}

In practice, if $\{\lambda_t, \lambda_{\min}, \lambda_{\max}\}$ are unavailable, we can set \cite{Lei2020MAMPTIT} 
  \BE
     \lambda_{\min}^{\rm low} = 0,\qquad
     \lambda_{\max}^{\rm up} =(N\lambda_{2\mathcal{T}})^{1/2\mathcal{T}},\label{Eqn:lambda_up}
 \EE
and $\{\lambda_t\}$ can be approximated by \cite{Lei2020MAMPTIT} 
        \BS\label{Eqn:b_t_app}\BE
           \lambda_{t}  \overset{\mr{a.s.}}{=} \lim_{N\to \infty} \|\bf{s}_{t}\|^2, \;\; {\rm for}\;\; t=1, 2,\dots 
        \EE
        where  
        \BE
        \bf{s}_{t}  =\left\{\!\!\!\begin{array}{ll}
            \bf{A}\bf{s}_{t-1}, &\;\; {\rm if}\; t {\rm \;is\; odd} \vspace{1mm}\\
            \bf{A}^{\rm H} \bf{s}_{t-1}, & \;\; {\rm if}\; t {\rm \;is\; even} 
        \end{array}\right.,\label{Eqn:y_t} 
        \EE   \ES 
and  $\bf{s}_0\sim \mathcal{CN}(\bf{0}_{N\!\times1},\tfrac{1}{N}\bf{I}_{N\!\times \!N})$. Let $\bf{B}  = \lambda^\dag\bf{I} - \bf{A}\bf{A}^{\mr H}$. For $t\ge 0$, we define\footnote{If the eigenvalue distribution $p_\lambda$ is available, $b_t$ can be calculated by $b_t=\int_{\lambda_{\min}}^{\lambda_{\max}}  (\lambda^\dag-\lambda)^t p_{\lambda} \, d \, \lambda.$}   
\BS\label{Eqn:sym_defs}\begin{align} 
 \bf{W}_t &\equiv  \bf{A}^{\rm H}\bf{B}^{t}\bf{A},\\
 b_t &\equiv \tfrac{1}{N}{\rm tr}\{\bf{B}^t\} = \textstyle\sum_{i=0}^{t} \binom{t}{i} (-1)^i(\lambda^\dag)^{t-i}\lambda_i,\\  
   w_t &\equiv  \tfrac{1}{N}{\rm tr}\{\bf{W}_t\}= \lambda^\dag b_{t}- b_{t+1}.\label{Eqn:b_w}
\end{align}  \ES 
For $1\le i\le t$,   
\BS\label{Eqn:orth_parameters} \begin{align}
   \vartheta_{t, i}  &\equiv \left\{\!\!\! \begin{array}{ll}
    \xi_t, & i=t   \vspace{0.1cm}\\
     \xi_i  \textstyle\prod_{1=i+1}^t\theta_\tau, & i<t
  \end{array}\right.,   \\ 
  p_{t, i}&\equiv - \vartheta_{t, i}   w_{t-i}, \\  {\varepsilon}_t^\gamma &\equiv - \textstyle\sum_{i=1}^t  p_{t, i},\\
   \bar{ w}_{i,j} &\equiv \lambda^\dag  w_{i+j}- w_{i+j+1} -  w_{i} w_{j}.\label{Eqn:bbar_w}
\end{align} \ES     
 Furthermore, $\vartheta_{t, i}=1$ if $i>t$.  The following is an SS-BO-MAMP algorithm. 
 
 \begin{framed}
 \emph{SS-BO-MAMP Algorithm:} Consider  
 \BS\label{Eqn:SS-MAMP}\BE
 {\bf{z}}_{t}= 
 \theta_t  \bf{B} {\bf{z}}_{t-1} +   \xi_t(\bf{y} - \bf{A}\bf{x}_t).  
 \EE 
Let $\tilde{\bf{R}}_{t} = [\tilde{\bf{r}}_{1} \dots \tilde{\bf{r}}_{t}]$. We construct an SS-BO-MAMP as: Starting with $t=1$ and ${\bf{x}}_{1}={\bf{z}}_{0}=\bf{0}$,  
\begin{alignat}{2}
 \tilde{\bf{r}}_t & = \gamma_t \left(\bf{y},\bf{X}_t\right) = \tfrac{1}{{\varepsilon}^\gamma_t}\big( \bf{A}^{\mr H}{\bf{z}}_{t} - \bf{X}_t \bf{p}_t \big), \label{Eqn:MLE}\\%
  \bf{r}_t  &= \tilde{\bf{R}}_t  \bf{\zeta}_t^{\gamma},\label{Eqn:dam_g_SS}  \\ 
  \bf{x}_{t + 1}  & =  \phi_t(\bf{r}_t), 
\end{alignat}	 
where 
\BE\label{Eqn:damp_SS}
 \bf{\zeta}^\gamma_{\mathcal{I}^\gamma_t} = \frac{  \big[\tilde{\bf{V}}^{\gamma}_{\mathcal{I}^\gamma_t}\big]^{-1} \bf{1}}{\bf{1}^{\rm T} \big[\tilde{\bf{V}}^{\gamma}_{\mathcal{I}^\gamma_t}\big]^{-1}\bf{1}}, \quad\;\qquad \bf{\zeta}^\gamma_{\bar{\mathcal{I}}^\gamma_t}=\bf{0}, 
\EE\ES
 and $\phi_t(\cdot)$ is the same as that in OAMP/VAMP (see \eqref{Eqn:OAMP/VAMP_NLE}). 
 \end{framed}

The convergence of SS-BO-MAMP is optimized with  relaxation  parameters $\{\theta_t\}$, weights $\{\xi_t\}$ and damping vectors $\{ \bf{\zeta}^\gamma_t\}$.  We provide some intuitive interpretations of SS-BO-MAMP below. Please refer to  \cite{Lei2020MAMPTIT} for the details of BO-MAMP.
\begin{itemize}
    \item The normalization parameter $\{{\varepsilon}^\gamma_t\}$ and  $\bf{X}_t \bf{p}_t$ in MLE is used for the orthogonality in \eqref{Eqn:Orth_MAMP}.  
    \item The relaxation parameter $\theta_t = 1/(\lambda^\dagger+v_{t,t}^\phi/\sigma^2)$ minimizes the spectral radius of $\theta_t\bf{B}$ to improve the convergence of SS-BO-MAMP and ensure that SS-BO-MAMP has a replica Bayes-optimal fixed point if converges. 
    
    \item The optimal damping vector $\bf{\zeta}^\gamma_{t}$   ensures the sufficient-statistic property (see Theorem \ref{The:SS-MAMP_con}), which ensures and further improves the convergence of SS-BO-MAMP.  
    
    \item The weight $\xi_t$  adjusts the contribution of  $\bf{x}_t$ to the estimation $\bf{r}_t$. Hence, the optimized $\xi_t$ accelerates the convergence of SS-BO-MAMP. However, the choice of $\xi_t$ does not affect the state-evolution fixed point (i.e., replica Bayes optimality) of SS-BO-MAMP. The optimal $\xi_t$ is given by  \cite{Lei2020MAMPTIT}: $\xi_1=1$ and  for $ t\ge 2$,
     \BE\label{Eqn:opt_xi}    
    \xi_t= \left\{\!\!\begin{array}{ll}
        \dfrac{c_{t,2}c_{t,0}+c_{t,3}}{c_{t,1}c_{t,0}+c_{t,2}}, &  \quad {\rm if}\; c_{t,1}c_{t,0}+c_{t,2} \neq 0 \vspace{3mm}\\ 
     +\infty, & \quad {\rm otherwise} 
    \end{array}\right.,
    \EE 
    where 
    \BS\label{Eqn:cs}\begin{align}
        c_{t,0} &= -\textstyle\sum\limits_{i=1}^{t-1} p_{t, i}/ w_0,\\
        c_{t,1}&=\sigma^2  w_0+ v_{t}^{\phi}\bar{ w}_{0,0},\\
        c_{t,2}&= - \textstyle\sum\limits_{i=1}^{t-1} \vartheta_{t, i}   (\sigma^2 w_{t-i}+ v_{t}^{\phi}\bar{ w}_{0,t-i} ), \\
        c_{t,3}&= \!\!\textstyle\sum\limits_{i=1}^{t-1}\textstyle\sum\limits_{j=1}^{t-1}  \vartheta_{t, i}\vartheta_{t,j}\big(\sigma^2  w_{2t-i-j} \!+\! v^{\phi}_{\max(i,j)}\bar{ w}_{t-i,t-j}\big),  
    \end{align}\ES
     and $\{\varepsilon_t^\gamma, p_{t, i}, \vartheta_{t, i}, w_i, \bar{w}_{i,j}\}$ in \eqref{Eqn:sym_defs} and \eqref{Eqn:orth_parameters}. { Furthermore, we find that $\xi_t = 1 / (\sigma^2 + v_t^{\gamma})$ yields the performance close to that of the optimal $\xi_t$.} 

\end{itemize}

\begin{remark}
  Note that the NLE in BO-MAMP is the same as that of BO-OAMP/VAMP and is a local MMSE estimator. From Lemma~\ref{Lem:SS_MAMP_MMSE}, damping is unnecessary for the NLE in BO-MAMP. This is why damping is not used for the NLE in SS-BO-MAMP. In contrast, the MLE in BO-MAMP is not a local LMMSE estimator, and thus damping is required for the MLE in SS-BO-MAMP.
\end{remark}

\subsection{State Evolution}
Following Theorem \ref{The:SS-MAMP_con}, we know that the orthogonality in \eqref{Eqn:Orth_MAMP} and the asymptotically IID Gaussianity hold for  SS-BO-MAMP. Therefore, we  obtain the state evolution of SS-BO-MAMP as follows.

\begin{lemma}[State Evolution]\label{Pro:SE_var}
Suppose that Assumptions \ref{ASS:x}-\ref{ASS:A}, \ref{Ass:bouned} and \ref{Ass:lambda} hold. Then,  SS-BO-MAMP is an SS-MAMP. Therefore, the covariance matrices $\{\bf{V}^{{\gamma}}_{t}\}$ and $\{\bf{V}^{{\phi}}_{t}\}$ in  SS-BO-MAMP are L-banded, and can be predicted by state evolution: Starting with $v_{1}^{{\phi}}=1$,
\BE\label{SE_SSMAMP} 
 v_t^{\gamma} = \gamma_t^{\mr{SE-SS}}(\bf{v}_t^{{\phi}}),\qquad
  v_{t+1}^{{\phi}}   ={\phi}_t^{\mr{SE}}({v}_t^{\gamma}),    
\EE
where $\{\bf{v}_{t}^{{\phi}}\}$ are defined in \eqref{Eqn:vars}, and ${\phi}_t^{\mr{SE}}(\cdot)$ is given in \eqref{Eqn:SE_OAMP_NLE} i.e., the same as that of BO-OAMP/VAMP. In addition, following \eqref{Eqn:damped_v_g}, $\gamma_t^{\mr{SE-SS}}(\cdot)$ is given by  
\BE\label{Eqn:v_gamma_damp}
   v^{\gamma}_{t} =  
             \dfrac{1}{\bf{1}^{\rm T} \big[\tilde{\bf{V}}^{\gamma}_{\mathcal{I}^\gamma_t}\big]^{-1}\bf{1}}. 
\EE
For $1\le t'\le t$,
\BE\label{Eqn:covar_LE_t}
    \tilde{v}^{\gamma}_{t,t'} \!=\!\! \frac{1}{\varepsilon^\gamma_{t}\varepsilon^\gamma_{t'}} \!\!\sum\limits_{i=1}^{t}\!\sum\limits_{j=1}^{t'}  \vartheta_{t, i}\vartheta_{t'\!,j}[\sigma^2  w_{t+t'-i-j} \!+\! v^{{\phi}}_{\max(i,j)}\bar{ w}_{t-i,t'\!-j}],  
\EE  
where ${w}_{i}$ is defined in \eqref{Eqn:b_w} and $\{\varepsilon^\gamma_t, \vartheta_{t, i}, \bar{w}_{i,j}\}$ in \eqref{Eqn:orth_parameters}.
\end{lemma}

Eqn. \eqref{Eqn:covar_LE_t} in Lemma \ref{Pro:SE_var}  follows the $\gamma_t^{\rm SE}$ in \cite{Lei2020MAMPTIT}. To avoid the high-complexity matrix inverse in \eqref{Eqn:v_gamma_damp}, we can calculate $\big[\tilde{\bf{V}}^{\gamma}_{\mathcal{I}^\gamma_t}\big]^{-1}$ based on $\big[\tilde{\bf{V}}^{\gamma}_{\mathcal{I}^\gamma_{t-1}}\big]^{-1}$ in the last iteration using the block matrix inversion in \eqref{Eqn:bm_inv}.

 \subsection{Improvement and Replica Bayes Optimality}
 The following lemma follows Theorem \ref{The:SS-MAMP_con} straightforwardly.
 
 \begin{lemma}[MSE Improvement of SS-BO-MAMP]\label{Lem:SS-BO-MAMP_better}
    Suppose that  Assumptions \ref{ASS:x}-\ref{ASS:A}, \ref{Ass:bouned} and \ref{Ass:lambda} hold. The MSE of SS-BO-MAMP is not worse than that of the original BO-MAMP in \cite{Lei2020MAMPTIT}. 
\end{lemma}


\begin{lemma}[Convergence and  Bayes Optimality]\label{Lem:Conv_SS-BO-MAMP}
Suppose that  Assumptions \ref{ASS:x}-\ref{ASS:A}, \ref{Ass:bouned} and \ref{Ass:lambda} hold. The state evolution of SS-BO-MAMP  converges to the same fixed point as that of OAMP/VAMP. That is, the SS-BO-MAMP is replica Bayes optimal if its state evolution has a unique fixed point. 
\end{lemma}
 
{ \begin{IEEEproof}
   Following Theorem \ref{The:SS-MAMP_con}(b), there exists a finite integer $I$ such that  $\lim_{t\to \infty} \bf{V}_{I\to  t}^\phi \overset{\rm a.s.}{=} v^\phi_{\star}\bf{1}$. The inputs $[\bf{x}_1 \dots \bf{x}_I]$ of $\gamma_t^{\rm SE}$ correspond to the high-order (i.e., $t-1$ to $t-I$) terms of the Taylor series, which tend to zero as $t\to \infty$ (see Appendix G-B in \cite{Lei2020MAMPTIT}). Then, following Theorem \ref{The:SS-MAMP_con}(b), we have
    \BE
        \tilde{v}_\star^{\gamma} = \lim_{t\to \infty} \tilde{v}_{t,t}^{\gamma} \overset{\rm a.s.}{=}  \lim_{t\to \infty} [\gamma^{\rm SE}_t(v^\phi_{\star}\bf{1})]_{t,t}.
    \EE
    Since $\phi_t$ is non-memory, following Theorem \ref{The:SS-MAMP_con}(b), we have
    \BE
        v^\phi_{\star} \overset{\rm a.s.}{=} \lim_{t\to \infty} \phi^{\rm SE}_t(v^\gamma_{\star}).
    \EE
    Furthermore, when SS-MAMP converges (i.e., $t\geq I$), the damping operation in \eqref{Eqn:SS-MAMP} follows a back-off mechanism. That is, $\bf{r}_t=\bf{r}_I, \forall t\geq I$. Therefore, $\bf{x}_t=\bf{x}_{I+1}=\bf{x}_\star, \forall t\geq I+1$. As a result, 
    \BE
        \lim_{t\to \infty}\gamma_t(\bf{y},\bf{X}_t)  \overset{\rm a.s.}{\to} \lim_{t\to \infty} \gamma_t(\bf{y},\bf{x}_\star\bf{1}_{1\times t}),
    \EE
    The right term is in fact a Taylor expansion of the LMMSE-LE $\gamma_t(\bf{y},\bf{x}_\star)$ of BO-OAMP/VAMP in \eqref{Eqn:BO-OAMP/VAMP}. Then,  following the second point in Lemma \ref{Lem:SS_MAMP_MMSE}, when $t\to \infty$, damping is unnecessary for SS-BO-MAMP, i.e., $ \tilde{v}_\star^{\gamma} = v^\gamma_{\star}$. Therefore, the state evolution of SS-BO-MAMP converges the fixed point given by  
    \BS
        \begin{align}\label{Eqn:SS-BO-FP}
             {v}_\star^{\gamma} &=   \lim_{t\to \infty} [\gamma^{\rm SE}_t(v^\phi_{\star}\bf{1})]_{t,t},\\
            v^\phi_{\star} &= \lim_{t\to \infty} \phi^{\rm SE}_t(v^\gamma_{\star}),
        \end{align}
    \ES
    which is the same as the fixed-point equations  of OAMP/VAMP (see Appendix G in \cite{Lei2020MAMPTIT}). In addition, the Bayes optimality of OAMP/VAMP is proven via the replica methods in \cite{Kabashima2006, Ma2016},  and lately rigorously for right-unitarily-invariant $\bf{A}$ when  its state evolution has a unique fixed point in~\cite{Li2022random}. Therefore, we obtain Lemma \ref{Lem:Conv_SS-BO-MAMP}. 
\end{IEEEproof}}

\subsection{Complexity of SS-BO-MAMP}
 We assume that $\bf{A}$ has no special structures, such as DFT or sparse matrices. Let $\mathcal{T}$ be the number of iterations.  
\begin{itemize}
    \item SS-BO-MAMP costs ${\cal O}(MN\mathcal{T} )$ time complexity for matrix-vector multiplications\footnote{The number of matrix-vector multiplications can be reduced by storing an intermediate variable for $\bf{A}\hat{\bf{r}}_t$ to prevent its double computation.} $\{\bf{A}\bf{A}^{\rm H}{\bf{z}}_t\}$ and $\{\bf{A}\bf{x}_t\}$,   ${\cal O}\big((N\!+\!M)\mathcal{T}^2 \big)$ for $\{\textstyle\sum_{i=1}^t p_{t,i}\bf{x}_i\}$ and ${\cal O}(\mathcal{T}^4)$ for calculating $\{ \bf{V}_{t}^{\gamma}\}$ (see \eqref{Eqn:covar_LE_t}). {  The number of matrix-vector multiplications can be reduced by storing an intermediate variable to avoid the recalculation of $\bf{A}{\bf{z}}_t$. } 
    
    \item SS-BO-MAMP needs ${\cal O}(MN)$ space  to store $\bf{A}$, ${\cal O}\big((M+N)\mathcal{T}\big)$ space for $\{\bf{x}_t, \bf{r}_t\}$, and ${\cal O}\big(\mathcal{T}^2\big)$ for $\{\bf{V}_{t}^{\gamma}\}$. 
\end{itemize}

\section{Simulation Results} \label{Sec:Simulation}
Consider a compressed sensing problem: $\forall i$,
 \BE
 x_i\sim \left\{\begin{array}{ll}
      0,&  {\rm probability} = 1- \mu  \\
      {\cal N} (0, \mu^{-1}),&  {\rm probability} = \mu
 \end{array}\right..
 \EE
The power of $x_i$ is normalized to 1. The SNR is defined as ${\rm SNR}=1/\sigma^2$. Let  $\bf{A}=\bf{U \Sigma V^{\rm H}}$ be the SVD of $\bf{A}$. We rewritten the system in \eqref{Eqn:unitary_sys} to
 ${\bf{y}} = {\bf{U \Sigma V}}^{\rm H}\bf{x} + {\bf{n}}$. Note that $\bf{U}^{\mr{H}}\bf{n}$ has the same distribution as $\bf{n}$. Thus, we can assume $\bf{U} =\bf{I}$ without loss of generality. To reduce the  complexity of OAMP/VAMP, unless otherwise specified, we approximate a large random unitary matrix by ${\bf{V}}^{\rm H} = {\bf{\Pi F}}$, where $\bf{\Pi}$ is a random permutation matrix and $\bf{F}$ is a DFT matrix. Note that all the algorithms involved here admit fast implementation for this matrix model. The entries $\{d_i\}$ of diagonal matrix $\bf{\Sigma}$ are generated as: $d_i/d_{i+1}=\kappa^{1/
J}$ for $i = 1,\ldots, J-1$ and $\sum_{i=1}^Jd_i^2=N$, where $J=\min\{M, N\}$ \cite{Vila2014}. Here, $\kappa\ge1$ controls the condition number of $\bf{A}$. Note that BO-MAMP does not require the SVD structure of $\bf{A}$. BO-MAMP only needs the right-unitarily invariance of $\bf{A}$.

\subsection{BO-OAMP/VAMP is Sufficient-Statistic}
\begin{figure}[t] 
\centering
\begin{tabular}{cc}
\includegraphics[width=.23\textwidth]{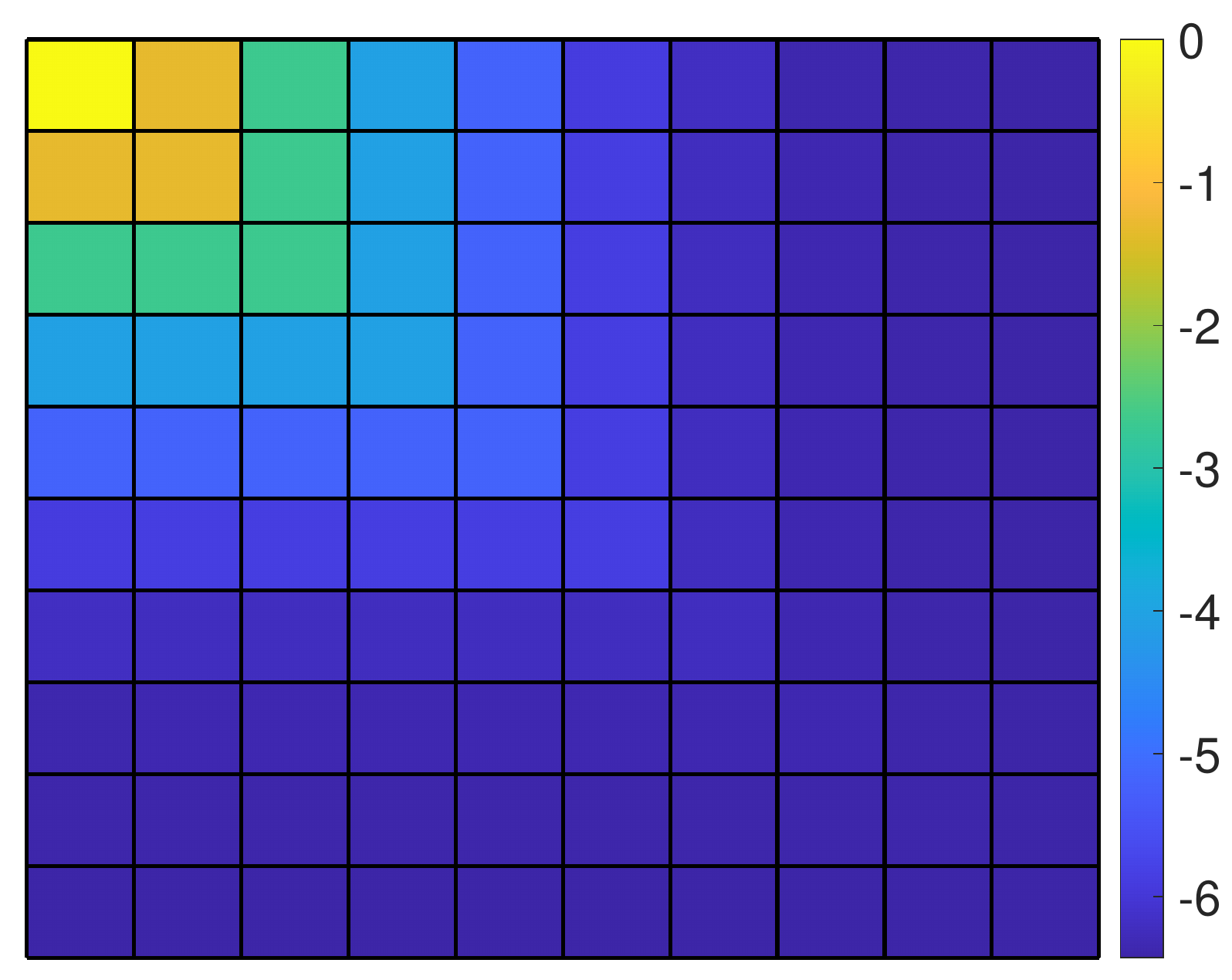} &
\includegraphics[width=.23\textwidth]{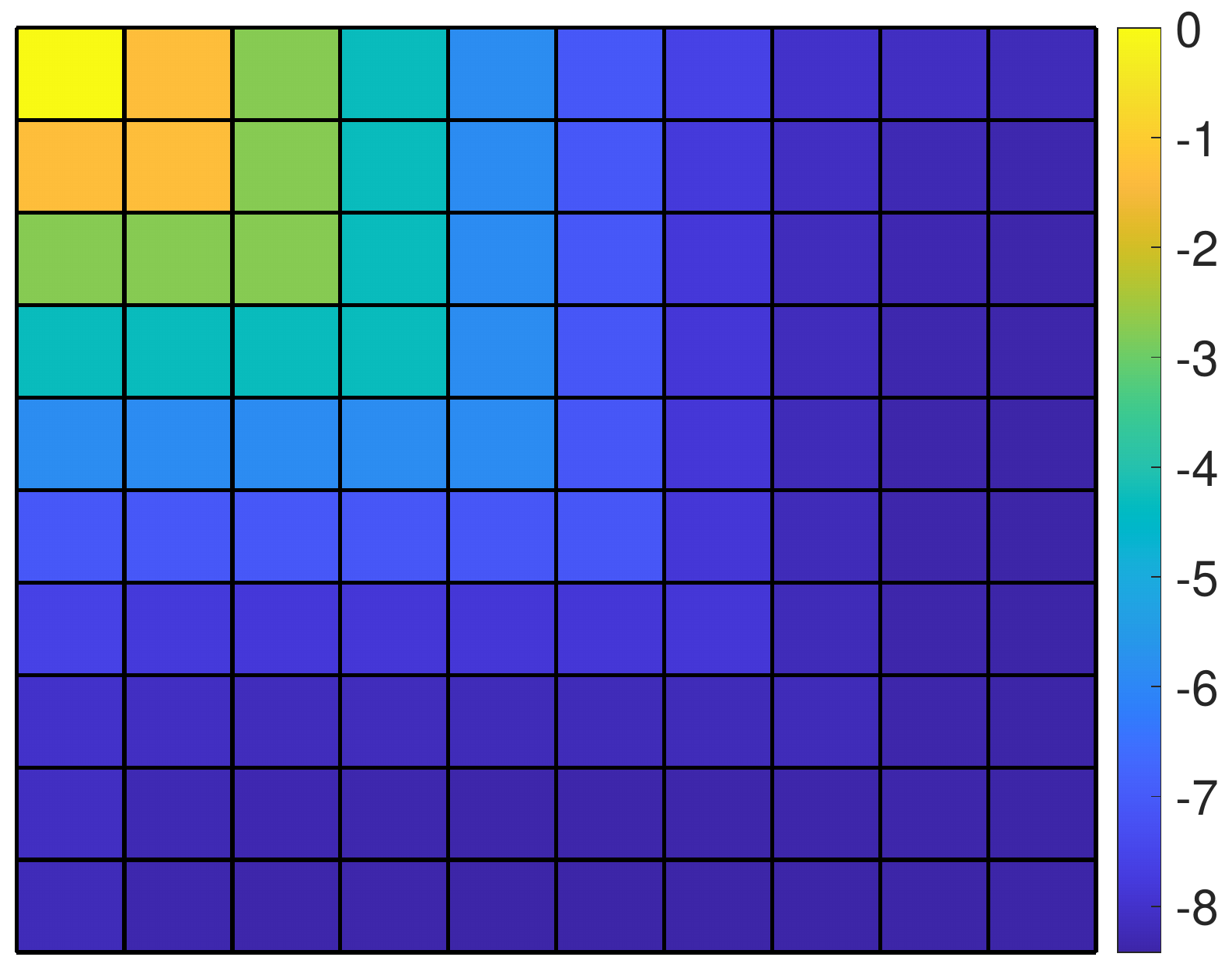} \\
\footnotesize{(a) $\log(\bf{V}_{10}^\gamma)$  in BO-OAMP/VAMP}  & \footnotesize{(b) $\log(\bf{V}_{10}^\phi)$  in BO-OAMP/VAMP}
\end{tabular}
\caption{Covariance matrices ${\bf{V}}^\gamma_t$ and ${\bf{V}}^\phi_t$ in BO-OAMP/VAMP. $N=16384$, $\delta=M/N=0.5$, $\mu=0.1$, $SNR=30$ dB and $\kappa=10$ and $t=10$. } \label{Fig:OAMP_L_V}
\end{figure}

To verify that BO-OAMP/VAMP is sufficient-statistic, we plot the covariance matrices ${\bf{V}}^\gamma_t$ and ${\bf{V}}^\phi_t$ of BO-OAMP/VAMP in Fig.~\ref{Fig:OAMP_L_V}. As can be seen, the covariance matrices in BO-OAMP/VAMP are L-banded, i.e., the elements in each L band of the covariance matrix are almost the same. Therefore, following Lemma \ref{Lem:SS}, we know that BO-OAMP/VAMP is sufficient-statistic,  which coincides with the theoretical result in Theorem \ref{The:OAMP_properties}.

\subsection{SS-BO-MAMP is Sufficient-Statistic}
\begin{figure}[t] 
\centering
\begin{tabular}{cc}
\includegraphics[width=.23\textwidth]{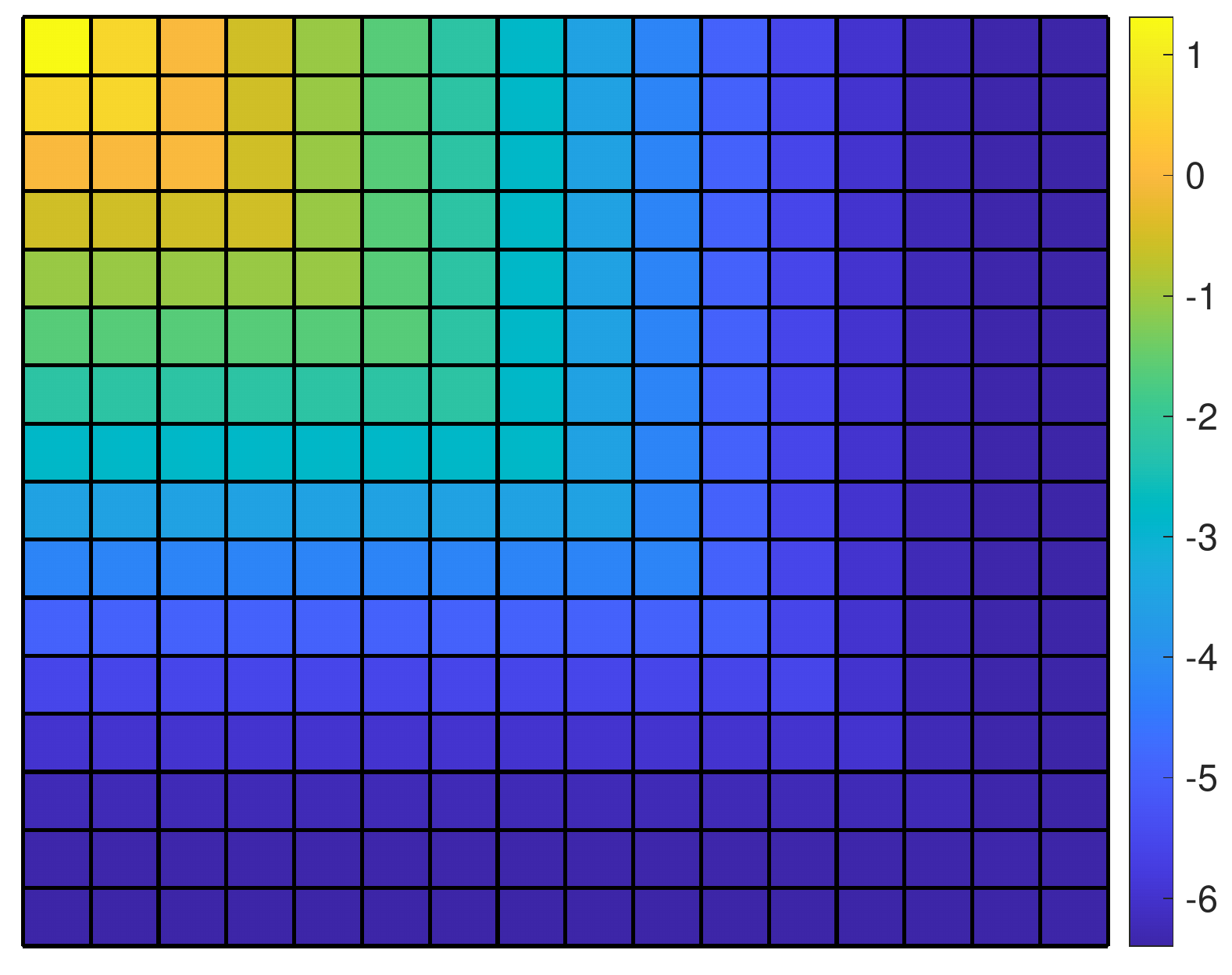}          &
\includegraphics[width=.23\textwidth]{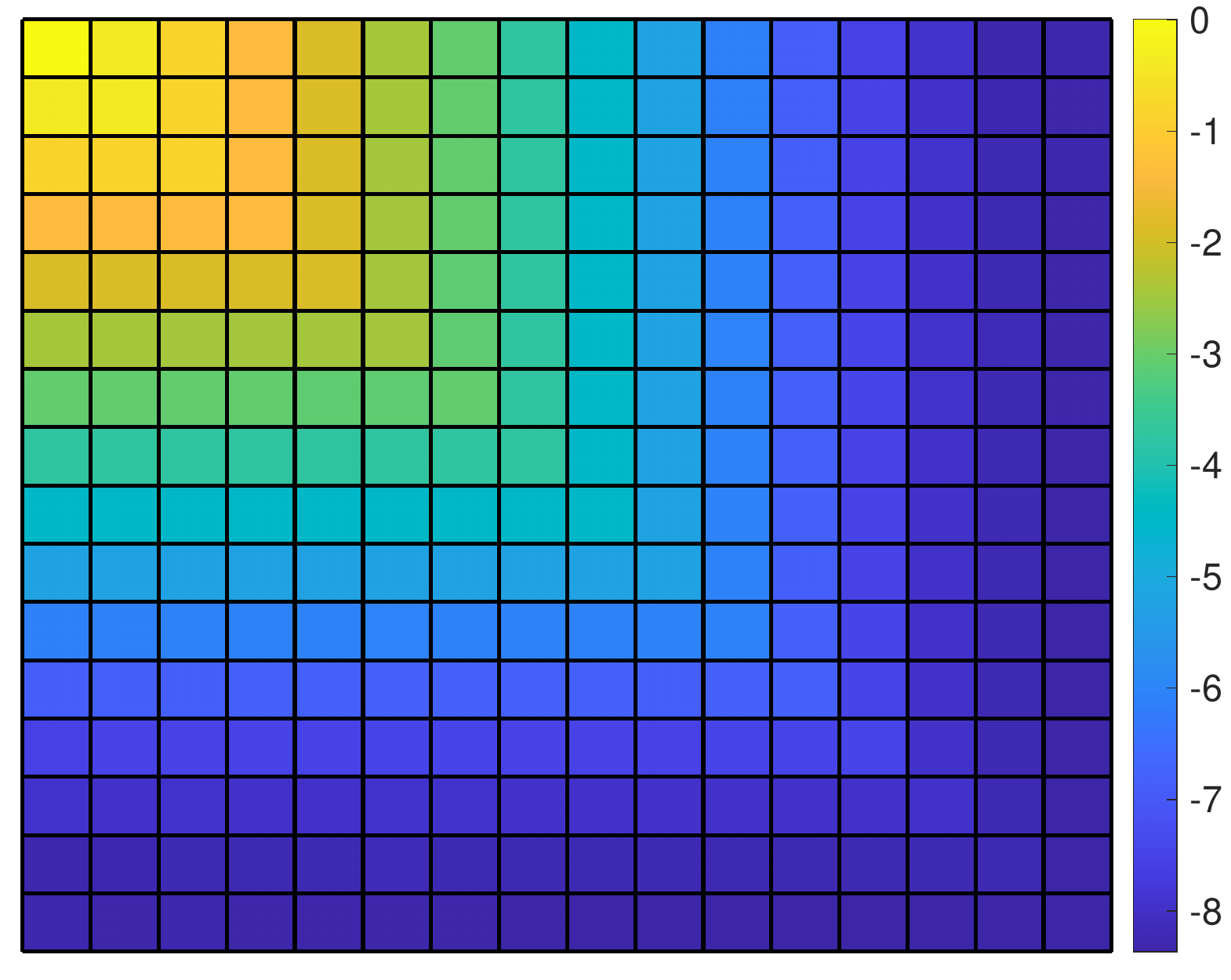} \\
\footnotesize{(a) $\log(\bf{V}_{16}^\gamma)$  in SS-BO-MAMP}  & \footnotesize{(b) $\log(\bf{V}_{16}^\phi)$  in SS-BO-MAMP}
\end{tabular}
\caption{Covariance matrices ${\bf{V}}^\gamma_t$ and ${\bf{V}}^\phi_t$ in SS-BO-MAMP. $N=16384$, $\delta=M/N=0.5$, $\mu=0.1$, $SNR=30$ dB, $\kappa=10$ and $t=16$.  } \label{Fig:MAMP_L_V} 
\end{figure}

To verify that SS-BO-OAMP/VAMP is sufficient-statistic, we plot the covariance matrices ${\bf{V}}^\gamma_t$ and ${\bf{V}}^\phi_t$ of SS-BO-MAMP in Fig.~\ref{Fig:MAMP_L_V}. As can be seen, the covariance matrices in SS-BO-MAMP are L-banded, i.e., the elements in each L band of the covariance matrix are almost the same. Therefore, following Lemma \ref{Lem:SS}, we know that SS-BO-MAMP is sufficient-statistic. 

  \begin{figure}[t]
  \centering 
  \includegraphics[width=.36\textwidth]{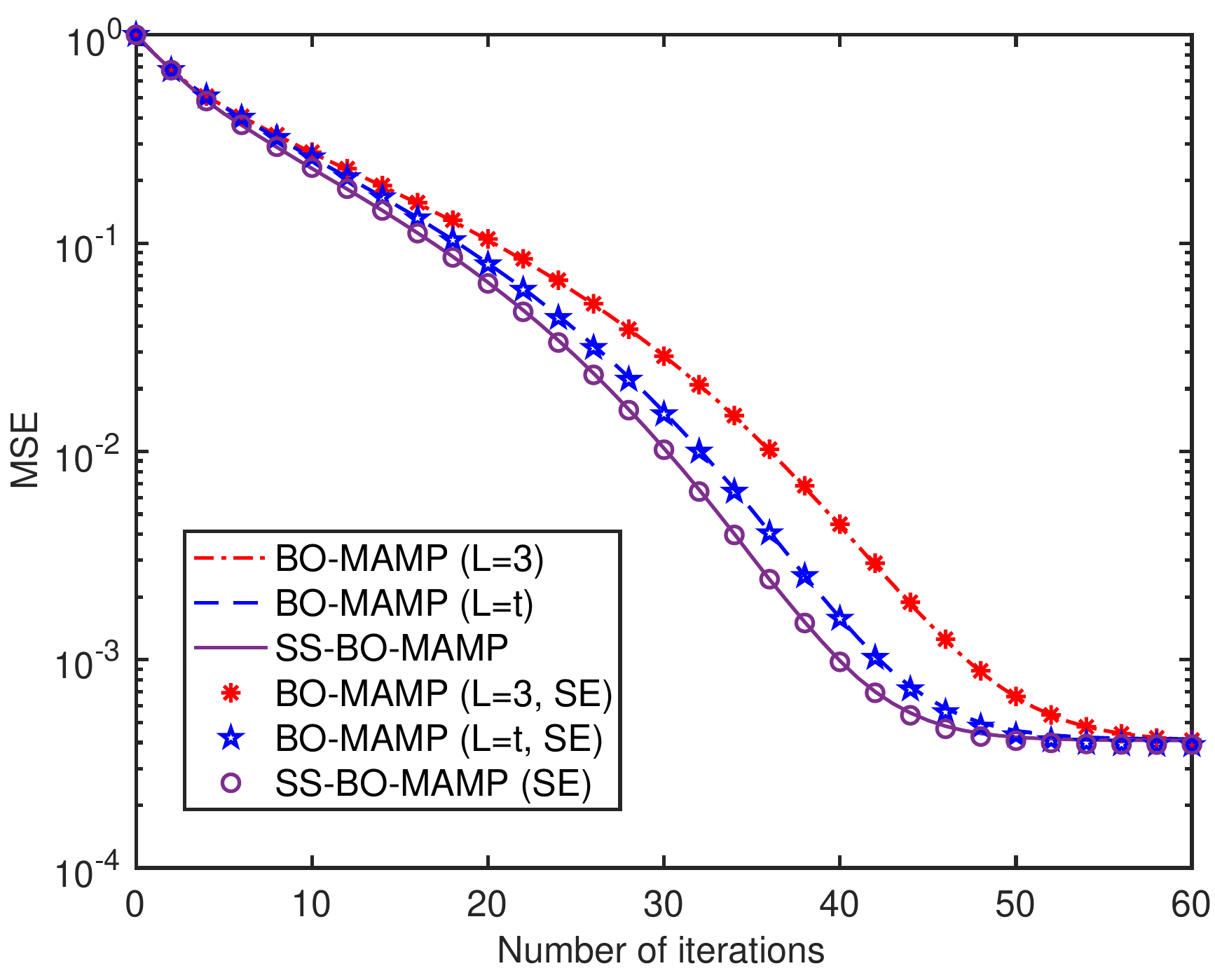}\\  
  \caption{MSE versus the number of iterations for SS-BO-MAMP and BO-MAMP in \cite{Lei2020MAMPTIT}.  $N=16384$, $\delta=M/N=0.5$, $\mu=0.1$, $SNR=30$ dB and $\kappa=100$. SE denotes state evolution.}\label{Fig:Compare}  
  \end{figure} 

\subsection{Comparison of BO-MAMP and SS-BO-MAMP}\label{Sec:sim_SS-BO-MAMP}
Fig.~\ref{Fig:Compare} shows the MSEs versus the number of iterations for SS-BO-MAMP and BO-MAMP. On the one hand, as can be seen in Fig.~\ref{Fig:Compare}, the MSE of SS-BO-MAMP is slightly better than those of BO-MAMPs in \cite{Lei2020MAMPTIT} (with damping length $L=3$ and $L=t$). This coincides with the theoretical result in Lemma \ref{Lem:SS-BO-MAMP_better} that the MSE of SS-BO-MAMP is not worse than that of BO-MAMP. Furthermore, since the BO-MAMP in \cite{Lei2020MAMPTIT} is replica Bayes optimal, SS-BO-MAMP is also replica Bayes optimal as it converges to the same performance as that of BO-MAMP. This coincides with the theoretical result in Lemma \ref{Lem:Conv_SS-BO-MAMP}. On the other hand, compared to the BO-MAMP with damping length $L=3$, the improvements of SS-BO-MAMP and the fully damped BO-MAMP (e.g., $L=t$) are negligible. Therefore, due to its low complexity and stability, the BO-MAMP with $L=3$ in \cite{Lei2020MAMPTIT} is more attractive in practice than SS-BO-MAMP and the fully damped  BO-MAMP. However, it would be meaningful to use the state evolution of SS-BO-MAMP as the limit (or a lower bound) for the performance of the damped BO-MAMP. 

 \emph{Instability of SS-MAMP}:  Based on our numerical results, for small-to-medium size systems or when the condition number of $\bf{A}$ is very large, the full damping in SS-MAMP is unstable. As a result, SS-MAMP may fail to converge. How to solve the instability of the full damping in SS-MAMP is an interesting and important future work.

\begin{figure}[t]
  \centering 
  \includegraphics[width=.37\textwidth]{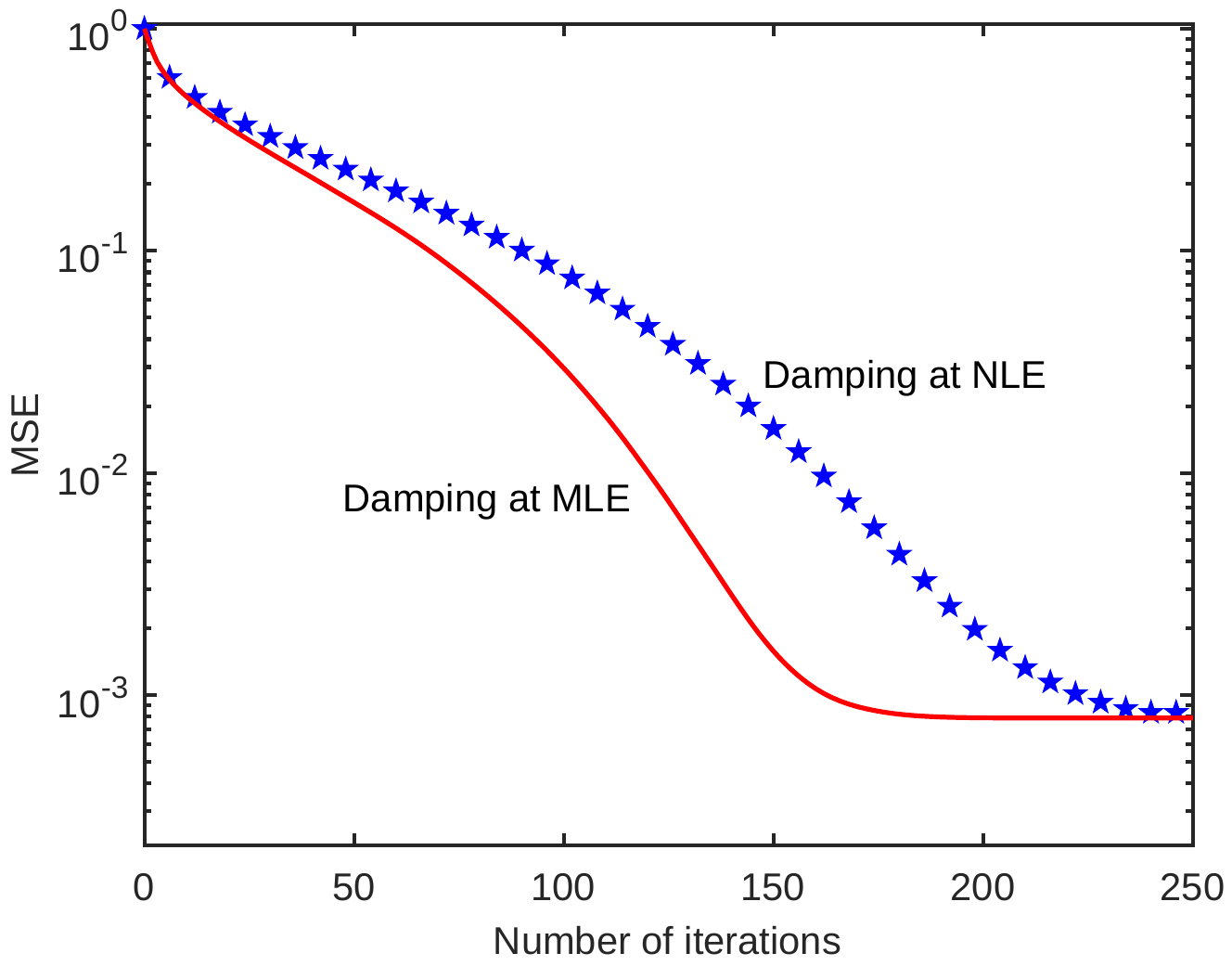}\\  
  \caption{MSE versus the number of iterations for BO-MAMP with short-memory damping ($L=2$).  $N=16384$, $\delta=M/N=0.5$, $\mu=0.1$, $SNR=30$ dB and $\kappa=500$.}\label{Fig:MLE_damping}  
  \end{figure}

{\subsection{Damping at MLE in BO-MAMP}\label{Sec:MLE_damp}
Lemma~\ref{Lem:SS_MAMP_MMSE} implies that when the NLE is locally Bayes-optimal, damping at MLE is preferable to damping at NLE in BO-MAMP, even for short-memory damping (i.e., $L<t$), which is commonly used in practice since MAMP is more stable with smaller $L$. To validate this claim, Fig. \ref{Fig:MLE_damping} compares the MSE of BO-MAMP with short-memory damping ($L=2$) at MLE and that at NLE. It shows that damping at MLE in BO-MAMP converges much faster than damping at NLE, while their per-iteration complexities are the same.
}
 
 
\section{Conclusion}

This paper proposes an optimal sufficient-statistic memory AMP (SS-MAMP) framework (from the local MMSE/LMMSE perspective) for AMP-type algorithms and solves the convergence problem of the state evolution for AMP-type algorithms in principle. We show that the covariance matrices of SS-MAMP are L-banded and convergent. Given an arbitrary MAMP, we construct an SS-MAMP by optimal damping. The dynamics of the constructed SS-MAMP can be rigorously described by state evolution.  

We prove that the BO-OAMP/VAMP is an SS-MAMP. As a result, we reveal two interesting properties of BO-OAMP/VAMP for large systems: 1) the covariance matrices of BO-OAMP/VAMP are L-banded and are convergent, and 2) damping and memory are not needed (i.e., do not bring performance improvement) in BO-OAMP/VAMP. These recover the main statements in \cite{Takeuchi2021OAMP}.

An SS-BO-MAMP is constructed based on the BO-MAMP. We show that 1) the SS-BO-MAMP is replica Bayes optimal, and 2) the MSE of SS-BO-MAMP is not worse than the original BO-MAMP in \cite{Lei2020MAMPTIT}.  Numerical results are provided to verify that BO-OAMP/VAMP and SS-BO-MAMP are sufficient-statistic, and verify the fact that SS-BO-MAMP outperforms the original BO-MAMP.  

The sufficient-statistic technique proposed in this paper  not only applies to OAMP/VAMP and MAMP, but also applies to more general iterative algorithms. It is an interesting future work to apply the sufficient-statistic technique in this paper to other well-known iterative algorithms, including AMP \cite{Donoho2009}, UTAMP \cite{UTAMPa, UTAMPb}, CAMP \cite{Takeuchi2020CAMP}, long-memory AMP \cite{Opper2016, Fan2020arxiv}, rotationally invariant AMP \cite{Ramji2021RIAMP}, GMAMP \cite{Tian2021GMAMP}, et al.

\section*{Acknowledgment}  
The authors thank Ramji Venkataramanan (Associate Editor) and the anonymous reviewers for their comments that have improved the quality of the manuscript greatly.  

\appendices

  \section{Proof of Lemma \ref{Lem:useless_damp}}\label{APP:damp_useless}
  Since $\bf{s}_t = x\bf{1}_{t\times 1}  + \bf{n}_t$ and  $\bf{n}_t$ is zero mean with covariance matrix $\bf{V}_t$, we have
    \BE
       \mathbb{E}\big\{\|\bf{\zeta}_{t}^{\rm T}\bf{s}_t   - x \|^2\big\} =  \bf{\zeta}_{t}^{\rm H} \bf{V}_t \bf{\zeta}_{t}.
   \EE
    The damping optimization is equivalent to solving the following quadratic programming problem.
\BS\begin{align}
    & \min_{\bf{\zeta}_t} \quad    \tfrac{1}{2} \bf{\zeta}_t^{\rm H} \bf{V}_t \bf{\zeta}_t,\\
   & \; {\rm s.t.} \quad \bf{1}^{\rm T} \bf{\zeta}_t =1.
\end{align}\ES 
Since $\bf{V}_t$ is positive semi-definite, it is a convex function with respect to $\bf{\zeta}_t$. We write the Lagrangian function as 
    \BE
    {\cal L}(\bf{\zeta}_t, c) =  \tfrac{1}{2} \bf{\zeta}_t^{\rm H} \bf{V}_t \bf{\zeta}_t + c (1-\bf{1}^{\rm T}\bf{\zeta}_t). 
    \EE
    Therefore, the solutions of the following equations minimize the objective function.  
    \BS\begin{align}
        \nabla_{\!\!\bf{\zeta}_t} {\cal L}(\bf{\zeta}_t, c) &= \bf{0},\\
        \partial {\cal L}(\bf{\zeta}_t, c)/\partial c &=0.
    \end{align}\ES 
     That is, 
    \begin{align}\label{Eqn:zeta_opt_cond}
         \bf{V}_t \bf{\zeta}_t    =  c  \bf{1},\qquad
        \bf{1}^{\rm T}\bf{\zeta}_t  =1.
    \end{align}
    Since $ v_{1,t} = v_{2,t} \dots =  v_{t,t}$, it is easy to verify that $\{\bf{\zeta}_t = [0\dots\, 0, 1]^{\rm T}, c=v_{t,t}\}$ is a solution of \eqref{Eqn:zeta_opt_cond}. Hence, $\bf{\zeta}_{t}=[0 \dots 0, 1]$ minimizes the MSE, i.e.,
    \BE
     [0\dots 0\;1]^{\rm T} = {\argmin}_{\bf{\zeta}_{t}^{\rm T}\bf{1}=1}\;  \|\bf{\zeta}_{t}^{\rm T}\bf{s}_t -{x} \|^2. 
    \EE 
     Therefore, we complete the proof of Lemma \ref{Lem:useless_damp}.

\section{Proof of Lemma \ref{Lem:SS_matrix_inv}}\label{APP:SS_matrix_inv}
 Define $\bf{D}\equiv \bf{V}_t^{-1}\bf{V}_t=\{d_{i,j}\}$. Following \eqref{Eqn:SS_cov}  and  \eqref{Eqn:SS_V_inv}, we have  the following.
      \begin{itemize}
          \item First, we consider the diagonal elements of $\bf{D}$.  For any $1<i<t$, 
              \BS\label{Eqn:dia1} \begin{align}
                  &d_{i,i} = v_{i,i-1}^\dagger v_{i-1,i} + v_{i,i}^\dagger v_{i,i} + v_{i,i+1}^\dagger v_{i+1,i} \nonumber\\
                  &=  \frac{v_i}{v_i\!-\!v_{i-1}} \! +\! \left[\frac{v_i}{v_{i-1}\!-\!v_i}\!+\!\frac{v_i}{v_{i}\!-\!v_{i+1}}\right] \!+\!  \frac{v_{i+1}}{v_{i+1}\!-\!v_{i}} \nonumber\\
                   &= 1.
              \end{align}\ES 
              In addition,
               \BS\label{Eqn:dia2} \begin{align}
                d_{1,1} &= v_{1,1}^\dagger v_{1,1} + v_{1,2}^\dagger v_{2,1}\nonumber\\
                &= \frac{v_1}{v_1-v_2}  +\frac{v_2}{v_2-v_1}  \nonumber\\
                &=1,  \\
                d_{ t, t} &= v_{ t, t-1}^\dagger v_{ t-1, t} + v_{ t, t}^\dagger v_{ t, t}\nonumber\\
                &= \frac{v_t }{v_t-v_{t-1}} + \left[\frac{v_t }{v_{t-1}-v_t} + \frac{v_t }{v_t} \right] \nonumber\\
                &=1. 
              \end{align}\ES
             
            Following \eqref{Eqn:dia1} and \eqref{Eqn:dia2}, we have
            \BE\label{Eqn:dia}
                d_{i,i} = 1, \quad \forall 1\le i\le  t.
            \EE
          \item Then, we consider the non-diagonal elements of $\bf{D}$.  For any $1<j\le  t$, 
                  \begin{align}\label{Eqn:non-dia1} 
                      d_{j,1} \!=\! d_{1,j}  
                     \! = \! v_{1,1}^\dagger v_{1,j} \!+\! v_{1,2}^\dagger v_{2,j} 
                      \!= \frac{v_j \! -\!   v_j}{v_1\!-\!v_2}
                       \! =\! 0.
                  \end{align}  
                For any $1<i<j\le t$, 
                   \begin{align}\label{Eqn:non-dia2}
                      & d_{j,i} = d_{i,j} \nonumber\\
                      &= v_{i,i-1}^\dagger v_{i-1,j} + v_{i,i}^\dagger v_{i,j} + v_{i,i+1}^\dagger v_{i+1,j} \nonumber\\
                      &=  \frac{v_j}{v_i-v_{i-1}} \!+\! \left[\frac{v_j}{v_{i-1}\!-\!v_i} \!+\! \frac{v_j}{v_{i}\!-\!v_{i+1}}\right] \!+\!  \frac{v_j}{v_{i+1}\!-\!v_{i}} \nonumber\\
                       &= 0.
                  \end{align}   
                  Following \eqref{Eqn:non-dia1} and \eqref{Eqn:non-dia2}, we have
                    \BE\label{Eqn:non-dia}
                        d_{i,j} = 0, \quad \forall i\neq j.
                    \EE 
      \end{itemize}
      From \eqref{Eqn:dia} and \eqref{Eqn:non-dia}, we have
      $\bf{D}=\bf{I}$.  That is, $\bf{V}_t^{-1}$ in \eqref{Eqn:SS_V_inv} is the  inverse of   $\bf{V}_t$. The remainder of Lemma 6 follows straightforwardly from \eqref{Eqn:SS_V_inv}. Thus,  we complete the proof of Lemma \ref{Lem:SS_matrix_inv}.

\section{Proof of Lemma \ref{Lem:SS_Est}}\label{APP:SS_Est}
     Since $\bf{s}_t = x\bf{1} + \bf{n}_t$ and $\bf{n}_t\sim \mathcal{CN}(\bf{0},\bf{V}_t)$, we have
    \BS\label{Eqn:ps}\begin{align}
     \!\!\! p(x|s_t) & = 
         \frac{p(x)p(s_t|x)}{p(s_t)} \\
                   & \propto   p(x)e^{-\frac{\|s_t - x\|^2}{v_{t,t}}} \\
                   &\propto p(x) e^{-(s_t - x)^{*} v_{t,t}^{-1}  (s_t - x)} \\ 
                   & = p(x) e^{-v_{t,t}^{-1} \|x\|^2 \,+\,  v_{t,t}^{-1}s_t^*x \,+\,   v_{t,t}^{-1}x^*s_t}, \end{align}\begin{align}
      \!\!\!\!\!\!  p(x|\bf{s}_t) & = \frac{p(x)p(\bf{s}_t|x)}{p(\bf{s}_t)}\\ 
                    & \propto p(x) e^{-(\bf{s}_t - x\bf{1})^{\rm H} \bf{V}_t^{{-1}} (\bf{s}_t - x\bf{1})} \\ 
                    &  \propto p(x) e^{-\bf{1}^{\rm T}\bf{V}_t^{{-1}}\bf{1} \|x\|^2 \,+\,  \bf{s}_t^{\rm H}\bf{V}_t^{{-1}}\bf{1} x \,+\,   x^{\rm H}\bf{1}^{\rm T}\bf{V}_t^{{-1}} \bf{s}_t}.
    \end{align}\ES
    From \eqref{Eqn:ps}, for any $\bf{s}_t$, $p(x|\bf{s}_t) =  p(x|s_t)$ if and only if
    \begin{align}
         \bf{V}_t^{{-1}}\bf{1}    = \big[0, \dots, 0, v_{t,t}^{-1}\big]^{\rm T},
    \end{align}
    which is equivalent to    
    \begin{align}
         \bf{V}_t \big[0, \dots, 0, 1\big]^{\rm T} =  v_{t,t}\bf{1},
    \end{align}
    i.e., 
    \BE
        v_{t,i}=v_{i,t}=v_{t,t}, \quad \forall 1\le i\le t,
    \EE 
    Therefore, we complete the proof of Lemma \ref{Lem:SS_Est}. 

{\section{Proof of Lemma \ref{Lem:LMMSE-SS}}\label{APP:LMMSE-SS}
Following Lemma \ref{Lem:IIDG_MIP}, we have
\BS\label{Eqn:ite2SE_phi}\begin{align}
    {\mr{lmmse}} \{{\bf{x}}|\bf{y}, {\bf{X}}_t\} \overset{\rm a.s.}{=} \overline{\mr{lmmse}}\{\bar{\bf{x}}|\bar{\bf{y}},\bar{\bf{X}}_t\} \\
    {\mr{lmmse}}\{{\bf{x}}|\bf{y},{\bf{x}}_t\} \overset{\rm a.s.}{=} \overline{\mr{lmmse}}\{\bar{\bf{x}}|\bar{\bf{y}},\bar{\bf{x}}_t\}
\end{align}\ES
 where $\bar{\bf{y}}=\bf{A}\bar{\bf{x}}+\bar{\bf{n}}$, $\bar{\bf{X}}_t=\bar{\bf{X}}+\bar{\bf{F}}_t$,  and $\bar{\bf{F}}_{t}$ is row-wise IID, column-wise jointly Gaussian, and independent of $\bar{\bf{n}}$, i.e., 
    \BE
        [\bar{f}_{i,1},\dots \bar{f}_{i,t}] \overset{\rm IID}{\sim} \mathcal{CN}{(\bf{0}, \bf{V}_t^\phi)},\;\; i=1,\dots t.
    \EE 

Define $\bf{\zeta}_t\equiv[\zeta_1,\dots,\zeta_t]^{\rm T}$ and  $\bf{\zeta}^{\star}_t\equiv[\zeta^{\star}_1,\dots,\zeta^{\star}_t]^{\rm T}$.

\begin{lemma}\label{Lem:LMMSE_all_damp}
Any optimal solution $\{\bf{Q}^{\star},\bf{P}^{\star},\{\zeta_i^{\star}\}\}$ of
\BS\begin{eqnarray}
      \mathcal{P}_1:\min_{\bf{Q},\bf{P},\{\zeta_i\}} &\mathbb{E}\big\{\|\bf{Q}\bar{\bf y}+\bf{P}\sum_{i=1}^{t}\zeta_i \bar{\bf{x}}_i-\bar{\bf{x}}\|_2^2\big\}\\
    \textit{s.t.} &\bf{QA}+\bf{P}=\bf{I},\\
    &\sum_{i=1}^{t}\zeta_i=1,
\end{eqnarray}\ES
derives an optimal solution of
\BS\begin{eqnarray}
    \mathcal{P}_2:\min_{\bf{Q},\bf{P}_{1:t}} & \mathbb{E}\big\{\|\bf{Q}\bar{\bf y}+\sum_{i=1}^{t}\bf{P}_i \bar{\bf{x}}_i-\bar{\bf{x}}\|_2^2\big\}\\
    \textit{s.t.} &\bf{QA}+\sum_{i=1}^{t}\bf{P}_i=\bf{I},
\end{eqnarray}\ES
by letting $\bf{Q}=\bf{Q}^{\star}$ and $\bf{P}_i=\zeta_i^{\star}\bf{P}^{\star}$. 
\end{lemma}
\begin{IEEEproof}
    See Appendix \ref{APP:proof_LMMSE_damp_all}.
\end{IEEEproof}

\begin{lemma}\label{Lem:P_1_decouple}
 Solving $\mathcal{P}_1$ is equivalent to solving
\BS\begin{eqnarray}
    \mathcal{P}_3:\min_{\bf{Q}, \bf{P}} & \mathbb{E}\big\{\|\bf{Q}\bar{\bf y}+\bf{P} \bar{\bf{X}}_t\bf{\zeta}^{\star}_t -\bar{\bf{x}}\|_2^2\big\}\\
    \textit{s.t.} &\bf{QA} + \bf{P}=\bf{I},
\end{eqnarray}\ES
where $\bf{\zeta}^{\star}_t$ is an optimal solution of
\BS\begin{eqnarray} 
    \mathcal{P}_4:\min_{\bf{\zeta}_t} & \bf{\zeta}_t^{\rm H}\bf{V}_t^\phi\bf{\zeta}_t \\
    \textit{s.t.} &\bf{\zeta}_t^{\rm T}\bf{1}=1,
\end{eqnarray}\ES
by letting $\bf{Q}=\bf{Q}^{\star}$ and $\bf{P}_i=\zeta_i^{\star}\bf{P}^{\star}$. 
\end{lemma}

\begin{IEEEproof}
   See Appendix \ref{APP:P_1_decouple}. 
\end{IEEEproof}

Following Lemma \ref{Lem:useless_damp} (see also Appendix \ref{APP:damp_useless}),  when $\{v^\phi_{t,i}=v^\phi_{i,t}=v^\phi_{t,t},  \forall 1\le i\le t\}$, $\bf{\zeta}^{\star}_{t}=[0 \dots 0, 1]$ is an optimal solution of $\mathcal{P}_4$. Therefore, $\bar{\bf{X}}_t\bf{\zeta}^{\star}_t=\bar{\bf{x}}_t$. Then, following Lemma \ref{Lem:LMMSE_all_damp} and Lemma \ref{Lem:P_1_decouple}, we know that any optimal solution $\{\bf{Q}^{\star},\bf{P}^{\star}\}$ of
\BS\begin{eqnarray}
    \mathcal{P}_5:\min_{\bf{Q}, \bf{P}} & \mathbb{E}\big\{\|\bf{Q}\bar{\bf y}+\bf{P} \bar{\bf{x}}_t -\bar{\bf{x}}\|_2^2\big\}\\
    \textit{s.t.} &\bf{QA} + \bf{P}=\bf{I},
\end{eqnarray}\ES
 derives an optimal solution of $\mathcal{P}_2$ by letting $\bf{Q}=\bf{Q}^{\star}$, $\bf{P}_t=\bf{P}^{\star}$ and $\bf{P}_1=\dots=\bf{P}_{t-1}=\bf{0}$. That is, the  minimums of the objective functions in problems $\mathcal{P}_2$ and $\mathcal{P}_5$ are the same. Following the definition of  $\overline{\mr{lmmse}}(\cdot)$ in \eqref{Eqn:EV_LMMSE}, we have
    \BE
           \overline{\mr{lmmse}}\{\bar{\bf{x}}|\bar{\bf{y}}, \bar{\bf{X}}_t\} =\overline{\mr{lmmse}}\{\bar{\bf{x}}|\bar{\bf{y}},\bar{\bf{x}}_t\}.
    \EE
Furthermore, following \eqref{Eqn:ite2SE_phi}, we have 
    \BE
           {\mr{lmmse}}\{{\bf{x}}|{\bf{y}}, {\bf{X}}_t\}\overset{\rm a.s.}{=} {\mr{lmmse}}\{{\bf{x}}|{\bf{y}},{\bf{x}}_t\}.
    \EE
Then, following Definition \ref{Def:LMMSE-SS}, ${\bf{x}}_t$ is an LMMSE sufficient statistic of ${\bf{x}}$ given ${\bf{X}}_t$. Hence, we complete the proof of Lemma \ref{Lem:LMMSE-SS}.

\subsection{Proof of Lemma \ref{Lem:LMMSE_all_damp}}\label{APP:proof_LMMSE_damp_all}
Before proving Lemma \ref{Lem:LMMSE_all_damp}, we first recall the following property of Kronecker products.
\begin{lemma} [Theorem 4.2.12 in \cite{Johnson1991}]\label{Lem:Kronecker}
    Let $\bf{A}\in \mathbb{C}^{m\times m}$ and $\bf{B}\in \mathbb{C}^{n\times n}$. Furthermore, let $\lambda_A$ be an eigenvalue of $\bf{A}$ with corresponding eigenvector $\bf{\xi}_A$, and let $\lambda_B$ be an eigenvalue of $\bf{B}$ with corresponding eigenvector $\bf{\xi}_B$. Then $\lambda_A\lambda_B$ is an eigenvalue of $\bf{A} \otimes \bf{B}$ with corresponding eigenvector $\bf{\xi}_A\otimes\bf{\xi}_B$, where $\otimes$ denotes the Kronecker matrix product. Any eigenvalue of $\bf{A} \otimes \bf{B}$ arises as such a product of eigenvalues of $\bf{A}$ and $\bf{B}$.
\end{lemma}
We denote $\bf{\mathbb{P}}_t$ as the vertical concatenation of $\{\bf{P}_i\}$, \textit{i.e.},
\BE
    \mathbb{\bf{P}}_t\triangleq[\bf{P}_1^{\rm H} \cdots \bf{P}_t^{\rm H}]^{\rm H}.
\EE
Let $\lambda\geq 0$ be a number that $\bf{V}_t^\phi-\lambda\bf{1}$ is positive semidefinite. Such $\lambda$ exists for any covariance matrix $\bf{V}_t^\phi$, since $\bf{V}_t^\phi$ is positive definite.

We simplify the objective function in $\mathcal{P}_2$ as follows. 
 \begin{align}
        &\mathbb{E}\big\{\|\bf{Q}\bar{\bf y} + \textstyle\sum_{i=1}^{t}\bf{P}_i \bar{\bf{x}}_i - \bar{\bf{x}}\|_2^2\big\} \nonumber\\
        &=\mathbb{E}\big\{\|\bf{Qn} + \textstyle\sum\limits_{i=1}^{t}\bf{P}_i \bar{\bf{f}}_i + (\bf{QA} + \sum\limits_{i=1}^{t}\bf{P}_i-\bf{I})\bar{\bf{x}}\|_2^2\big\} \nonumber\\
        &=\sigma^2{\rm tr}\big\{\bf{Q}^{\rm H}\bf{Q}\big\} + {\rm tr}\big\{\textstyle\sum_{i=1}^t\textstyle\sum_{j=1}^tv_{ij}^\phi\bf{P}_i^{\rm H}\bf{P}_j\big\}\nonumber\\
        &=\sigma^2{\rm tr}\big\{\bf{Q}^{\rm H}\bf{Q}\big\} + {\rm tr}\big\{\mathbb{\bf P}^{\rm H}_t[\bf{V}_t^\phi\otimes\bf{I}]\mathbb{\bf P}_t\big\}\nonumber\\
        &=\sigma^2{\rm tr}\big\{\bf{Q}^{\rm H}\bf{Q}\big\} + {\rm tr}\big\{\lambda\mathbb{\bf P}_t^{\rm H}[\bf{1}\otimes\bf{I}]\mathbb{\bf P}_t\big\} \nonumber\\
         &\hspace{2.5cm} + {\rm tr}\big\{\mathbb{\bf P}^{\rm H}_t[(\bf{V}_t^\phi-\lambda\bf{1})\otimes\bf{I}]\mathbb{\bf P}_t\big\}\nonumber\\
        &=\sigma^2{\rm tr}\big\{\bf{Q}^{\rm H}\bf{Q}\big\}+\lambda\cdot {\rm tr}\big\{(\textstyle\sum_{i=1}^{t}\bf{P}_i)^{\rm H}(\textstyle\sum_{i=1}^{t}\bf{P}_i)\big\} \nonumber\\
         &\hspace{2.5cm} +{\rm tr}\big\{\mathbb{\bf P}^{\rm H}_t[(\bf{V}_t^\phi-\lambda\bf{1})\otimes\bf{I}]\mathbb{\bf P}_t\big\}.
    \end{align}
Therefore, the following problem $\mathcal{P}_6$ is equivalent to $\mathcal{P}_2$.
\begin{equation}\begin{aligned}
    \mathcal{P}_6:\min_{\bf{Q},\bf{P},\bf{P}_{1:t} } \quad &\sigma^2{\rm tr}\big\{\bf{Q}^{\rm H}\bf{Q}\big\}+\lambda{\rm tr}\big\{\bf{P}^{\rm H}\bf{P}\big\} \\
     & \hspace{0.5cm} + {\rm tr}\big\{\mathbb{\bf P}^{\rm H}_t[(\bf{V}_t^\phi-\lambda\bf{1})\otimes\bf{I}]\mathbb{\bf P}_t\big\}\\
    \textit{s.t.}\quad  &\qquad \bf{QA}+\bf{P}=\bf{I}, \\ 
    &\qquad \bf{P}=\textstyle\sum_{i=1}^{t}\bf{P}_i.
\end{aligned}\end{equation}

Since $\bf{V}_t^\phi-\lambda\bf{1}$ is positive semidefinite, from Lemma \ref{Lem:Kronecker}, $(\bf{V}_t^\phi-\lambda\bf{1})\otimes\bf{I}$ is also positive semidefinite and its any eigenvectors corresponding to eigenvalue 0 can be decomposed to $\bf{w}\otimes\bf{\eta}$ where $\bf{w}\in \mathbb{C}^{t\times 1}$ is an eigenvector of $\bf{V}_t^\phi-\lambda\bf{1}$ corresponding to eigenvalue 0 and $\bf{\eta}\in \mathbb{C}^{N\times 1}$ is an arbitrary vector. Therefore, we have
\begin{equation}
    {\rm tr}\big\{\mathbb{\bf P}^{\rm H}_t[(\bf{V}_t^\phi-\lambda\bf{1})\otimes\bf{I}]\mathbb{\bf P}_t\big\}\geq 0,
\end{equation}
where the equality takes when each column of $\mathbb{\bf P}_t$ is an eigenvector of $\bf{V}_t^\phi-\lambda\bf{1}$ corresponding to eigenvalue 0.

For any given $\bf{P}$, let $\bf{P}_i=\zeta^{\star}_i\bf{P}$ where $\sum_{i=1}^t\zeta^{\star}_i=1$ and $\bf{\zeta}^{\star}$ is an eigenvector of $\bf{V}_t^\phi-\lambda\bf{1}$. Obviously, it satisfies the conditions of $\mathcal{P}_6$, and the objective function takes the minimum regarding $\{\bf{P}_i\}$. Moreover, the objective function also takes the minimum at $\bf{\zeta}^{\star}$ regarding $\bf{\zeta}$ when $\bf{P}_i=\zeta_i\bf{P}$ and $\sum_{i=1}^t\zeta_i=1$. As a result, each solution to problem $\mathcal{P}_1$ derives a solution to problem $\mathcal{P}_2$ by $\bf{P}_i=\zeta_i\bf{P}$, which completes the proof of Lemma \ref{Lem:LMMSE_all_damp}.

\subsection{Proof of Lemma \ref{Lem:P_1_decouple}}\label{APP:P_1_decouple}
We demonstrate that $\mathcal{P}_1$ can be solved by independently solving $\mathcal{P}_3$ and $\mathcal{P}_4$.  The objective function of $\mathcal{P}_1$ can be transformed to
\begin{equation}
    \begin{aligned}
    &\mathbb{E}\big\{\|\bf{Q}\bar{\bf y}+\bf{P}\textstyle\sum_{i=1}^{t}\zeta_i \bar{\bf{x}}_i-\bar{\bf{x}}\|_2^2\big\}\\
    &=\sigma^2{\rm tr}\big\{\bf{Q}^{\rm H}\bf{Q}\big\} + {\rm tr}\big\{\bf{\zeta}_t^{\rm H}\bf{V}_t^\phi\bf{\zeta}_t\bf{P}^{\rm H}\bf{P}\big\}\\
    &=\sigma^2{\rm tr}\big\{\bf{Q}^{\rm H}\bf{Q}\big\} + \bf{\zeta}_t^{\rm H}\bf{V}_t^\phi\bf{\zeta}_t{\rm tr}\big\{\bf{P}^{\rm H}\bf{P}\big\}.
\end{aligned}\end{equation}
As we always have ${\rm tr}\big\{\bf{P}^{\rm H}\bf{P}\big\}\geq 0$, problem $\mathcal{P}_1$ can be decomposed to two problems regarding $\{\bf{Q},\bf{P}\}$ (problem $\mathcal{P}_3$) and $\bf{\zeta}_t$ (problem $\mathcal{P}_4$). 
}

\section{Proof of Lemma \ref{Lem:LMMSE_orth}}\label{APP:orth_LMMSE}

For simplicity, define $\hat{\bf{r}}_t\equiv  \mr{LMMSE}\{\bar{\bf{x}}|\bar{\bf{y}},\bar{\bf{x}}_{t}\} = \bar{\bf{x}} + \hat{\bf{g}}_t$.  Following Lemma \ref{Lem:IIDG_MIP}, we have 
     \begin{align}\label{Eqn:ite_se_lmmse}
          & \big\langle   \mr{LMMSE}\{\bf{x}|\bf{y},\bf{x}_{t}\} \! - \!\bf{x}\big|  \mr{LMMSE}\{\bf{x}|\bf{y},\bf{x}_{t}\} \!- \! \gamma(\bf{y}, \bf{X}_t) \big\rangle \nonumber\\
         & \!\overset{\rm a.s.}{=} \mathbb{E}\big\{ \big\langle   \hat{\bf{r}}_t - \bar{\bf{x}} \big|  \hat{\bf{r}}_t - \gamma(\bar{\bf{y}}, \bar{\bf{X}}_t) \!\big\rangle\big\},
     \end{align}     
   where $\bar{\bf{y}}=\bf{A}\bar{\bf{x}}+\bar{\bf{n}}$ and $\bar{\bf{X}}_t=\bar{\bf{X}}+\bar{\bf{F}}_t$, and $\bar{\bf{F}}_{t}$ is independent of $\bar{\bf{n}}$ and for $i=1,\dots t$, 
    \BE
        [\bar{f}_{i,1},\dots \bar{f}_{i,t}] \overset{\rm IID}{\sim} \mathcal{CN}{(\bf{0}, \bf{V}_t^\phi)}.
    \EE 
     
  Let $\hat{\bf{r}}_t = \bar{\bf{x}} + \hat{\bf{g}}_t$. Consider the following unbiased linear estimation
    \BS\begin{align}
        \tilde{\gamma}(\bar{\bf{y}},\bar{\bf{X}}_t) & = (\alpha+1) \hat{\bf{r}}_t - \alpha\gamma(\bar{\bf{y}},\bar{\bf{X}}_t) \\
        &=  \bar{\bf{x}} +(\alpha +1)   \hat{\bf{g}}_t -\alpha \bar{\bf{g}}_t,
    \end{align}\ES
    whose MSE is given by       
      \begin{align}\label{Eqn:const_MSE}
          \tfrac{1}{N} \mathbb{E}\big\{  \|  \alpha (\hat{\bf{g}}_t -\bar{\bf{g}}_t) +\hat{\bf{g}}_t   \|^2\big\}. 
    \end{align}
    Since $\bf{x}_{t}$ is an LMMSE sufficient statistic of $\bf{x}$ given $\bf{X}_t$, from Definition \ref{Def:LMMSE-SS}, we know that $\mr{LMMSE}\{\bar{\bf{x}}|\bar{\bf{y}},\bar{\bf{x}}_{t}\}$ achieves the lowest MSE among all the unbiased linear estimates $\gamma(\bar{\bf{y}},\bar{\bf{X}}_t)$. Therefore,  $ \mathbb{E}\big\{ \|  \alpha (\hat{\bf{g}}_t -\bar{\bf{g}}_t) +\hat{\bf{g}}_t   \|^2\}\geq     \mathbb{E}\big\{\| \hat{\bf{g}}_t   \|^2\big\}$
   holds for any $\alpha\in \mathbb{C}$, i.e., $\alpha=0$ minimizes \eqref{Eqn:const_MSE}, which happens if and only if (see Appendix \ref{Sec:comp_min})
   \BE\label{Eqn:se_orth}
       \mathbb{E}\big\{ \big\langle \hat{\bf{g}}_t\big|  \hat{\bf{g}}_t - \bar{\bf{g}}_t\big\rangle\big\} = 0.
   \EE
   Following \eqref{Eqn:se_orth} and \eqref{Eqn:ite_se_lmmse}, we get the desired  \eqref{Eqn:orth_mmse_g}. Otherwise, we can always find a $\alpha\in \mathbb{C}$ such that $  \mathbb{E}\big\{ \|  \alpha (\hat{\bf{g}}_t -\bar{\bf{g}}_t) +\hat{\bf{g}}_t   \|^2\} <  \mathbb{E}\big\{ \| \hat{\bf{g}}_t   \|^2\}$, which is contradictory to the condition that ``$\mr{LMMSE}\{\bar{\bf{x}}|\bar{\bf{y}},\bar{\bf{x}}_{t}\}$ achieves the lowest MSE among all the unbiased linear estimates  $\gamma(\bar{\bf{y}},\bar{\bf{X}}_t)$". Hence, we complete the proof of  Lemma \ref{Lem:LMMSE_orth}. 
   
\subsection{Proof of \texorpdfstring{\eqref{Eqn:se_orth}}{TEXT}}\label{Sec:comp_min}
First, we show ``$\alpha = 0$ minimizes $ \mathbb{E}\big\{  \| \alpha (\hat{\bf{g}}_t -{\bf{g}}_t) +\hat{\bf{g}}_t \|^2\big\}$ if $\mathbb{E}\big\{ \big\langle \hat{\bf{g}}_t\big|  \hat{\bf{g}}_t - {\bf{g}}_t\big\rangle\big\}= 0$". Following $\mathbb{E}\big\{ \big\langle \hat{\bf{g}}_t\big|  \hat{\bf{g}}_t - {\bf{g}}_t\big\rangle\big\} =0$, we have
    \begin{align*}
       \mathbb{E}\big\{ \!\| \alpha (\hat{\bf{g}}_t - {\bf{g}}_t) \!+\!\hat{\bf{g}}_t \|^2\big\} =
        \mathbb{E}\big\{ \| (\hat{\bf{g}}_t - {\bf{g}}_t) \|^2 \|\alpha\|^2  \!+ \!\| \hat{\bf{g}}_t \|^2\big\},
    \end{align*}
    which is obviously minimized by $\alpha=0$.

    Then, we prove ``$\mathbb{E}\big\{ \big\langle \hat{\bf{g}}_t\big|  \hat{\bf{g}}_t - {\bf{g}}_t\big\rangle\big\} = 0$ if $\alpha = 0$ minimizes $\mathbb{E}\big\{ \| \alpha (\hat{\bf{g}}_t -{\bf{g}}_t) +\hat{\bf{g}}_t \|^2\big\}$".  Let $\alpha = \Re(\alpha) + \Im(\alpha)\mathrm{i}$, where $\Re(\alpha)$ and $\Im(\alpha)$ denote the real and imaginary parts of $\alpha$, respectively. Then,
    \BS    \begin{align}
     \!\!\! \!\!\!& h(\alpha)   \equiv\mathbb{E}\big\{ \| \alpha (\hat{\bf{g}}_t -{\bf{g}}_t) +\hat{\bf{g}}_t \|^2 -  \| \hat{\bf{g}}_t \|^2\big\}\\
    \!\!\!  \!\!\!  &= \mathbb{E}\big\{  \| \alpha ( \hat{\bf{g}}_t - {\bf{g}}_t) \|^2 + 2\Re\{\alpha\hat{\bf{g}}_t^{\rm H}(\hat{\bf{g}}_t-\bf{g}_t)\}\big\}\\
   \!\!\!  \!\!\!   &=\mathbb{E}\big\{  \| \hat{\bf{g}}_t-\bf{g}_t \|^2 \Re(\alpha)^2 + 2N\Re\big\{\big\langle \hat{\bf{g}}_t\big| \hat{\bf{g}}_t - {\bf{g}}_t\big\rangle\big\}\Re(\alpha) \nonumber\\
    \!\!\! \!\!\!   & \quad +  \| \hat{\bf{g}}_t\!-\!\bf{g}_t \|^2 \Im(\alpha)^2\! - \!2N\Im\big\{\big\langle \hat{\bf{g}}_t\big| \hat{\bf{g}}_t\! -\! {\bf{g}}_t\big\rangle\big\}\Im(\alpha)\big\}.
    \end{align} \ES
     If $\mathbb{E}\big\{ \big\langle \hat{\bf{g}}_t\big| \hat{\bf{g}}_t - {\bf{g}}_t\big\rangle \}\neq 0$  , then $\mathbb{E}\big\{ \Re\{\langle \hat{\bf{g}}_t\big| \hat{\bf{g}}_t - {\bf{g}}_t\rangle\}\big\}$ and $\mathbb{E}\big\{ \Im\{\langle \hat{\bf{g}}_t\big| \hat{\bf{g}}_t - {\bf{g}}_t\rangle\}\big\}$ do not all equal to zeros, and $\mathbb{E}\big\{  \| \hat{\bf{g}}_t-\bf{g}_t \|^2\big\} > 0$. Thus, there always exists $\alpha \in \mathbb{C}$ such that $h(\alpha) < 0 = h(0)$. That is, $\alpha = 0$ does not minimize $h(\alpha)$, and thus it does not minimize $\mathbb{E}\big\{  \| \alpha (\hat{\bf{g}}_t - {\bf{g}}_t) +\hat{\bf{g}}_t \|^2\big\}$. Therefore,  $\mathbb{E}\big\{ \!\big\langle \hat{\bf{g}}_t\big|  \hat{\bf{g}}_t - {\bf{g}}_t\big\rangle\!\big\} {=} 0$ if $\alpha\! =\! 0$ minimizes $\mathbb{E}\big\{  \| \alpha (\hat{\bf{g}}_t -{\bf{g}}_t) +\hat{\bf{g}}_t \|^2\big\}$.

\section{Proof of Lemma \ref{Lem:SS_MAMP_MMSE}}\label{APP:SS_MAMP_MMSE}  
From Lemma \ref{Lem:SS}, to prove Lemma \ref{Lem:SS_MAMP_MMSE}, it is sufficient to prove:
\begin{itemize}
    \item[\textcircled{1}] $\bf{V}^\gamma_t$ is L-banded and $\bf{r}_{t}  = \bf{r}_{t-1}$ if ${\bf{V}}^{\gamma}_{\{{\mathcal{I}}^{\gamma}_{t-1}, t\}}$ is singular given \eqref{Eqn:MAMP_MMSE_g}, and 
    \item[\textcircled{2}] $\bf{V}^\phi_{t+1}$ is L-banded and $\bf{x}_{t+1}  = \bf{x}_{t}$ if ${\bf{V}}^{\phi}_{\{{\mathcal{I}}^{\phi}_{t}, t+1\}}$ is singular given \eqref{Eqn:MAMP_MMSE_f},
\end{itemize}
under the condition that $\{\bf{V}^\gamma_{t-1}, \bf{V}^\phi_{t}\}$ are L-banded and  
 \BS\begin{align} 
   &    \bf{r}_{i+1}  = \bf{r}_{i}, \;{\rm if}\;    {\bf{V}}^{\gamma}_{\{{\mathcal{I}}^{\gamma}_{i}, i+1\}} {\rm \;is\; singular}, \; 1\le i\le t-2, \label{Eqn:rep_r}\\
 &   \bf{x}_{i+1}  = \bf{x}_{i}, \; \!{\rm if}\;   {\bf{V}}^{\phi}_{\{{\mathcal{I}}^{\phi}_{i}, i+1\}} {\rm \;is\; singular}, \; 1\le j\le t-1.\label{Eqn:rep_x}
 \end{align} \ES
Due to the symmetry of \textcircled{1} and \textcircled{2}, we only prove \textcircled{1}, and the same proof applies to \textcircled{2}.

In the following,  given \eqref{Eqn:MAMP_MMSE_g}, we  first prove   $\bf{V}^\gamma_t$ is L-banded  in 1), and then prove $\bf{r}_{t}  = \bf{r}_{t-1}$ if ${\bf{V}}^{\gamma}_{\{{\mathcal{I}}^{\gamma}_{t-1}, t\}}$ is singular in 2).
    
1)  For simplicity, we let
\BE
    \hat{\bf{g}}_t  \equiv \hat\gamma_t(\bf{y},\bf{x}_t) -\bf{x}.
\EE
Then, for all $1\le t'\le t$, 
\BS\label{Eqn:V_t_SS}\begin{align}
     v^{\gamma}_{t',t}  & = v^{\gamma}_{t,t'}\\
     & =   \langle \bf{g}_{t}| \bf{g}_{t'} \rangle  \\
     &  =    \langle c_{t}^\gamma \hat{\bf{g}}_{t} + (1-c_{t}^\gamma) \bf{f}_{t}| \bf{g}_{t'} \rangle \\
    &  \overset{\rm a.s.}{=} c_t^\gamma   \langle  \hat{\bf{g}}_{t}| \bf{g}_{t'} \rangle \label{Eqn:V_t_SS_d} \\
    & = c_{t}^\gamma  \langle   \hat{\bf{g}}_{t}|\gamma_{t'}(\bf{y},\bf{x}_{t'}) - \bf{x} \rangle \\ 
    & \overset{\rm a.s.}{=}  c_{t}^\gamma  \langle   \hat{\bf{g}}_{t}|  \hat\gamma_t(\bf{y},\bf{x}_t) - \bf{x} \rangle \label{Eqn:V_t_SS_f}\\
    &  = c_{t}^\gamma  \langle   \hat{\bf{g}}_{t}|  \hat{\bf{g}}_{t} \rangle \\
    &  \overset{\rm a.s.}{=}  v^{{\gamma}}_{t}, \label{Eqn:V_t_SS_h}
\end{align}\ES
where \eqref{Eqn:V_t_SS_d} follows $  \langle \bf{f}_{t} |\bf{g}_{t'} \rangle \overset{\rm a.s.}{=} 0$ (see \eqref{Eqn:Orth_MAMP}), \eqref{Eqn:V_t_SS_f} follows $\langle\hat{\bf{g}}_{t}|\hat\gamma_t(\bf{y},\bf{x}_t) - \gamma_{t'}(\bf{y},\bf{x}_{t'})\rangle  \overset{\rm a.s.}{=} 0$ based on Lemma \ref{Lem:LMMSE_orth} and the L-banded $\bf{V}^\phi_{t}$,  \eqref{Eqn:V_t_SS_h} follows $ \langle   \hat{\bf{g}}_{t} | \hat{\bf{g}}_{t} \rangle \overset{\rm a.s.}{=}\overline{\mr{lmmse}}\{\bf{x}|\bf{y}, \bf{x}_{t}\} =  v^{{\gamma}}_{t}/ c_{t}^\gamma $. From \eqref{Eqn:V_t_SS} and the L-banded $\bf{V}^\gamma_{t-1}$, we have that $\bf{V}^\gamma_{t}$ is L-banded.  

2) Since ${\bf{V}}^{\gamma}_{{\mathcal{I}}^{\gamma}_{t-1}}$ is invertible (following the definition of ${{\mathcal{I}}^{\gamma}_{t-1}}$) and ${\bf{V}}^{\gamma}_{\{{\mathcal{I}}^{\gamma}_{t-1}, t\}}$ is L-Banded (following \eqref{Eqn:V_t_SS}), we have $v^{\gamma}_{t,t}=v^{\gamma}_{t-1,t-1}$ if ${\bf{V}}^{\gamma}_{\{{\mathcal{I}}^{\gamma}_{t-1}, t\}}$ is singular.
Since $\hat{\gamma}_t$ is a local MMSE function, the variance transfer function $\gamma_t^{\rm SE}$ is a strictly monotonic increasing function \cite[Lemma 2]{Ma2016}. That is, $v^{\gamma}_{t,t}=v^{\gamma}_{t-1,t-1}$ means  $v^{\phi}_{t,t}=v^{\phi}_{t-1,t-1}$, which means that ${\bf{V}}^{\phi}_{\{{\mathcal{I}}^{\phi}_{t-1}, t\}}$ is singular since ${\bf{V}}^{\phi}_t$ is L-Banded. Finally, from \eqref{Eqn:rep_x}, we have $\bf{x}_{t}  = \bf{x}_{t-1}$ and  $\bf{r}_{t}  = \bf{r}_{t-1}$ following \eqref{Eqn:MAMP_MMSE_g}.  


\end{document}